\documentclass[twocolumn,prb,superscriptaddress,showpacs,footinbib]{revtex4}

\usepackage[utf8]{inputenc}
\usepackage[T1]{fontenc}
\usepackage{times}
\usepackage[american]{babel}
\usepackage{graphicx}
\usepackage{xcolor}
\usepackage{booktabs}
\usepackage{amsmath, amssymb}
\usepackage{braket}    
\usepackage{xspace}
\usepackage[unicode]{hyperref}
\renewcommand{\vec}{\boldsymbol}

\renewcommand{\vec}{\boldsymbol}
\newcommand{\db}[2][]{\text{d}^{#1}#2}

\newcommand{\avr}[1]{\braket{#1}}

\newcommand{\abs}[1]{|#1|}
\DeclareMathOperator{\Tr}{Tr}

\begin{document}
\title{Collective and single-particle excitations in 2D dipolar Bose gases}
\date{\today}

\author{A.~Filinov}
\email{filinov@theo-physik.uni-kiel.de}
\affiliation{Institut für Theoretische Physik und Astrophysik,
Christian-Albrechts-Universitat, Leibnizstr. 15, D-24098 Kiel, Germany}
\affiliation{Joint Institute for High Temperatures RAS, Izhorskaya Str.~13, 125412 Moscow, Russia}
\author{M.~Bonitz}
\affiliation{Institut für Theoretische Physik und Astrophysik,
Christian-Albrechts-Universitat, Leibnizstr. 15, D-24098 Kiel, Germany}

\begin{abstract}
The Berezinskii-Kosterlitz-Thouless transition in 2D dipolar systems has been studied recently by path integral Monte Carlo (PIMC) simulations [A.~Filinov {\em et al.}, PRL {\bf 105}, 070401 (2010)]. Here, we complement this analysis and study temperature-coupling strength dependence of the density (particle-hole) and single-particle (SP) excitation spectra both in superfluid and normal phases. The dynamic structure factor, $S(q,\omega)$, of the longitudinal excitations is rigorously reconstructed with full information on damping. The SP spectral function, $A(q,\omega)$, is worked out from the one-particle Matsubara Green's function. A stochastic optimization method is applied for reconstruction from imaginary times. In the {\em superfluid regime} sharp energy resonances are observed both in the density and SP excitations. The involved hybridization of both spectra is discussed. In contrast, in the {\em normal phase}, when there is no coupling, the density modes, beyond acoustic phonons, are significantly 
damped. Our results generalize previous zero temperature analyses based on variational many-body wavefunctions~[F. Mazzanti {\em et al.}, PRL {\bf 102}, 110405 (2009), D. Hufnagl {\em et al.}, PRL {\bf 107}, 065303 (2011)], where the underlying physics of the excitation spectrum and the role of the condensate has not been addressed. 
\end{abstract}


\maketitle

\section{Introduction}\label{intro}

Dipolar bosonic systems are of increasing interest for recent experiments studying the onset of superfluidity in nonideal Bose systems and its connection with correlation and quantum degeneracy effects. Examples include dipolar gases, as in recent studies of ultra-cold dipolar gases,~\cite{atomic1,atomic2,pfau,baranov,dysp,erbium} indirect excitons in coupled quantum wells~\cite{exciton,d1,d2,d3} in external electric fields or exciton-polaritons in quantum wells embedded in optical micro-cavities.~\cite{polariton1,polariton2} Due to significant experimental achievements, a superfluid regime of polar molecules is expected to be observed in the near future.~\cite{hmol1,hmol2,hmol3,hmol4,hmol5,41K,rydberg_mol} There is also an increasing activity on alkali atom gases where the long-long range dipolar interactions can be generated by excitations to high energy atomic Rydberg states (one electron is excited to a very high principal quantum number). The applied moderate electric field can result in a large 
polarizability and dipole moment.~\cite{saffman,pupilo} All these achievements have a direct impact on understanding the properties of Bose-condensates dominated by long-range correlation effects. 

From the theoretical side, the static properties of dipolar bosons have been quite extensively analyzed in recent years. Strongly correlated phases of dipolar bosons confined in a harmonic potential have been studied from first principle simulations by Lozovik {\em et al.},~\cite{loz} Nho {\em et al.},~\cite{nho} Pupillo {\em et al.,}~\cite{pupilo} Golomedov {\em et al.}~\cite{golomedov} and Jain {\em et al.}~\cite{cinti} 2D homogeneous systems have been analyzed by Astrakharchik {\em et al.}~\cite{ast1} and B\"uchler {\em et al.}~\cite{buch} with the prediction that above a critical density a superfluid dipolar gas undergoes a crystallization transition. Properties of a dipole solid have been addressed by Kurbakov {\em et al.}~\cite{kurbakov} The presence of a finite fraction of vacancies and interstitials, in incommensurate crystal, leads to a superfluid response and quasi-equilibrium supersolid phase. In PIMC simulations by Filinov {\em et al.}~\cite{fil2010} the density-dependence of the transition 
temperature from a superfluid to a normal gas phase has been analyzed. The composite dipolar bosons, such as indirect excitons (pairs of electrons and holes spatially separated in semiconductor bilayers) have been recently studied by B\"oning {\em et al.}~\cite{filex} with focus on the effective exciton-exciton interaction and its consequence for the complete phase diagram of excitonic system. Relevant experimental parameters to observe exciton crystallization, in a semiconductor quantum well, have been predicted. All these results provides a reliable estimate of the critical temperature and degeneracy parameters required to observe a combined effect of inter-particle correlations and Bose statistics in experiment.

On the other side, theoretical predictions for dynamic properties, in particular excitation spectra of density fluctuations, are more challenging if performed on a microscopic level. For 3D dipolar superfluids at low densities the rotonization of the excitation spectrum has been predicted from mean-field analyses.~\cite{santos2003,fisher2006,sron,wilson} A cigar-shaped trap is a promising candidate for observation of a roton minimum, where the dipole-dipole interaction is anisotropic, and partially attractive. In a quasi 2D trap (a pancake geometry), a roton has a different physical nature and is due to a strong in-plane repulsion of similar oriented dipoles. This regime assumes a high density and goes beyond the mean-field predictions. This calls for a more accurate treatment. The correlated basis functions theory (CBF), originally developed for strongly correlated systems like $^4$He, has been renewed recently by Mazzanti {\em et al.}~\cite{Mazz} to analyze in detail the phonon-roton dispersion of an 
infinite 2D dipolar gas at $T=0$. The density range has been identified, where the CBF dispersion (which includes three-phonon interactions) deviates from the Bijl-Feynman~\cite{ast1,huf2} and Bogolubov predictions. The  upper bound for the phonon-maxon-roton dispersion, derived from the frequency-sum rules,~\cite{Plaz,lifbook,sting92} is mainly in a good agreement.~\cite{fil2010} Important extension of the CBF theory to 3D geometry and full anisotropy of the dipole-dipole interaction has been performed by Hufnagl {\em et al.}~\cite{huf} This study builds an important bridge between the nature of two kinds of rotons caused either by attractive or repulsive tail of the dipole interaction. Still the current analyses of the excitation spectrum lack the important regime of finite temperatures (including $T\sim T_c$) and treat excitation damping in an approximate way.~\cite{cbf1,cbf2,cbf3} With the present analysis we aim to fill this gap and, in addition, discuss the nature of excitations in Bose-condensed 
systems and the role played by a condensate.

We start the discussion about the nature of excitations from the model developed for a one-component neutral BEC.

The properties of the density fluctuation spectrum have been discussed extensively in relation with superfluid liquid $^4$He. In particular, the temperature ($T$) dependence of $S(q,\omega)$ from inelastic-neutron scattering experiments was compared vs. Glide-Griffin (GG) model based on the dielectric function formalism.~\cite{noz,grif1990} The relation with the dynamic structure factor is written as
\begin{align}
 &S(q,\omega)=-\frac{1}{\pi} \frac{1}{1-e^{-\beta \omega}} \Im [\chi(q,\omega)]\label{sq},\\
 &\chi(q,\omega)=\frac{\tilde\chi(q,\omega)}{1-V(q)\tilde\chi(q,\omega)},\label{chi}\\
 &\tilde\chi(q,\omega)=\Lambda(q,\omega) G_1(q,\omega)\Lambda(q,\omega)+\tilde\chi_{\rho}(q,\omega)\label{tchi}
\end{align}
where $\tilde\chi$ is the irreducible part of dynamic susceptibility $\chi$, separated into a singular condensate part (related to single-particle excitations) and a regular thermal (multiparticle) component, present also above $T_c$. The coupling to the SP spectrum is described by the vertex function $\Lambda \propto \sqrt{n_0(T)}$ and the single-particle Green function $G_1$. The first term in (\ref{tchi}) vanishes in the normal phase ($T>T_c$), with the result $\tilde\chi=\tilde\chi_{\rho}$, with $\tilde\chi_{\rho}$ being the density response of weakly interacting particle-hole excitations, e.g. treated via a Lindhard function. 

The excitation spectrum described by $\chi(q,\omega)$ is expected to include for {\em small $q$} a zero sound and acoustic phonons, for {\em intermediate $q$} a maxon-roton branch, and for {\em large $q$} a combination of sharp SP resonances and a broad distribution close to the recoil energy $\epsilon_q=q^2/2m$ due to the multiple scattering of weakly interacting particle-hole excitations. 

To fit the experimental observations for $^4$He  by Eqs.(\ref{sq})-(\ref{tchi}), i.e. a {\em single phonon-maxon} dispersion, one suggests that the imaginary parts of $\chi$ and $G_1$ share a common denominator due to the vertex $\Lambda(q,\omega)$.~\cite{glyde92} As a result the SP and collective excitations can not be distinguished in $S(q,\omega)$ in the superfluid phase. For $T>T_c$ the SP spectrum decouples and only the thermal part remains ($\chi=\tilde\chi_{\rho}$). The dispersion gets broadened and remains well defined only in the phonon (low-$q$) part of the spectrum. 

In its simple form the dielectric function model failed to predict the double roton feature observed in $^4$He. Therefore, it was necessary to include the mechanism of the quasiparticle-decay processes. This has lead to development of theories beyond the roton minimum.~\cite{pit59,zaw0,zaw,jac,bedell,rot2,rot3} In his pioneer work, Pitaevskii~\cite{pit59} considered a semi-emperical ansatz for a single-particle Green's function 
\begin{align}
 G_1(q,\omega)\sim \frac{1}{\omega^2 - \omega_q^2 -2 \omega_q \Sigma^{12}(q,\omega)}\label{gp}
\end{align}
which includes the unhybridized Feynman-Cohen phonon-maxon-roton branch,~\cite{fc} $\omega_q=\omega_q^{\text{FC}}$, and the self energy, $\Sigma^{12}= n_0 J\chi_2 J$, expressed in terms of the two-(quasi)particle response function $\chi_2$ and the three-point interaction vertex $J$. The self energy accounts for the typical repulsion by anticrossing of two modes, i.e. the interaction between a single and a pair of quasiparticle excitations. As a result the SP dispersion, beyond the roton, is renormalized and bends to the energy of two roton minimum. This concept has been further successfully developed~\cite{zaw0,jac,bedell,rot2,zaw,sakhel} by improving both $\omega_q^{\text{FC}}$ and $\Sigma^{12}$.

In particular, Zawadowski et al.~\cite{zaw} developed the theory of a bound two-roton state. Due to hybridization with this state the dispersion splits into two branches. The upper branch consists of heavily damped density excitations with the energy close to the recoil energy $\epsilon_q$. The low-energy branch is due to the SP excitations and saturates slightly below the energy of twice the roton minimum. At low $T$ with increasing $q$ its weight is transfered to the upper branch. While Zawadowski's model provides a good description for low-temperature $^4$He data by using several fit parameters, it becomes unsatisfactory at high temperatures. The expected decrease of the intensity of the lower (SP) branch, by vanishing of the condensate $n_0(T)$, is not compensated by contribution of atoms above a condensate. This is in a contradiction with the sum rule, $S_L(q)+S_H(q)=S(q)$, as the static structure factor $S(q)$ for $^4$He was found to be nearly temperature independent for $T \sim T_c$.~\cite{griffinbook}
 Further, in some range of $q$-wavevectors a fit to the $T$-dependent experimental $S(q,\omega)$-data resulted in the repulsive interaction between two roton excitations,~\cite{fak} opposite to the negative coupling constant suggested in the original model. This point, has been resolved by a more elaborated theory~\cite{bedell} using a T-matrix which allowed the coupling constant to oscillate with $q$.

The dielectric function models interpret the roton and the two-roton state as a specific feature of the SP spectrum coupled in $S(q,\omega)$ in a superfluid phase. The standard procedure to follow is to fit the involved key parameters ($\omega_q,J,\chi_2$) of the Green's function~(\ref{gp}) to be consistent with the low temperature $S(q,\omega)$-data. The $T$-dependence is neglected and enters only via $n_0(T)$ and the Bose function in Eq.~(\ref{sq}). In a similar way, the density response function $\tilde{\chi}_{\rho}$~(\ref{tchi}) is treated also as $T$-independent, and, therefore, is fitted by high temperature data, where the singular condensate part does not contribute.  

The weak point of this model is the discrepancy with experiments near $T_c$. The intensity of the sharp peak (two-roton state) assigned with the SP excitations (poles of $G_1$) becomes too low as its spectral weight vanishes with $n_0(T)$. In contrast, a typical experiment on $^4$He demonstrates a continuous broadening of this peak and nearly conserved integrated spectral weight. Also a broad multi-excitation background, related with $\tilde{\chi}_{\rho}$,  extends significantly to low frequencies, as $T$ is increased, being in contradiction with the model assumption on its temperature independence. These discrepancies have been reviewed in Refs.~\cite{fak,sakhel,sven1,sven2,raman} 

The experiments~\cite{sven1,sven2} provided no indication of a well-defined mode (SP excitations) suddenly appearing in $S(q,\omega)$ for $T < T_c$ due to the first term in Eq.~(\ref{prove1}). In contrast, they demonstrated that temperature and density dependence of the spectral density in the phonon and roton part of the spectrum can be fitted with just a single-mode susceptibility. In particular, the roton mode demonstrates a rapid attenuation when approaching $T_c$ from below and continues as an overdamped diffusive mode of zero frequency above $T_c$.  The authors concluded ``our results cannot be explained using the GG model, unless of course the two components in the GG model hybridize into one (having one lifetime and excitation energy) at all temperatures and pressures, independent of the value of $n_0$''.~\cite{remark} This suggests, that the phonon-roton spectrum must be considered as {\em unified branch} as was done in the original papers by Landau, Feynman and Bogolubov.~\cite{oneb1,oneb2,oneb3} However the origin of the excitations 
and their structure can be different. By rigorous considerations, Nepomnyashchy~\cite{nepom} has demonstrated that a simultaneous presence in the spectrum of both zero sound (ZS) and SP branch is not possible. At $T< T_c$ the ZS is suppressed and replaced by the SP branch defined by poles of the SP Green function. The phonons become identical with the elementary excitations. The SP branch, however, accumulates some properties of the ZS branch, such as linear spectrum and independence on $n_0$. This is in a principle contradiction with the ansatz~(\ref{tchi}) when one uses the substitution $\Lambda \propto \sqrt{n_0(T)}$. Finally, we note that the Bogolubov type spectrum can be also derived without involving the gauge symmetry breaking, i.e. also for a normal phase with the conserved gauge symmetry.~\cite{yukalov} This implies that the phonon-maxon-roton dispersion, observed in a superfluid, should be also preserved in the normal phase, once the damping is small.  
 
For a 2D dipolar gas, we found that its properties near $T_c$ qualitatively follow  that of 3D $^4$He discussed above. From the comparison of $S(q,\omega)$ at $T\sim 0.7 T_c$, we can confirm that the intensity of the two-roton ``plateau'' is unexpectedly too large and does not scale with the zero-temperature condensate fraction once the density is varied. The condensate demonstrates a broad variation~\cite{ast1} ($n_0 =0.70\ldots 0.025$) and this implies, that for $\Lambda \propto \sqrt{n_0(T)}$ in~(\ref{tchi}), the two-roton peak should be strongly suppressed at high densities, if we assume a roton being solely a SP excitation. In our simulations we do not find any evidence of this behavior. The intensity of the peak is not much reduced when approaching $T_c$ either. Formally, in 2D geometry, as considered here, at $T\neq 0$ the condensate fraction should vanish in the thermodynamic limit and no coupling to the SP spectrum is expected. Still the two-roton state is observed for $T\leq T_c$ and even for $T> 
T_c$ at high densities (but strongly damped).  The finite size effects are negligible for the considered system sizes ($N\sim 165, 576$) and, therefore, do not influence the imaginary time correction functions used for the reconstruction of the spectral densities (Sec.~\ref{qcorrf}). All observed characteristic spectrum features should remain also in the thermodynamic limit (except the limit $q \rightarrow 0$). This brings us to a preliminary conclusion that if the two-roton state is a feature of SP spectrum the sufficient condition for the hybridization in $S(q,\omega)$ is the presence of at least off-diagonal quasi-long range order and a non-zero superfluid density.

It is important to mention that, there exists a second viewpoint on roton and different combinations (roton+roton, roton+maxon, maxon+maxon) as the intrinsic density excitations related with the short-range particle correlations.~\cite{roton1,roton2,roton-kalman}

To clarify these important issues one needs to independently access $S(q,\omega)$ and the SP spectral density $A(q,\omega)$ for temperatures below and above $T_c$ to indentify how the excitation branches of both spectra are coupled in the superfluid phase and lead to rotonization. This gives motivation for the present analysis. The path integral Monte Carlo simulations in combination with the stochastic optimization method are used to reconstruct collective and single-particle excitations {\em at finite temperatures} from first principles (in the linear response regime). Both should be coupled or form {\em a unified branch} in the density fluctuation spectrum $S(q,\omega)$ once a system has a condensate or the Bose broken symmetry.~\cite{pines1959} Whether this prediction remains also valid for 2D Bose systems, with the off-diagonal quasi-long range order and only local Bose condensate, remains an open issue and calls for further analysis on the level of microscopic simulations.

Such analyses in application to 2D dipolar gas will be presented in Sec.~\ref{disc_spec}. In particular, we found that at weak and intermediate coupling  (Sec.~\ref{weak},\ref{intermediate}) the single-particle and the collective (density fluctuations) modes have similar dispersion relations in the normal phase (when they are independent). As a result, in the superfluid phase due to the coupling (Eq.~\ref{tchi}) both branches become hybridized. In general, the dispersion relation $\omega_{q}$ observed in $S(q,\omega)$ splits into two branches, but the high energy (multiparticle) part gets significant broadening in the normal phase. At intermediate $q$ (the maxon-roton region) the lower branch is shifted to lower energies and its damping is enhanced for $T > T_c$.

For strongly correlated dipolar gas ($D=12.5$) our results support the viewpoint on the roton as a collective density mode. While, the SP spectrum $A(q,\omega)$ shows a well pronounced roton minimum (denoted as $\tilde{E}_R$), the corresponding energies $\tilde{E}_R\sim 10$ and $2 \tilde{E}_R \sim 20$ are larger than those of the roton-minimum ($E_R\sim 5$) and two-roton state (Pitaevskii ``plateau'' $2 E_R\sim 10$) observed in $S(q,\omega)$. According to the Zawadowski's model, this suggests a strong binding mechanism which currently lacks experimental confirmation in other strongly correlated Bose systems ($^4$He). 

Independent on the presence of the two-roton state, the frequency sum rules (see Appendix~\ref{app}) predict a branch near the recoil energy $\epsilon_q$. Indeed, the simulations reproduce this branch, however, it is strongly damped, as expected from weakly interacting density quasiparticles. This multiparticle feature is present both below and above $T_c$. In contrast to $^4$He, in a dipolar system at high density (e.g. $D=12.5$) the Pitaevskii ``plateau'' appears to be well separated from this multiparticle continuum. Similar free-particle like branch is observed in the SP spectrum $A(q,\omega)$.

Based on our results we interpret the two-roton state as a combination of two density rotons (the poles of the dynamic susceptibility) as we can observe it also for $T> T_c$. The detailed discussion will be presented in Sec.~\ref{intermediate},\ref{strong}.

In Sec.~\ref{model} we introduce the model of a 2D dipolar gas. In Sec.~\ref{physrel} physical realizations are listed and experimental parameters are compared with our model. In Sec.~\ref{intro2} the dynamic structure factor $S(q,\omega)$ is introduced for a system with Bose condensate. In Sec.~\ref{qcorrf} we analyze the structure of the imaginary time correlation functions and their relation with the SP spectral density $A(q,\omega)$ and $S(q,\omega)$. In Sec.~\ref{sto} reconstruction of $A(q,\omega)$ and $S(q,\omega)$ using the stochastic optimization procedure is discussed. In Sec.~\ref{specsec} we analyze temperature/density dependence of the momentum distribution and the spectral densities. In Sec.~\ref{disc_spec} we give our interpretation of the different excitation branches observed in $A(q,\omega)$, $S(q,\omega)$. In Sec.~\ref{sum} we list our conclusions.

\section{Model}\label{model}

Dipolar BEC has been first realized in atomic systems of magnetic dipoles.~\cite{atomic1,atomic2,pfau,baranov,dysp,erbium} New exciting results have been recently reported for dipolar molecules~\cite{hmol1,hmol2,hmol3,hmol4,hmol5} with the major advantage of significantly larger electric dipole moments. The mean field description of such systems predicts that the dipolar condensates in 3D geometry become dynamically unstable when the dipole interaction dominates the $s$-wave scattering contact interaction.~\cite{santos2003,fisher2006,sron,wilson} This effect is a consequence of the rotonization of the Bogolubov-type spectrum~\cite{santos2003,odell,fisher2006} and leads to an instability of the time-dependent Gross-Pitaevskii equation.~\cite{sron} 

To avoid this problem and study the behavior of dipolar systems well beyond the mean-field regime with a stable condensate and/or a superfluid phase, we consider a quasi-2D geometry. The dipole moments are oriented along the direction of the strong trap confinement and produce only repulsive interaction $p^2/r^3$. The dipolar forces are assumed to completely dominate the contact interaction, therefore, the latter is neglected in the present model. We assume 2D spatial homogeneity by considering an infinite pancake trap geometry. The evaluated spectrum for the in-plane momentum $q$ will be valid for the excitation wavelengths, $\lambda_q=2\pi/q$, bounded by the perpendicular $L_z$ and the in-plane $L_{\rho}$ system size, i.e $L_z < \lambda_q < L_{\rho}$. Otherwise, the excitations acquire 3D character. For a quasi-2D model of dipolar condensate,~\cite{fisher2006} the perpendicular extension can be estimated by $L_z\sim l_z=\sqrt{\hbar/m \omega_z}$.

The dipole moment $p=ed$ and particle density $n=1/a^2$ are free parameters controlled by external fields and in-plane trap frequencies. The system Hamiltonian
\begin{equation}
{\hat H} = - \sum_{i=1}^N \frac{\hbar^2\nabla_i^2}{2m} + \frac{1}{2}\sum_{i \ne j} \frac{p^2}{\epsilon_b |\vec{r}_i-\vec{r}_j|^3},
\label{h}
\end{equation}
can be made dimensionless by using the length and energy units: $a=1/\sqrt{n}$ (the mean interparticle distance) and $E_0=\hbar^2 /m a^2$. In the canonical ensemble, the thermodynamic properties of~(\ref{h}) are completely defined by the dipole coupling $D=p^2/\epsilon_b a^3 E_0$, temperature $T=k_B T/E_0$ and particle number $N$. The Bose broken-symmetry condition, i.e. $\avr{\hat{\Psi}(\vec{r})},\avr{\hat{\Psi}^{+}(\vec{r})}\neq 0$, and evaluation of imaginary-time Green's functions suggest to use the grand-canonical ensemble. In this case, the worm algorithm~\cite{prokof} has been employed by fixing a value of the chemical potential $\mu$ and system volume $V$ instead of $N$ (see note~\cite{mudef}). Indeed, existence of the order parameter, $\Phi_0(\vec{r})=\avr{\hat{\Psi}(\vec{r})}$, is related with the fluctuations of a particle number in a condensate state. For a weakly interacting dipolar gas ($D=0.1$), we observed significant particle number fluctuations, $\avr{\Delta N^2}\sim 0.17 \avr{N}$, with a 
condensate fraction, $n_0 \sim 0.7$. This can lead to some differences between the canonical ensemble (experimental BEC realized in traps) and the grand-canonical results. 

The parameters of PIMC simulations are from Ref.~\cite{fil2010} The dipole coupling (or density), $D\sim n^{1/2}$, was varied in the range $0.1 \leq D \leq 12.5$ and temperature $0.7 \leq T \leq 3.3$. With the known $D$-dependence of the critical temperature~\cite{fil2010} ($T_c\approx 0.70 \ldots 1.4$) we scan both the superfluid and the normal phase. 

\section{Physical realizations}\label{physrel}

Next we discuss what $D$ values are accessible in current experimental realizations. Consider the effective radius of dipole-dipole interaction, $a_d=m p^2/\epsilon_b\hbar^2$, corresponding to the distance when the dipolar energy reaches the value of the zero-point kinetic energy, i.e. $p^2/\epsilon_b a_d^3=\hbar^2/m a_d^2$. The introduced dipolar parameter equals the ratio of two characteristic length scales, $D=a_d/a$. Given a value of a magnetic/electric dipole moment (which enters in $a_d$) we can estimate the density  required for a specific $D$. The effective dipole interaction radius can be expressed as
\begin{align}
 a_d[\text{\AA}]=\frac{149.36}{\epsilon_b} m[\text{u}]\, p^2[\text{Debye}^2],\label{adex}
\end{align}
where $m$ is the mass in the unified atomic mass unit, $p$ is the dipole moment in the debye. An external electric field aligns dipolar particles leading to a repulsive $1/r^3$-interaction. Several examples are considered below.

\begin{itemize}
 \item {\em Cold bosonic atoms} with a permanent magnetic moment~\cite{atomic1,atomic2,pfau} in tight pancake traps.
Magnetic dipoles are aligned by a magnetic field. A dipolar gas of $^{52}$Cr with $p=6 \mu_B$ and $n\sim 10^{11}$cm$^{-2}$ ($a\sim 316$~\AA) has $a_d\sim 24$~\AA. The dipolar interaction will dominate, when the s-wave scattering length ($a_s \sim 100 a_B$) is suppressed to $a_s<a_d$ using a high magnetic field Feshbach resonance. The resulted coupling, $D\sim 7.6 \cdot10^{-2}$, is close to the analyzed value $D=0.1$. The superfluid transition temperature is estimated~\cite{fil2010} to be $T_c\sim 12 \mu$K  (in a 2D geometry).

Even higher coupling can be achieved in the Bose gas of dysprosium~\cite{dysp} ($^{164}$Dy) with $p=10 \mu_B$. The evaluation for $n\sim 10^{9}\ldots 10^{11}$cm$^{-2}$ ($a\sim 3160\ldots 316$~\AA) results in $a_d\sim 210.5$~\AA, $D\sim 6.6 \cdot10^{-2} \ldots 0.67$ and $T_c \sim 0.04 \ldots 4 \mu$K. An ultracold dysprosium gas does not suffer from chemical reactions, like certain polar molecules,~\cite{molloss} and, therefore, is stable at high densities.

Very recently, a Bose-Einstein condensate of erbium ($^{168}$Er) has been reported,~\cite{erbium} being another promising candidate for experimental realization of strongly correlated dipolar quantum gases. Similar to dysprosium, erbium has a large magnetic moment $p=7 \mu_B$ and atomic mass which enhance the dipole interaction radius~(\ref{adex}). A large dipole moment also allows to reach a high efficiency both in a magneto-optical trapping and subsequent evaporative laser cooling to temperatures $T \sim 200$mK. 

 \item {\em Polar molecules}~\cite{hmol3,hmol4,hmol5,41K,rydberg_mol} (Rb$_2$, Cs$_2$, $^{41}$K$^{87}$Rb, $^{40}$K$^{87}$Rb) transfered to the rovibrational ground state with a permanent electric dipole moment, $p=0.05 \ldots 0.6$~Debye,~\cite{hmol3} aligned by an electric field. For $^{41(40)}$K$^{87}$Rb gas with $n\sim 10^{9}\ldots 10^{11}$cm$^{-2}$  and $p=0.57$~Debye, we estimate $a_d\sim 6118$~\AA, $D\sim 1.9 \ldots 19$, $T_c(D=1.9)\sim 0.053\mu$K and $T_c(D=15)\sim 3.4\mu$K. High densities and $D$-values can be achieved by localizing Feshbach molecules in an optical lattice to suppress their collisions and then quenching to a ground state via stimulated Raman adiabatic passage.~\cite{41K} Besides molecule collisions, an upper bound for quantum degeneracy and the dipole coupling $D$ is limited by chemical reactions and associated molecule loss rates.~\cite{molloss}

\item {\em Rydberg-dressed atoms} excited to high principal quantum numbers $k$ and confined in two dimensions.~\cite{pupilo,saffman} A large dipole moment can be reached in the linear Stark regime with the result $p=(3/2)e a_B k(k-1)\sim 1450$~Debye for $k=20$. We consider 
an ensemble of ground state atoms excited to high Rydberg states with the single-atom excitation probability $P=(\Omega/\Delta)^2\sim 0.01$ (where $\Omega$ and $\Delta$ are a Rabi-frequency and a detuning laser frequency). The effect of a local dipole blockade,~\cite{dipblock} i.e. a zero probability to excite more than one atom within a distance $R_s$, can be included via the reduced Hamiltonian of ``superatoms'',~\cite{dipblock,dipblock2} with the average superatom interaction $V(r)|_{r>R_s}=p^2/N_s^2 r^3$, where $N_s=n R_s^2$ is a typical number of atoms (in 2D) within a superatom radius $R_s$. The latter depends on the properties of the excitation laser ($P, \Omega, \Delta$), the gas density $n$ and the interaction potential $V(r)$. For estimation, we treat an excitation probability $P(n,R_s)$ of a many-body system (modified by correlation effects and defined by $P(n,R_s) N_s \sim 1$) as that of an uncorrelated gas, i.e. $P(n,R_s)\sim P \sim 0.01$. For a gas of rubidium Rydberg atoms ($^{87}$Rb) with the 
number density $n\sim 10^{9}$cm$^{-2}$ we estimate $N_s\sim 100$ ($R_s \sim 3.1\mu$m), but only a half of the superatoms are interacting as they perform Rabi oscillations between their ground and excited states. This results in the superatom spacing $a\sim \sqrt{2} R_s$, and, correspondingly, $a_d\sim 272 \mu$m and $D= a_d/a \sim 61$. For low enough temperature a Rydberg gas will undergo a crystallization transition, as the coupling is above the critical value $D_c\sim 17$.~\cite{ast1,buch,fil2010} Quantum effects between Rydberg-dressed atoms will be relevant for $T< 1$~nK, due to low densities.

\item {\em Composite bosons}, e.g. spatially indirect excitons in coupled QWs with the exciton temperature in the range $0.5-4$~K. For typical experimental parameters~\cite{exciton,d1,d2,d3} ($n\sim 10^{10}$cm$^{-2}, m\sim 0.2 m_e, \epsilon_b\sim 10$, the inter-well distance $L\sim 200$~\AA, and $p \sim eL \sim 960$~Debye) from Eq.~(\ref{adex}) we estimate $a_d \sim 1500$~\AA\, and $D\sim 1.5$. A maximum dipolar coupling ($D\lesssim 7.5$) can be reached below the excitonic Mott transition density $n \sim 1/L^2\sim 2.5 \cdot 10^{11}$cm$^{-2}$.
\end{itemize}

These estimations show that the systems of ultracold dysprosium (erbium), polar molecules and indirect excitons are good candidates for experimental validations of the present analyses for $0.1 \leq D\leq 12.5$. We assume that in an experiment a long-lived quantum gas is created and, hence, a thermodynamic description remains valid. Further, there should be no noticeable heating, e.g. due to breaking of molecular bonds or recombination of excitonic states. By this assumption, experimental parameters which correspond to different $D$-values are listed in Tab.~\ref{tab0}.
 \begin{table}[h]
\caption{Relevant parameters for 2D systems of the dysprosium $^{164}$Dy (Dy), $^{41}$K$^{87}$Rb polar molecules (M) and indirect excitons (Ex).
The assumed dipole moments are, correspondingly, $p^{\text{Dy}}=10 \mu_B$, $p^{\text{M}}=0.6$~Debye and $p^{\text{Ex}}=960$~Debye. The dipole coupling $D$ is expressed in terms of the in-plane density $n$ [cm$^{-2}$]. For $T\leq T_c$ a system is in a superfluid phase.~\cite{fil2010} The frequency $\omega_0$ ($\hbar \omega_0\equiv E_0$) denotes a characteristic energy scale of collective and single-particle excitations.}
  \label{tab0}
 \begin{tabular}{c|c c c|c c c|c c c}
 \hline
 \hline
  $D$ &$n^{\text{Dy}}$ & $T_c^{\text{Dy}}$ & $\omega_0^{\text{Dy}}$ &$n^{\text{M}}$ & $T_c^{\text{M}}$ & $\omega_0^{\text{M}}$ & $n^{\text{Ex}}$ & $T_c^{\text{Ex}}$ & $\omega_0^{\text{Ex}}$ \\
  &$10^{11}$ & $\mu$K & MHz & $10^{8}$ & nK & KHz & $10^{9}$ & K & GHz \\
 \hline
 \hline
 0.1 & 0.023 & 0.087& 0.009&  0.027 & 0.13 & 0.013  & 0.044 & 0.0025 & 0.25\\    
 0.5 & 0.56 & 0.23 &0.22   &  0.67  & 3.4  & 0.33   & 1.1  & 0.065 & 6.33\\
 1.75 & 6.9 & 28 & 2.7     & 8.2    & 43   & 4.1    & 13.4  & 0.82 & 77\\
 7.5 & 127 & 458 & 49      & 150    & 695  & 75     & 246  & 13 & 1424 \\
 \hline
 \hline
 \end{tabular}
 \end{table}

\section{Single particle and density response excitations in a Bose liquid}\label{intro2}

Existence of a Bose-Einstein condensate is introduced by the Bose broken symmetry condition, i.e. by a finite value of the order parameter, $\avr{\hat{a}_0^{(+)}(\vec{r})}=\sqrt{N_0} e^{i \phi}$, where the phase $\phi$ can be set to zero for a homogeneous condensate. As introduced by Beliaev~\cite{beliaev1958} the number density operator in momentum space can be decomposed as
\begin{align}
\hat{\rho}(\vec{q})=\hat{a}^+_0 \hat{a}_{\vec{q}}+\hat{a}^+_{-\vec{q}} \hat{a}_0+\sum_{\vec{k}\neq 0} \hat{a}^{+}_{\vec{k}} \hat{a}_{\vec{k}+\vec{q}}.
\end{align}
The first two terms describe the scattering process which creates or destroys an atom with zero momentum. The last term includes all thermal atoms outside the condensate ($\vec{k}\neq 0$). In the presence of a BEC the density operator is factorized
\begin{align}
\hat{\rho}(\vec{q})=\sqrt{N_0} \hat{A}_{\vec{q}}+\tilde{\rho}_{\vec{q}}, \quad \hat{A}_{\vec{q}}=\hat{a}_{\vec{q}}+\hat{a}^{+}_{-\vec{q}},
\end{align}
which directly demonstrates the coupling of the single-particle operator, $\hat{A}_{\vec{q}}$, and the operator of particle-hole excitations of thermal atoms, $\tilde{\rho}_{\vec{q}}$. The result of this coupling in the density fluctuation spectrum was written down by Hugenholtz and Pines.~\cite{pines1959} The dynamic structure factor $S(\vec{q},\omega)$ describing propagation of the particle-hole excitations
\begin{align}
S(\vec{q},\omega)=\mathcal{F}\{\avr{\hat{\rho}_{\vec{q}}(t) \hat{\rho}_{-\vec{q}}(0)}\}, \,  \mathcal{F}=\frac{1}{2 \pi N} \int_{-\infty}^{\infty} dt\, e^{i \omega t}
\end{align}
decouples into three parts
\begin{align}
 S(\vec{q},\omega)=S_{\text{A}}(\vec{q},\omega)+S_{\text{I}}(\vec{q},\omega)+\tilde{S}(\vec{q},\omega),\label{prove1}
\end{align}
where
\begin{align}
 &S_{\text{A}}(\vec{q},\omega)=\mathcal{F}\{N_0\avr{\hat{A}_q(t)\hat{A}_{-q}(0)}\}\label{11},\\
 &S_{\text{I}}(\vec{q},\omega)=\mathcal{F}\{\sqrt{N_0}[\, \avr{\tilde{\rho}_q(t)\hat{A}_{-q}}+\avr{\hat{A}_{q}(t)\tilde{\rho}_{-q}}]\},\\
&\tilde{S}(\vec{q},\omega)=\mathcal{F}\{\avr{\tilde{\rho}_{\vec{q}}(t) \tilde{\rho}_{-\vec{q}}(0)}\}.\label{13}
\end{align}
Correspondingly, $S_{\text{A}}$ describes the density fluctuations involving condensate atoms, $S_{\text{I}}$ is a result of the condensate-induced intermixing of single-particle and density excitations, and $\tilde{S}$ defines the dynamic response of thermal atoms.

In the long wavelength limit and $T=0$ Gavoret and Nozi{\`e}res~\cite{noz} showed explicitly that the acoustic phonons are the poles of both $S_A(q,\omega)$ and $S(q,\omega)$. At finite temperatures these analyses have been extended by Griffin and Glide~\cite{grif1990} based on the dielectric function formalism. As a Bose liquid is cooled down to a critical temperature and a Bose condensate is formed ($N_0 \neq 0$), the single-particle excitations, i.e. the poles of $S_A(q,\omega)$, get a finite weight in the density response spectrum. Independent on the representation~(\ref{11})-(\ref{13}), as discussed in Sec.~\ref{intro}, the microscopic calculations~\cite{nepom} predict that $S_A(q,\omega)$ and $S(q,\omega)$ have the same poles in the presence of the condensate and that the weight of the SP excitations should be independent on the absolute value of the condensate fraction. 

Hence, the theory allows to interpret the sharp component observed beyond the maxon region in superfluid helium as a single-particle excitation branch, which vanishes in the normal phase with the condensate $N_0$. This in turn allows to indirectly probe the single-particle spectrum of $^4$He atoms by $S(q,\omega)$, as the latter can be measured experimentally by neutron-scattering. The development of this concept was stimulated by a set of high-resolution measurements.~\cite{talbot}

 Our goal is to verify this interpretation and identify the SP excitations in $S(q,\omega)$. The reconstructed spectra contain possible errors coming from the statistical noise in the evaluated imaginary time correlation functions and the stochastic reconstruction procedure discussed in Sec.~\ref{sto}.

\section{Quantum correlation functions}\label{qcorrf}

In linear response theory an external field produces a weak perturbation of a system in thermodynamic equilibrium. In this regime dynamical properties can be evaluated via time-correlation functions (TCF) of the corresponding dynamical operators. For classical many-body systems, TCF can be efficiently computed in the canonical ensemble by molecular dynamics (MD) simulations. A brute-force solution of classical equations of motion nowadays is possible for large ensembles ($N\sim 10^6$) of particles. This allows to make accurate extrapolation to the thermodynamic limit, study in detail topological phase transitions~\cite{hartmd} (e.g. BKT in 2D) involving large-scale density fluctuations and divergence of the correlation length near a critical point. 

In contrast, for quantum systems first principle time-dependent simulations can be done only for few-particle atomic and molecular systems, e.g. by time-dependent multi-configurational Hartree Fock and CI methods.~\cite{david} For condensed matter systems the continuous-time Monte Carlo method proved to be an efficient and powerful technique.~\cite{werner} While predictions from time-dependent methods for equilibrium can be accurately checked by quantum Monte Carlo (QMC) methods for ground state and finite temperatures,  the resolved dynamic correlations are more difficult to control for self-consistency. In particular for Bose-condensed systems, the importance to accurately treat dynamic correlations between thermal and condensate atoms has been  discussed recently by Griffin, Nikuni and Zaremba.~\cite{zarembabook} 

Here, we follow the approch based on quantum correlation functions evaluated via path integral Monte Carlo.

The correlation function of single-particle operators in Eq.~(\ref{prove1}) (for real time $t>0$) contains the contributions of two normal and two anomalous Green's functions
\begin{align}
 &\avr{\hat{A}_{\vec{q}}(t)\hat{A}_{-\vec{q}}(0)}=\label{aa} \\
 &\avr{\hat{a}_{\vec{q}}(t)\hat{a}^+_{\vec{q}}}+\avr{\hat{a}^+_{-\vec{q}}(t)\hat{a}_{-\vec{q}}}+\avr{\hat{a}_{\vec{q}}(t)\hat{a}_{-\vec{q}}}+\avr{\hat{a}^+_{-\vec{q}}(t)\hat{a}^+_{\vec{q}}} \nonumber \\
 &\equiv -i G_{11}(\vec{q},t) -i G_{22}(\vec{q},t) -i G_{12}(\vec{q},t) -i G_{21}(\vec{q},t) \nonumber.
\end{align}
The poles of $G_{nm}$ are given by zeros of the common denominator as follows from the formal solution of the Dyson-Beliaev equation in terms of the self-energies.~\cite{pines1959,fetter} Hence, characteristic single-particle excitations can be extracted from one (normal) Green's function $G_{11}$
\begin{align}
 A(\vec{q},\omega)\equiv A_{11}(\vec{q},\omega)=\mathcal{F}\{\avr{\hat{a}_{\vec{q}}(t)\hat{a}^+_{\vec{q}}}\},\label{Adyn}
\end{align}
as other spectral densities $A_{nm}$ share the common poles. Therefore, we can concentrate on comparison between $A(\vec{q},\omega)$ and $S(\vec{q},\omega)$, and their possible hybridization in a Bose condensed phase.

\begin{figure}
\begin{center}
\includegraphics[width=0.51\textwidth]{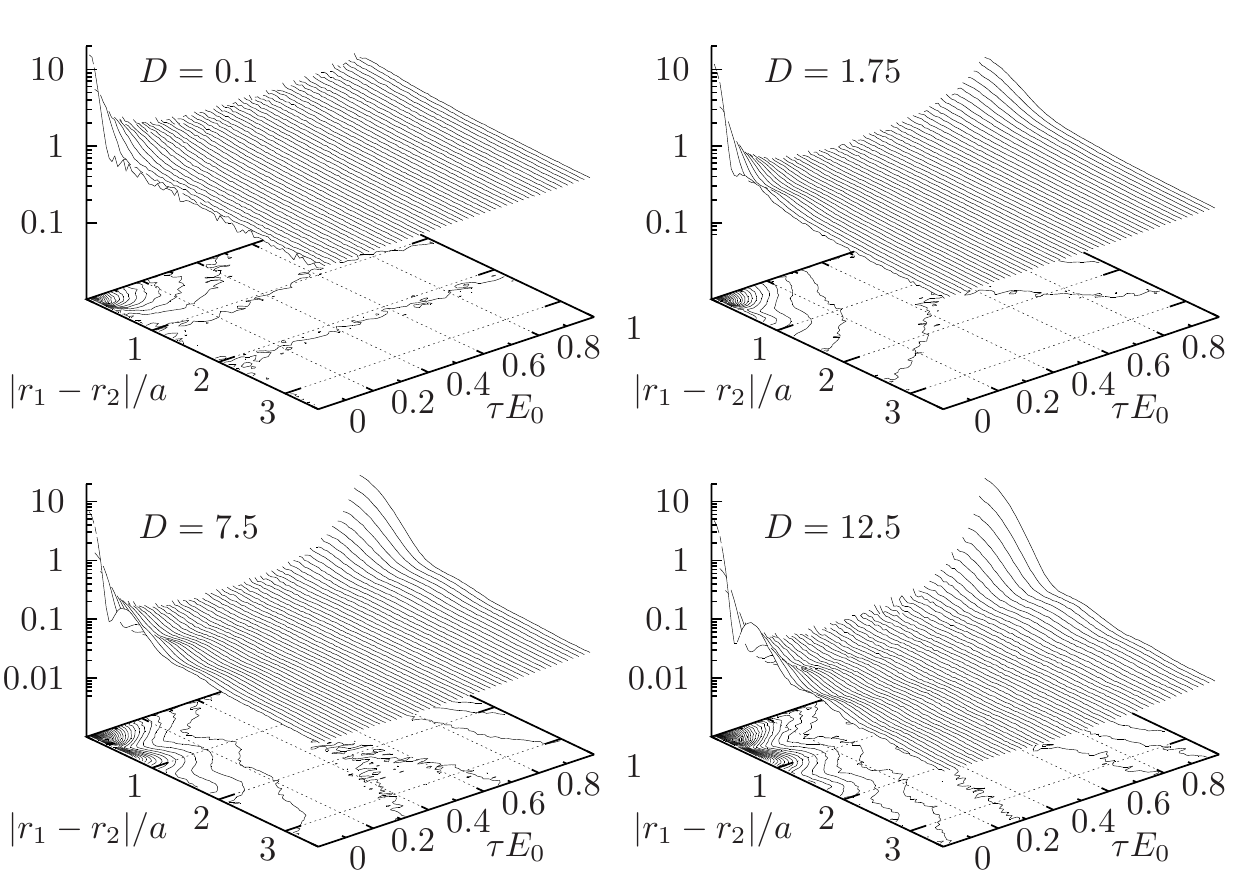}
\end{center}
\vspace{-0.50cm}
\caption{Matsubara Green's function $\avr{\Psi(\vec{r}_2,t_2)\Psi^+(\vec{r}_1,t_1)}$ (in units $a^2$) for a set of $D$-values. Simulation parameters ($D,\mu,V$) are specified in Tab.~\ref{tab1}. The temperature ($T=1.0$) is below to the BKT transition. The superfluid fraction is $\rho_s/\rho >0.80$. Due to spatial homogeneity we plot the dependence on the relative spatial distance $\abs{\vec{r}_2-\vec{r}_1}$ and relative imaginary time $\tau=t_2-t_1$.}
\label{fig:g1}
\end{figure}
\begin{figure}
\begin{center}
\vspace{-0.30cm}
\includegraphics[width=0.51\textwidth]{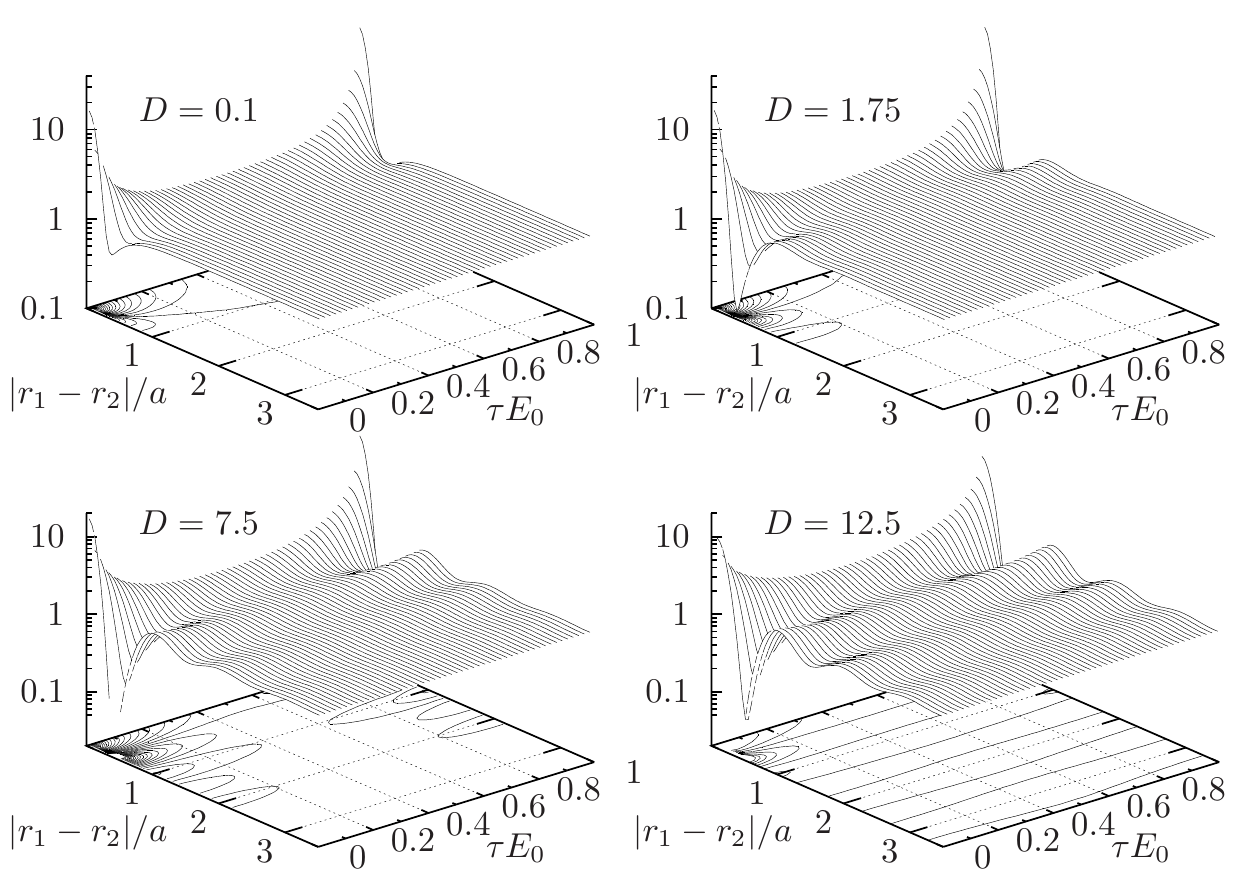}
\end{center}
\vspace{-0.50cm}
\caption{The same as in Fig.~\ref{fig:g1} but for the density-density correlation function  $\avr{\rho(\vec{r}_2,t_2)\rho(\vec{r}_1,t_1)}$ (in units $a^4$).}
\label{fig:g2}
\end{figure}

The general problem with the time-dependent ensemble averages using stochastic methods, like QMC, is the evaluation of the high-dimensional integrals from rapidly oscillating exponentials in real-time propagators. This well known ``dynamical sign problem''~\cite{werner} is the main obstacle for first principle evaluation of TCF in Eq.~(\ref{Adyn}) in the framework of the {\em particle orbital-free} continuous phase-space methods. 

Another possibility to compute the spectral function is the analytical continuation of an imaginary time correlation function to real frequencies.~\cite{fetter} The imaginary-time formulation of QMC (e.g. DMC, PIMC, GFPIMC) is the best suited tool. In particular, the Matsubara Green's function
 \begin{align}
  G_1(\vec{r}_2,\vec{r}_1,\tau)=\frac{1}{Z}\Tr \left[ e^{\beta \mu \hat{H}} e^{-(\beta-\tau)\hat{H}}\, \hat{\Psi}(\vec{r}_2) \, e^{-\tau\hat{H}} \, \hat{\Psi}^{+}(\vec{r}_1) \right],\label{g1pimc}
 \end{align}
can be efficiently evaluated via PIMC in the grand canonical ensemble on a set of imaginary time points $\tau \in (0,\beta=1/k_B T]$. Below we present several examples.

Fig.~\ref{fig:g1} (Fig.~\ref{fig:g2}) shows the changes in the structure of the Matsubara Green's function (density correlation function) with the increase of the dipole coupling $D$ on the spatial and imaginary time axis. The simulation parameters correspond to a superfluid (the superfluid fraction is $\rho_s/\rho >0.80$). In both cases, we observe that an enhancement of the inter-particle interaction leads to a more complicated structure with oscillations. These oscillations decay both in space and time, which is a general feature related to damping of one-particle or density (collective) excitations of a specific wave-length and energy. As we will see later, the interaction also leads to additional high-energy excitation branches which are absent in the weakly coupled regime, e.g. $D=0.1$.

The behavior of the spatially isotropic function $G_1(r,\tau)$, with $r=\abs{\vec{r}_2-\vec{r}_1}$, can be easily understood in two limiting cases. As $\tau \rightarrow 0$ and $r\rightarrow 0$, the many-body effects have only a small influence on the one-particle propagator, which is then close to the free-particle density matrix, $e^{-(\vec{r}_2-\vec{r}_1)^2/2 \lambda^2_{\tau}}$ with $\lambda^2_{\tau}=\hbar^2 \tau/m$. Indeed, in Fig.~\ref{fig:g1} near the origin we observe a similar Gaussian-shaped peak irrespective on the coupling $D$. After few oscillations in real space it evolves into a flat distribution. Another recognizable behavior is recovered as $\tau \rightarrow \beta$. The Green's function $G_1(r,\beta)$ becomes the one-particle density with two well-known limits, i.e.  $G_1(r=0,\beta)=n$, where $n=\avr{N}/V$ is the average particle density, and  $G_1(r=L/2,\beta)$ being an upper bound for the condensate density $n_0=\avr{N_{q=0}}/V$ which decays with a power law with system size (in 2D and 
$T\neq 0$). Both values depend on $\mu,V,T$ and can be read out from Tab.~\ref{tab1}. The listed values explain the observed off-set in the region where $G_1(r,\tau)$ is almost flat. For temperatures above the critical, $T_c$, the off-diagonal quasi-long range order is lost and $G_1(r,\tau)$ decays exponentially with $r$. With our choice~\cite{mudef} of the chemical potential $\mu(D)$ (see Tab.~\ref{tab1}) the particle density $n$ and the Green's function $G_1(0,\beta)=n$ take a similar value, i.e. $n\approx 1$, independent on $D$.
\begin{table}
 \caption{The $D$ and $T$-dependence of the average particle number $\avr{N}$ and the zero momentum occupation $\avr{N_{q=0}}$ in the volume $V_1=165$. Simulation parameters: $T_1=1.0$, $T_2=0.714$ and chemical potential $\mu_1$. The critical temperature $T_c$ is from Ref.~\cite{fil2010} The statistical error is given in the brakets or if not specified is in the last significant digit.}
\label{tab1}
\begin{tabular}{c|c|c|c c|c c}
\hline
\hline
 $D$ & $\mu_1$& $T_c$& $\avr{N}_{T_1}$ & $\avr{N}_{T_2}$ & $\avr{N_{q=0}}_{T_1}$ & $\avr{N_{q=0}}_{T_2}$  \\
\hline
\hline
0.1 & 4.7    &1.30&163.85(2)  & 164.82(2) &107.4(1) &115.4(2) \\    
0.5 & 12.55  &1.35&163.175(6) & 163.308(4)&77.24(2) &79.52(4) \\
1.75& 32.85  &1.39&169.379(3) & 169.403(2)&44.35(1) &44.95(1)\\
7.5&  105.0  &1.22&165.063(4) & 165.075(4)&9.70(1)  &9.885(6) \\
12.5& 163.53 &1.01&164.469(4) & 164.492(4)&3.605    &3.885(5)\\
\hline
\hline
\end{tabular}
\end{table}
\begin{table}
\caption{The same as in Tab.~\ref{tab1} for the system volume $V_2=576$.}
\label{tab1a}
\begin{tabular}{c|c|c|c|c}
\hline
\hline
 $D$ & $\mu_1$& $T_c$& $\avr{N}_{T_1}$ & $\avr{N_{q=0}}_{T_1}$  \\
\hline
\hline
0.1    & 4.8    &1.30&577.85(7)  & 342.2(1) \\
0.5    & 12.70   &1.35& 567.72(6)& 239.3  \\
1.75    &31.40 & 1.39 & 563.21(3) & 132.8 \\
7.5    & 103.20 & 1.22 & 558.06(2) & 27.06 \\
12.5   & 165.53 & 1.01 &566.4(5) & 9.048\\    
\hline
\hline
\end{tabular}
\end{table}

The behavior of the density correlator $\avr{\rho(\vec{r}_2,t_2)\rho(\vec{r}_1,t_1)}$ (Fig.~\ref{fig:g2}) has one simple limit. At $\tau=0$ it is a superposition of a $\delta$-function (at $r=0$) and of the pair distribution function. At finite but small $\tau$ the $\delta$-function turns into the free-particle like density matrix, which at strong coupling (large $D$) gets more localized due to the inter-particle interactions.     

Next we consider the reconstruction problem of spectral density from the imaginary-time. 

\section{Stochastic optimization method}\label{sto}

In this section we present the stochastic optimization method for reconstruction of the spectral densities. The method is free of difficulties involved in the analytic continuation of imaginary time correlation functions. As application, the dispersion relations for a 2D dipolar Bose system will be presented in Sec~\ref{specsec}.

We start from the general relation between the spectral density and the single-particle/density-density propagator Fourier transformed to the momentum-space
\begin{align}
 &G_1(q,\tau)=\avr{\hat{a}_{q}(\tau) \hat{a}^{+}_{q}(0)}= \int_{-\infty}^{\infty} \frac{d \omega}{2 \pi} \frac{e^{-\tau \omega}}{1- e^{-\beta \omega}} A(q,\omega), \label{g1}\\
 &G_2(q,\tau)=\frac{1}{\avr{N}}\avr{\hat{\rho}_{q}(\tau) \hat{\rho}_{-q}(0)}=\int_{-\infty}^{\infty} d\omega \, e^{-\tau \omega} S(q,\omega) \label{g2}.
\end{align}
The spectral densities satisfy two normalization conditions
\begin{align}
 \int_{-\infty}^{\infty} \frac{d \omega}{2 \pi} \, A(q,\omega)=1, \quad \int_{-\infty}^{\infty}  d\omega\, S(q,\omega) =S(q). \label{norm}
\end{align}
The inversion of equations similar to (\ref{g1})-(\ref{g2}) is known to be an ill-conditioned problem and results in a non-uniqueness of solution. By Monte Carlo simulations the reconstructed spectral densities get affected by the statistical errors present in $G_n(q,\tau)$. 
The standard tool used to partially overcome this problem is the Maximum Entropy (ME) method.~\cite{me} However, the reconstructed spectral densities for Bose liquids appear to be too smooth and the important information on a sharp $\delta$-like quasi-particle feature present in the excitation spectra is typically lost.~\cite{bonin96} Recently, the Stochastic Optimization (SO) method has been introduced by Mishchenko {\em et al.,}~\cite{Mich} which allows to overcome this difficulty. Its main advantages compared to the standard regularization methods, like the ME, are the continuous parametrization in frequency space, instead of a predefined mesh, and non-suppression of high derivatives of the spectral function, performed by the regularization methods. As a result, sharp peaks and edges are not lost during the reconstruction. In its core, with the SO one solves by stochastic sampling the minimization problem of the least deviation
\begin{align}
 &D_n[\tilde{G}_n]=\sum_{\tau_i} \abs{1-\tilde{G}_n(q,\tau_i)/G(q,\tau_i)}\, \tilde{w}(q,\tau_i), \label{dn}
\end{align}
where $\delta G(q,\tau_i)$ is the statistical error of $G(q,\tau_i)$ at the imaginary time $\tau_i$ ($\Delta \tau=\tau_{i+1}-\tau_i$, with $i=1,\ldots N$ and $N\Delta \tau=\beta$), the weight factor $\tilde{w}(q,\tau_i)$ is chosen in the form 
\begin{align}
& \tilde{w}(q,\tau_i)=\left(N/{N_{\tau}}\right) w(q,\tau_i), \quad N_{\tau}=\sum_{\tau_i} w(q,\tau_i),\\ 
&w(q,\tau_i)=\min[10,\abs{G(q,\tau_i)/\delta G(q,\tau_i)}], \\
&w(q,\tau_i)=\max[1,\abs{G(q,\tau_i)/\delta G(q,\tau_i)}],
\end{align}
and $\tilde{G}_n$ is generated from 
\begin{align}
\tilde{G}_n(q,\tau)=\int_{-\infty}^{\infty} e^{-\tau \omega} \tilde{S}_n(q,\omega) \, d\omega, \label{gngen}
\end{align}
with $\tilde{S}_n(q,\omega)$ being a trial spectral density  parameterized into some basis set. The deviation $D_n[\tilde{G}_n]$ is optimized by a random sequence of updates which can change both the parameters of the basis set and its size. By increasing $n$ (the number of independent solutions) we evaluate the corresponding variance $\sigma^2_{D_n}=\avr{D_n^2}$ (with the zero mean). In the end, we select only ``good'' solutions from the whole sample which satisfy the condition $D_n \leq D_{\text{min}}$ with $D_{\text{min}} =1.5 \,\sigma_{D_n}$. The final solution is constructed from their linear combination ($100\leq M \leq 400$)
\begin{align}
 S_{\text{est}}(q,\omega,M)=\frac{1}{M} \sum\limits_{n=1}^{M} \tilde{S}_n(q,\omega), \label{finals}
\end{align}
to take advantage of a self-averaging of the noise.

For the parametrization of trial $\tilde{G}_n$ we use a set of rectangles~\cite{Mich}
\begin{align}
 &\{P_i\}_{i=1,N}=\{h_i,w_i,c_i\}_{i=1,N},\label{rec}\\
 &\tilde{S}_n(q,\omega)=\sum\limits_{i=1}^N h_i(\omega),\label{srec}\\
 & h_i(\omega) = \begin{cases}
\neq 0,& \omega \in [c_i-w_i/2,c_i+w_i/2] ,\\
0,& \text{otherwise}.
\end{cases}
\end{align}
with height $h$, width $w$ and center $c$ being the optimization parameters for a fixed value of the $q$-vector. The basis size $N$ was also varied during the optimization in the range $80 \lesssim N \lesssim 400$.

For the reconstruction of the dynamic structure factor $S(q,\omega)$ we have used only positive frequencies, $0 < \omega< \omega_{\text{CO}}$, with the cut-off frequency $\omega_{\text{CO}} \approx 400$, by taking into account explicitly the relation between negative and positive energy transfers, i.e. $S(-\vec{q},-\omega)=e^{-\beta \omega} S(\vec{q},\omega)$. This results in 
 \begin{align}
\tilde{G}_n(q,\tau)=\int_{0}^{\infty} e^{-\tau \omega} \tilde{S}(q,\omega) \left(1+e^{-\beta \omega}\right) \, d\omega.\label{g2_sym}
\end{align}
In the basis of the rectangular functions the trial imaginary time density correlation function takes the form
\begin{align}
 \tilde{G}_2(q,\tau)=2\sum\limits_{i=1}^N  h_i \left(\sum_{t=\tau,\beta-\tau} \frac{1}{t}  e^{-c_i t} \sinh \frac{w_i t}{2}\right), \label{recbasis}
\end{align}
which is symmetric with respect to the mid-point $t=\beta/2$. Therefore, it is sufficient to evaluate $G_2(q,\tau)$ (see Eq.~(\ref{g2}) for the imaginary times $\tau \in [0,\beta/2]$. The normalization condition~(\ref{norm}) results in the constraint
\begin{align}
&G_2(q,\tau=0)=G_2(q,\beta)=S(q),\\
&\sum \limits_{i=1}^N h_i w_i+ S^{\prime}=S(q),\\
&S^{\prime}=2\sum\limits_{i=1}^N  \frac{h_i}{\beta}  e^{-c_i \beta} \sinh \frac{w_i \beta}{2},
\end{align}
where the factor $S^{\prime}$ corresponds to the occupation of the negative frequencies with the meaning of the energy transfer $\hbar \omega$ from a system to a scattered probe particle, i.e. from the excitations existing in the system. Typically, this contribution is important in the $q$-region of the acoustic phonons and rotons, and one can start the optimization first by neglecting $S^{\prime}$ and find an optimized solution $\tilde{S}_1(q,\omega)$. Then the normalization of a second solution is corrected by 
\begin{align}
 \sum \limits_{i=1}^N h_i^{(n)} w_i^{(n)}=S(q)-S^{\prime}(S_{\text{est}}(q,\omega,M)),\label{siter}
\end{align}
i.e. the correction $S^{\prime}$ is evaluated based on the spectral densities obtained in the previous iterations ($M \geq 1$). With $n$ increasing  the on fly estimation of both $S_{\text{est}}(q,\omega,M)$ and $S^{\prime}$ is improved. This results in a fast convergence of $S^{\prime}$ within few ($n\sim 5$) iterations.

\begin{figure}
\begin{center}
\vspace{-1cm}
\includegraphics[width=0.5\textwidth]{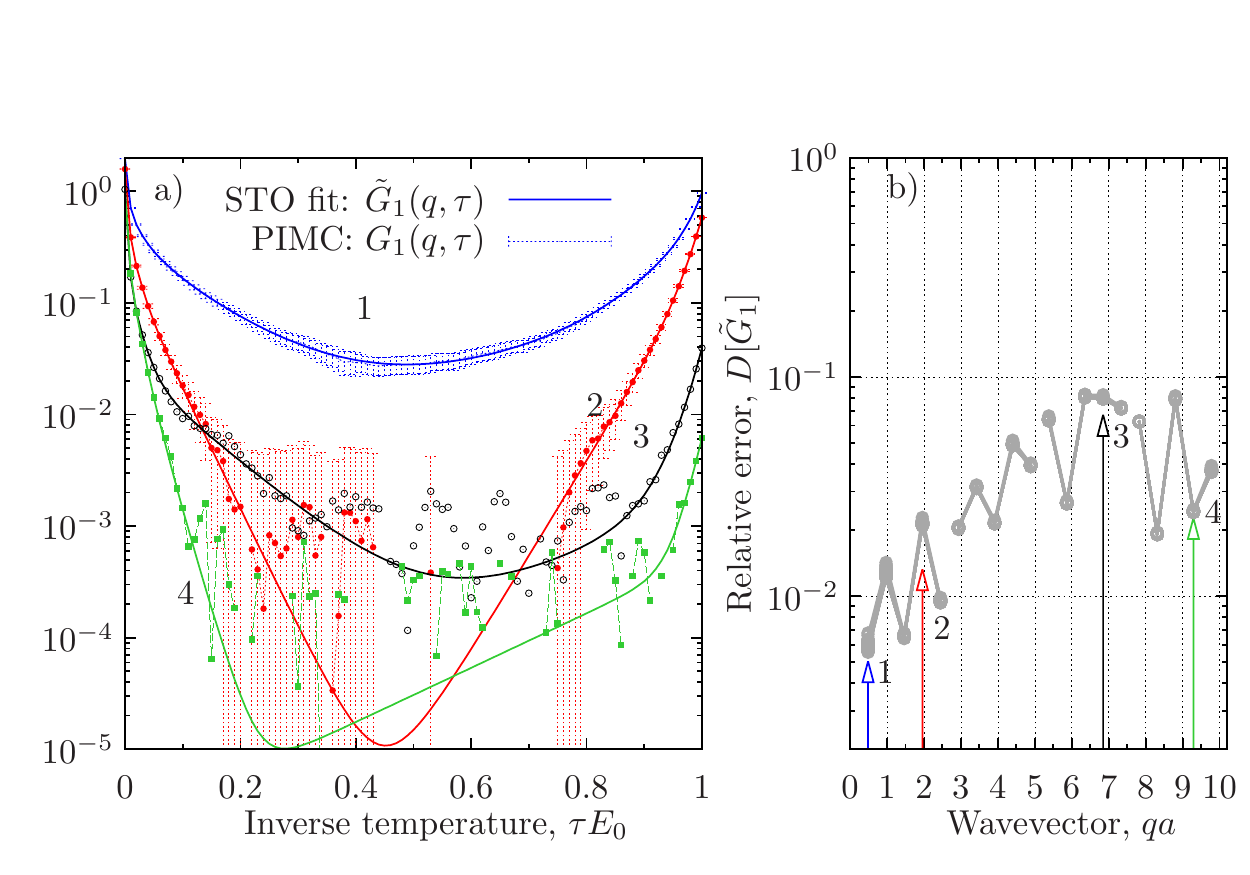}
\end{center}
\vspace{-0.90cm}
\caption{(Color online) (a) Fourier transform of the one-particle Green function $G_1$~(\ref{g1pimc}): PIMC data (symbols) and the stochastic optimization fit (solid lines). Vertical dashed error-bars demonstrates the statistical error (shown only for two curves). Colors (and numbers) correspond to the chosen $q$-vectors shown on the left panel.  (b) Relative deviation~(\ref{dn}) after optimization. Simulation parameters: $D=12.5$ and $T=1.25$ (see Tab.~\ref{tab1}). Temperature is close to $T_c$. Superfluid fraction $\rho_s/\rho=0.38(2)$.}
\label{fig:optgreenq}
\end{figure}

The reconstruction of the spectral density $A(q,\omega)$ of the one-particle Green's function~(\ref{g1}) gets a bit more involved. First, there are no simple relations between the densities at positive ($A_>(q,\omega)|_{\omega \geq 0} \neq 0, A_>(q,\omega)|_{\omega < 0}= 0$)  and negative ($A_<(q,\omega)|_{\omega \leq 0} \neq 0, A_<(q,\omega)|_{\omega > 0} = 0$) frequencies. They should be worked out independently in the frequency range, $-\omega_{\text{CO}} < \omega < \omega_{\text{CO}}$. We found that using $\omega_{\text{CO}}\sim 600$ is sufficient to fit a fast drop of $G_1(q,\tau)$ near $\tau=0$, see Fig.~\ref{fig:optgreenq}.  Second, $G_1(q,\tau)$ is not symmetric relative to $\tau=\beta/2$ and should be evaluated on the whole interval, $0< \tau \leq \beta$. Third, the Laplace transform in Eq.~(\ref{g1}) contains the additional Bose factor which leads to the rigorous result,~\cite{Pitaev} $A_{<}(q,\omega)\leq 0$. 

A simple solution which allows to use the same SO procedure as for $S(q,\omega)$ is to introduce two spectral densities which are both positive and defined for $\omega \geq 0$
\begin{align}
 &\tilde{A}_{>}(q,\omega)=A_{>}(q,\omega)/(1-e^{-\beta \omega}) \geq 0,\\
 &\tilde{A}_{<}(q,\omega)=A_{<}(q,\omega)/(e^{-\beta \omega}-1) \geq 0.
\end{align}
This results in the following decomposition
\begin{align}
 G_1(q,\tau)=\int_{0}^{\infty} \frac{d \omega}{2 \pi} \,\left[  e^{-\tau \omega} \tilde{A}_>(q,\omega) +e^{-(\beta-\tau) \omega} \tilde{A}_<(q,\omega) \right].
\end{align}
Using Eqs.~(\ref{g1}),(\ref{norm}) we end up with two normalization constrains
\begin{align}
 \int\limits_{0}^{\infty} \frac{d \omega}{2 \pi} \, \tilde{A}_>(q,\omega)&=G_1(q,0)-\int\limits_{0}^{\infty} \frac{d \omega}{2 \pi} \, e^{-\beta \omega} \tilde{A}_<(q,\omega),\label{tilde1}\\
 \int\limits_{0}^{\infty} \frac{d \omega}{2 \pi} \, \tilde{A}_<(q,\omega)&=G_1(q,\beta)-\int\limits_{0}^{\infty} \frac{d \omega}{2 \pi} \, e^{-\beta \omega} \tilde{A}_>(q,\omega),\label{tilde2}\\
 G_1(q,0)&=1+G_1(q,\beta).
\end{align}
At low temperature and/or high excitation energies, $\beta \omega \gg 1$, the integral terms in Eqs.~(\ref{tilde1}),(\ref{tilde2}) can be neglected. Then both spectral densities become independent with the normalization given by the 
momentum distribution
\begin{equation}
 n(q)= \sum\limits_{m,n} \frac{e^{-\beta E_n}}{Z} |\langle m| a_q^+ | n \rangle |^2=\int\limits_{-\infty}^{\infty} \frac{d \omega}{2 \pi}   \frac{A(q,\omega)}{e^{\beta \omega}- 1}=G_1(q,\beta). \label{nq}
\end{equation}
The latter can be directly evaluated via Fourier transform of the one-particle density matrix
\begin{align}
 & n(q)=\avr{\hat{a}_q \hat{a}_q^{+}}=\int\int_V\, \db \vec{r} \, \db\vec{r}^{\prime}\, e^{i \vec{q}(\vec{r}^{\prime}-\vec{r})} \langle \hat{\Psi}(\vec{r}^{\prime},t) \hat{\Psi}^{+}(\vec{r},t) \rangle, \label{pmom}\\
 & \hat{a}_q=\int_V \db \vec{r} \, e^{-i \vec{q}\vec{r}}\, \hat{\Psi}(\vec{r},t).
 \end{align}
In contrast, at high temperatures, the normalizations~(\ref{tilde1})-(\ref{tilde2}) are mutually dependent and can be treated iteratively, similarly to Eq.~(\ref{siter}), i.e.
\begin{align}
 &\sum \limits_{i=1}^{N_>} h_{i, >}^{(n)} w_{i, >}^{(n)}=n(q)+1-A^{\prime}(A_{\text{est},<}(q,\omega,M)),\label{aiter1}\\
 &\sum \limits_{i=1}^{N_<} h_{i, <}^{(n)} w_{i, <}^{(n)}=n(q)-A^{\prime}(A_{\text{est},>}(q,\omega,M)).\label{aiter2}
\end{align}
The procedure converges within few iterations. 

For the detailed description of the stochastic optimization method and types of the update we refer to Ref.~\cite{Mich} 
We found of particular importance to implement the annealing allowing to escape from local minimum and minimization of the deviation~(\ref{dn}) using parabolic interpolation. During the reconstruction, the updates involving changes in two rectangles~(\ref{rec}) (i.e. change a weight of two rectangles, add a new rectangle, remove a rectangle) correctly redistribute the total spectral density $A(q,\omega)$ between $A_>(q,\omega)$ and $A_<(q,\omega)$, even if initially they are chosen to be equal. The decay of spectral weight $A_<(q,\omega)$ at large $q$-vectors should follow that of the momentum distribution and, therefore, practically vanishes beyond the roton region ($qa\sim 6$).

To speed up the optimization process and convergence of the finite-temperature corrections in the normalizations~(\ref{siter}),(\ref{aiter1}),(\ref{aiter2}) the centers of the basis set rectangles $\{c_i\}$, representing $\tilde{S}(q,\omega)$, $A_>(q,\omega)$, $A_<(q,\omega)$, have been initially  normally distributed around the frequency $\omega_{\chi}(q)$, i.e. the upper bound of the lower excitation branch, see Appendix~\ref{app}. The optimization process was started with $N,N_>, N_< \approx 30$ basis functions with the initial width, $w=w_{\text{min}}$, i.e. equal to the frequency resolution $\hbar w_{\text{min}}/E_0=0.5$. The optimization was performed in several iterations each of $120000$ steps. In each step one of the following update types has been randomly chosen: (1) shift of a rectangle, (2) change a rectangle-width, (3) change the heights of two rectangles, (4) add a new rectangle, (5) remove a rectangle, (6) split a rectangle into two, (7) glue two rectangles. For each update involving a change 
in one or couple of parameters $\{h_i\},\{w_i\},\{c_i\}$ we try to converge to a local minimum by parabolic interpolation. In the first 60000 steps we perform several annealing sequences with a duration $1000-5000$ steps and temporary accept the updates increasing the deviation $D_n$~(\ref{dn}). Once an iteration is finished, the actual deviation $D_n$ and the optimized solution are saved and checked for the acceptance criterion, $D_n \leq D_{\text{min}}$. If not accepted, we proceed to the next iteration simultaneously increasing the minimal deviation by some factor, $D_{\text{min}} \rightarrow \gamma D_{\text{min}}$, e.g. $\gamma= (1.1+\nu)$, with the random number $\nu \in [0,1)$. The initial value $D_{\text{min}}$ was chosen based on the statistical noise in the quantum correlation functions, Eqs.~(\ref{g1}),(\ref{g2}). Typically we start from $D^0_{\text{min}}\sim 10^{-5} \ldots 10^{-4}$ and can reach the acceptance error $D_{\text{min}}\sim 10^{-4} \ldots 10^{-3}$ within $5-6$ iterations, depending on 
the level of noise in the simulation 
data. 

\begin{figure}
\begin{center}
\vspace{-1cm}
\includegraphics[width=0.5\textwidth]{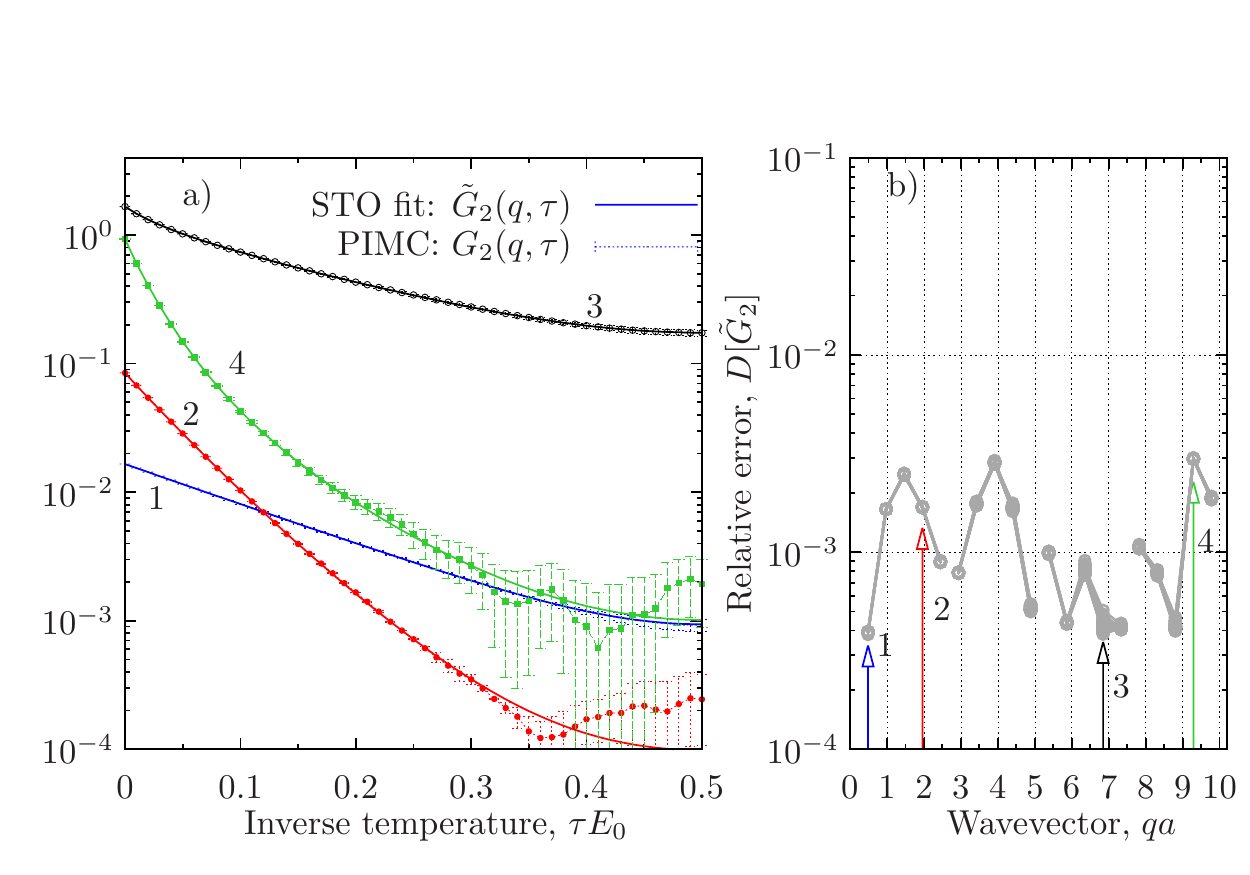}
\end{center}
\vspace{-0.90cm}
\caption{(Color online) The same as in Fig.~\ref{fig:optgreenq} for the density correlation function $G_2$~(\ref{g2}). Simulation parameters: $D=12.5$ and $T=1.0$. Superfluid fraction $\rho_s/\rho=0.81(2)$. $G_2(q,\tau)$ is evaluated only for $\tau \in [0,\beta/2]$ due to the symmetry with respect to the mid-point $\tau=\beta/2$, see Eq.~(\ref{g2_sym}).}
\label{fig:optskdyn}
\end{figure}
Figs.~\ref{fig:optgreenq} and~\ref{fig:optskdyn} demonstrate two examples of the performance of the optimization procedure. The symbols show the PIMC data Fourier transformed to the momentum space and demonstrate the level of statistical noise. For the density correlation function we get a typical error, $\delta G_2 \sim 10^{-4}$, which is one order of magnitude smaller compared to the single-particle Green's function, $\delta G_1 \sim 10^{-3}$. The solid line is the result of the optimization, i.e. the correlation function~(\ref{gngen}) evaluated from one of the solutions $\tilde{S}_n(q,\omega)$ or $\tilde{A}_n(q,\omega)$ which enters in the final estimation~(\ref{finals}). The right panel shows the relative error for a set of solutions $\tilde{G}_n$ ($n=1,100$). The difficulty to distinguish individual curves demonstrates the convergence of the optimization process. All solutions are obtained from independent (randomly chosen) initial spectral densities. The relative error depends on the $q$-vector and 
its value is mainly 
determined not by the optimization result but by the statistical fluctuations near $\tau=\beta/2$, when the value of the correlation function significantly drops. Still the optimization process converges. A well-behaved decay at short times (for $\tau$ or $\beta-\tau$) fixes some of excitation energies $\omega$ in the spectral density and, hence, partially the decay at large times when approaching the mid-point $\tau=\beta/2$. As follows from Eq.~(\ref{dn}) the fitted points contribute to the least deviation $D_n$ not equally but with the weight factor determined by the statistical error. Hence, the middle points with large statistical fluctuations are less important for the fit.

The influence of the statistical noise on the accuracy of the reconstruction procedure is discussed in detail in Appendix~\ref{acc_resonst}.

\section{Static and dynamic properties}\label{specsec}

\subsection{Momentum distribution}

\begin{figure}
\begin{center}
\includegraphics[width=0.49\textwidth]{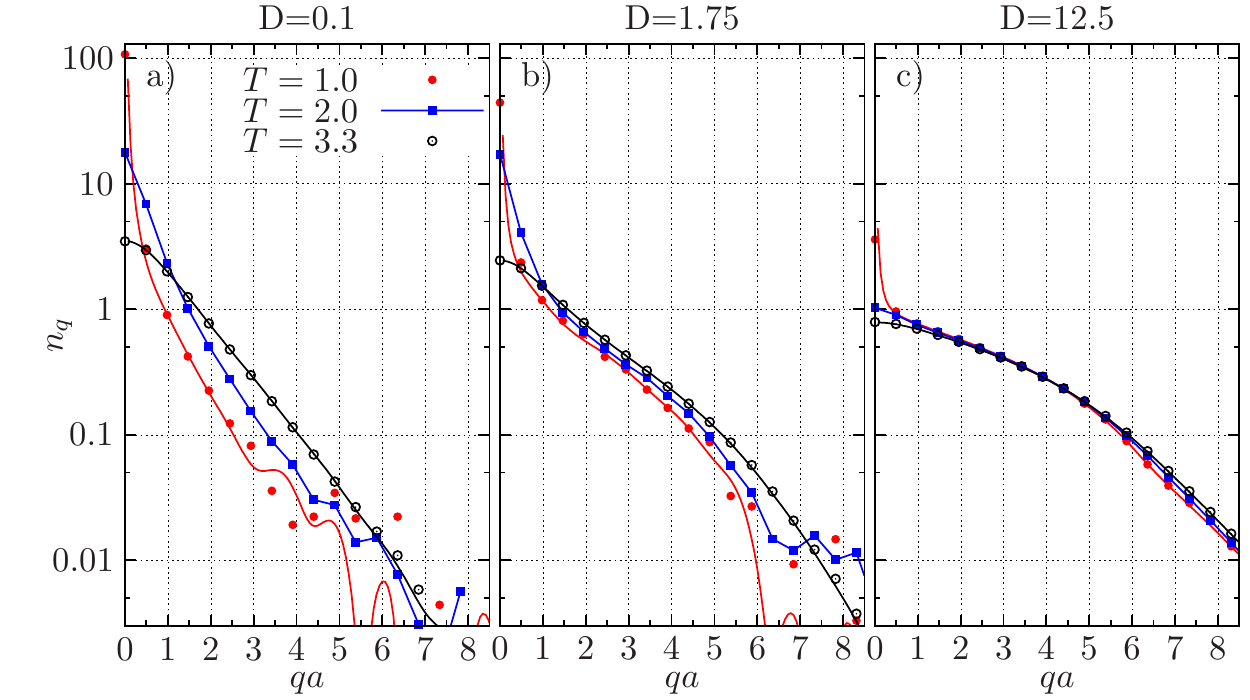}\\
\vspace{-0.3cm}
\includegraphics[width=0.49\textwidth]{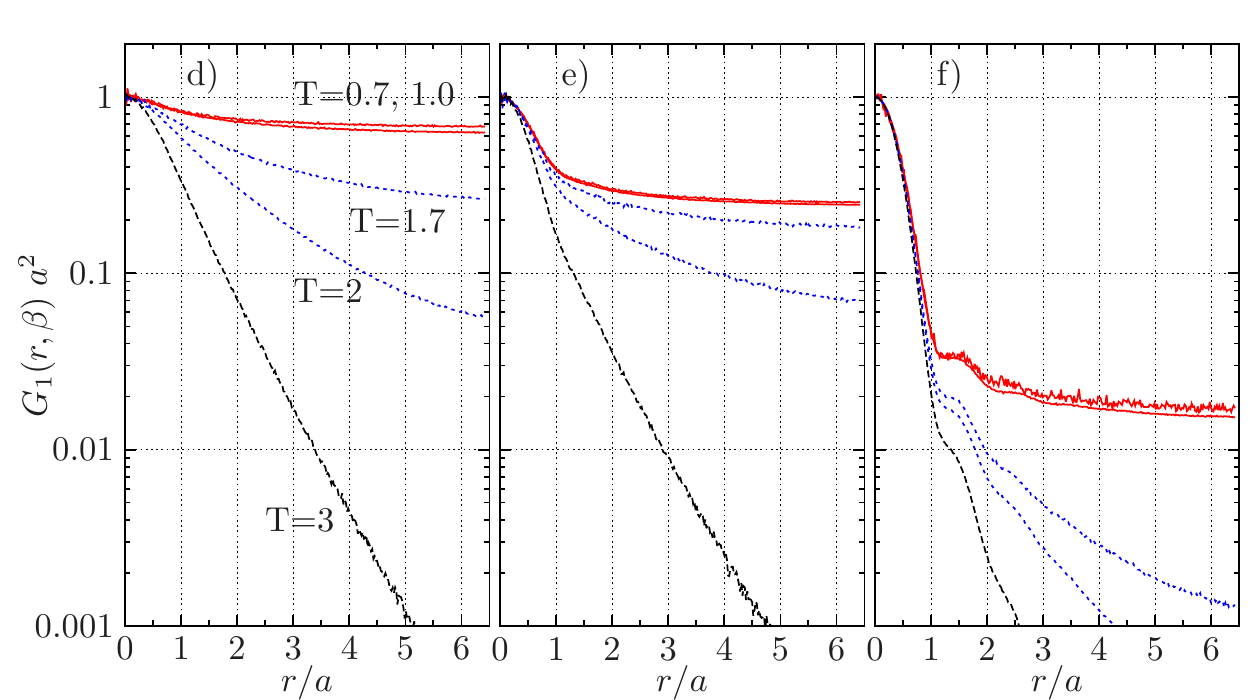}
\end{center}
\vspace{-0.50cm}
\caption{(Color online) Temperature dependence of the momentum distribution $n_q$ (a-c) and one-particle density matrix $G_1(r,\beta)$ (d-f) in the log-scale for dipole coupling $D=0.1, 1.75, 12.5$. In the superfluid phase, the decay of $G_1(r,\beta)$ is characterized by: 1)
a fast short-range decay to a condensate fraction $n_0(T)$; 2) off-diagonal quasi-long-range order which depends on temperature and the superfluid density. Simulation parameters are $V=165$, chemical potential $\mu(D)=4.7,32.85,163.53$, the average particle number $\avr{N} \approx 164$. The used wavevectors: $qa =2 \pi n (a/L), \; (n=0,1,\ldots)$. The solid lines are obtained by interpolation (see text). The system is partially superfluid for $T \leq 1.67$ ($D=0.1, 1.75$) and  $T \leq 1.0$ ($D=12.5$).}
\label{fig:pmom}
\end{figure}
Before analyzing of the spectral densities, we now discuss the momentum distribution which enters in the normalization~(\ref{aiter1})-(\ref{aiter2}). These data will characterize the normal and superfluid phase of 2D dipolar gas.

In Fig.~\ref{fig:pmom}a,b,c the low ($T < T_c$) and high ($T> T_c$) temperature momentum distribution is shown for $D=0.1, 1.75, 12.5$. These are referenced in the following as {\em weak}, {\em intermediate} and {\em strong} coupling, correspondingly. The $T$-dependence of the one-particle density matrix~(\ref{pmom}) is shown in Fig.~\ref{fig:pmom}d,e,f. For low temperatures ($T=0.714,1.0$) in the superfluid phase it demonstrates only minimal changes. Our system has a finite volume and satisfies the periodic boundary conditions (PBC), therefore, the momentum distribution is evaluated at a discrete set of wavevectors, $\vec{q}=2 \pi \vec{n}/L\; (n\in Z)$, shown by symbols in Fig.~\ref{fig:pmom}. For comparison, the solid line is the result (shown for $T=1.0, 3.3$) obtained by extension of Eq.~(\ref{pmom}) to the limit $V\rightarrow \infty$ with the assumption that $G_1(r,\beta)$ decays as a power-law (exponent) below (above) $T_c$ beyond the simulation box size $L=\sqrt{V}$. The fitting parameters for $T<T_c$ 
are given in Tab.~\ref{tabnu}. This method agrees with the finite-size results (shown by symbols). As the system size $L$ gets larger, more discrete values of $q$ will get occupied dwelling on this interpolation curve.

We get a full agreement in the normal phase, when $G_1(r,\beta)$ decays fast to a small value at $r=L/2$ (e.g. to $\sim 10^{-6}$ for $T=3.3$ for $D=12.5$), and finite-size effects are of minor importance for systems with $V \geq 165$. 

\begin{table}
 \caption{The $D$ and $T$-dependence ($T_1=1.0$, $T_2=0.714$) of the critical exponent by fitting the long-range asymptotic of the one-particle DM, $G_1(r)/n\sim a r^{-\nu(T)}$, and $\tilde{G}_1^{L/2}\equiv G_1(L/2)/n$ for the volume $V_1=165$.}
\label{tabnu}
\begin{tabular}{c|c c|c c|c c}
\hline
\hline
 $D$ & $a(T_1)$& $a(T_2)$& $\nu(T_1)$ &$\nu(T_2)$ & $\tilde{G}_1^{L/2}(T_1)$  & $\tilde{G}_1^{L/2}(T_2)$ \\
\hline
\hline
0.1 & 0.738(4) &0.762(6)&0.088(3)& 0.063(5)&0.628(4) &0.678(3) \\
0.5 & 0.526(4) &0.523(4)&0.085(4) & 0.061(5)&0.449(3) &0.467(4) \\
1.75& 0.294(2) &0.285(1)&0.103(3)  & 0.075(2)&0.245(2) &0.250(2)\\
7.5&  0.0656(5) &0.066(1)&0.159(4)   & 0.155(8)&0.0495(2)  &0.050(1) \\
12.5& 0.0243(5) &0.025(1)&0.258(5)  & 0.22(3)&0.0153(1) &0.0172(6)\\
\hline
\hline
\end{tabular}
\end{table}

Some discrepancies appear in the superfluid phase. In particular, for $T=1$, $D=0.1$ (Fig.~\ref{fig:pmom}a) and $qa >5$ we observe some statistical noise in $n_q$ related with the Fourier transform of $G_1(r,\beta)$. In the long wave-length limit ($qa<0.5$) the discrepancies are induced by the interpolation. The density matrix is influenced by the PBC and deviates from the expected power-law decay $r^{-\nu(T)}$ near $r=L/2$. Therefore, the fit with the critical exponent $\nu(T)$ was applied in the range $4 \leq r \leq 6$ for $V_1=165$ and $4 \leq r \leq 10$ for $V_2=576$. This partially allows to exclude the effects of the short-range correlations and of the finite-size errors on the expected asymptotic decay of $G_1$. As predicted by the interpolation curve, in a system with off-diagonal quasi-long range order, the momentum distribution diverges when approaching zero momentum. For a finite system and discrete $q$ we can only see the onset of this regime. For $V=165$, the smallest wavevector ($qa=0.486$) is 
too 
large to capture the divergence. The low-momentum behavior will be discussed in more detail later.

The information about condensate or occupation of the zero-momentum state is given by the value of the Matsubara Green's function~(\ref{nq}), $G_1(q=0,\beta)$. The momentum distribution in Eq.~(\ref{pmom}) is normalized by the average particle number $\avr{N}$ in volume $V$. Therefore, the number of particles at zero momentum $\avr{N_{q=0}}$ also depends on $V$ and $\avr{N}$, see Tab.~\ref{tab2}. For a macroscopic system this value will diverge as $V \rightarrow \infty$, in agreement with the interpolation curves in Fig.~\ref{fig:pmom}a,b,c. On the other hand, at a finite temperature the condensate fraction should vanish in the thermodynamic limit,~\cite{lifbook}  $\lim_{V\rightarrow \infty}\avr{N_{q=0}}/V=0$, as the zero momentum state will be depleted by thermal fluctuations. A number of occupied phonon modes with $q=2\pi n/L \leq q_c$ gets larger with increasing $L$. 

How the thermal depletion of the low-momentum states proceeds can be analyzed for $T=1.0,2.0$ and $3.3$. For $D=0.1$ broadening of $n_q$ is observed in a wide range of momenta. The formation of the condensate feature at small $q$ is accompanied by suppression of the high-momentum states. For $D=1.75$ this is noticeable for $qa <1$, and for $D=12.5$ for $qa <0.5$. Here, formation of the condensate is due to suppression of the states with $qa>6$. We conclude that the increase of the coupling/density narrows the interval of momenta where a fast divergence can be observed in the superfluid phase. This, certainly, will complicate an experimental detection of a superfluid transition in a strongly correlated Bose system based only on the specific features of the momentum distribution. In particular, for $D=12.5$, in a broad range of momenta ($1<qa<6$) the distribution is practically temperature independent ($0\leq T\lesssim 3.3$). Here the main depletion mechanism (also at $T=0$) is a strong interparticle 
interaction. Only a small 
fraction of particles occupies the $q=0$ state. The depletion out of the condensate is enhanced by the presence of low-energy excitations, e.g. rotons. More insight should be given by the spectral densities of single-particle excitations and their dependence on the interaction strength. 

\begin{table}
 \caption{The $D$ and $T$-dependence of the condensate fraction $n_0(T)$ (for system size $V_1=165$ and $V_2=576$) at $T_1=1.0$ and $T_2=0.714$ evaluated from Tab.~\ref{tab1},~\ref{tab1a}. compared with the zero-temperature condensate $n_0(0)$.~\cite{ast1,ast2,ast3}}
\label{tab2}
\begin{tabular}{c|c c |c |c}
\hline
\hline
 $D$ & $n_0^{V_1}(1)$ &$n_0^{V_1}(0.714)$ & $n_0^{V_2}(1)$ &$n_0(0)$  \\
\hline
\hline
0.1 & 0.655(5)&0.700(5)& 0.592(4) &0.72\\
0.5 & 0.473  &0.487(1) &  0.422(2) &0.50  \\
1.75& 0.262  &0.265   & 0.236 &0.28 \\
7.5&  0.0587 & 0.0599  &0.0485 &0.062 \\
12.5& 0.0219 &0.0236   & 0.016 &0.025\\
\hline
\hline
\end{tabular}
\end{table}

The average particle number in the zero-momentum state $\avr{N_{q=0}}$ and the condensate fraction $n_0(T)$ for $T=0.714,1.0$  ($T<T_c$) are given in Tabs.~\ref{tab1},\ref{tab2}. The PIMC results are compared with the DMC.~\cite{ast1,ast2,ast3} The zero-temperature condensate fraction stays the upper bound for the PIMC values, $n_0(T) < n_0(0)$.
Both results strongly deviate from the predictions~\cite{schick} based on the perturbation expansion in the gas parameter $n a_s^2$. As shown in Refs.~\cite{ast2,fil2010} the s-wave scattering model for the interaction fails at the densities $n a_s^2\gtrsim 10^{-3}$ ($D\gtrsim 0.01$) significantly lower then considered here ($D \geq 0.1$).   

The  $T$-dependence of the condensate is more pronounced for $D=0.1$ and $12.5$. In the first case, it is due to occupation of the higher-momentum states (see Fig.~\ref{fig:pmom}a). In the second case, $T_1$ and $T_2$ are close to the critical temperature of an infinite system ($T_c=1.01$, Tab.~\ref{tab1}) and thermal fluctuations play an important role.

The condensate fraction is also estimated for a larger system ($V_2=576$) to demonstrate the finite size effect. For $D=1.75,7.5$ and $12.5$ we find $\tilde{G}_1^{L/2}(T_1)=0.220(5), 0.0438(8)$ and $0.0136(5)$, correspondingly ($\tilde{G}_1\equiv G_1/n$). Compared to $n_0(T)$ in Tab.~\ref{tab2}, the relation, $n_0(T) \gtrsim \tilde{G}_1^{L/2}(T)$, always holds and connects the reduction of the condensate with the boundary value of $G_1$. The difference of both decreases with $\nu(T)$. For $D=0.1,0.5$ and $1.75$ (Tab.~\ref{tabnu}) the difference is within few percents. For $D=7.5$ and $12.5$ for a better agreement $T$ should be lowered to decrease the critical exponent $\nu(T)$.

Taking $\tilde{G}_1^{L/2}$ as the estimation of the condensate fraction at $T \neq 0$ (when $\nu$ is small), some predictions can be made for experimental systems with a number of bosons beyond direct numerical treatment (see Appendix~\ref{exci}).

Next, we consider the divergence of the momentum distribution as $q \rightarrow 0$. Small system size limits the resolution in this important region to $q\geq 2\pi/L$. To overcome this limitation we employ the ''$1/p^2$ sum rule''~\cite{bog,mart,grif}
\begin{align}
 \lim_{p\rightarrow 0} -G_{11}(p,\omega=0)=\int\limits_{-\infty}^{+\infty}  \frac{\db\omega}{2\pi}\, \frac{A(p,\omega)}{\omega}=\frac{m n_0}{\rho_s p^2} \label{sump2}
\end{align}
which holds independently on the coupling strength.
In the collisionless ($T \ll T_c$, $\rho_s=n$) and low-frequency (hydrodynamic) regime we can apply the ansatz by Gavoret and Nozi{\`e}res~\cite{noz}
 \begin{align*}
   A(q,\omega)=2 \pi (Z(q)+1) \delta(\omega-c q)- 2 \pi Z(q) \delta(\omega+cq),\label{nozas}
 \end{align*}
where $c$ is the compressional sound speed. Substitution of this ansatz in ($\ref{sump2}$) yields both the spectral weight $Z$ and the $T$-dependence of the momentum distribution~\cite{noz}
\begin{align}
 &Z(q)=\frac{m n_0 c}{2 \rho_s q}, \quad N(\omega)=\frac{1}{e^{\beta \omega}-1},\\
 &n_q=Z(q)\left[2 N(cq)+1\right].\label{nqtheor}
\end{align}
The sound speed $c(T)$ can be evaluated from the particle number fluctuations
\begin{align}
&\langle N^2\rangle-\langle N \rangle^2 =\left[\avr{N}^2 k_B T \, \kappa_T \right]/V ,\\
&c=\left[m n \; \kappa_T\right]^{-1/2},\label{sound} 
\end{align}
where $\kappa_T$ is the isothermal compressibility. The sound speed $c$ and the superfluid fraction are given in Tab.~\ref{tab3}.
\begin{figure}
\begin{center}
\vspace{0.0cm}
\hspace{-0.5cm}\includegraphics[width=0.51\textwidth]{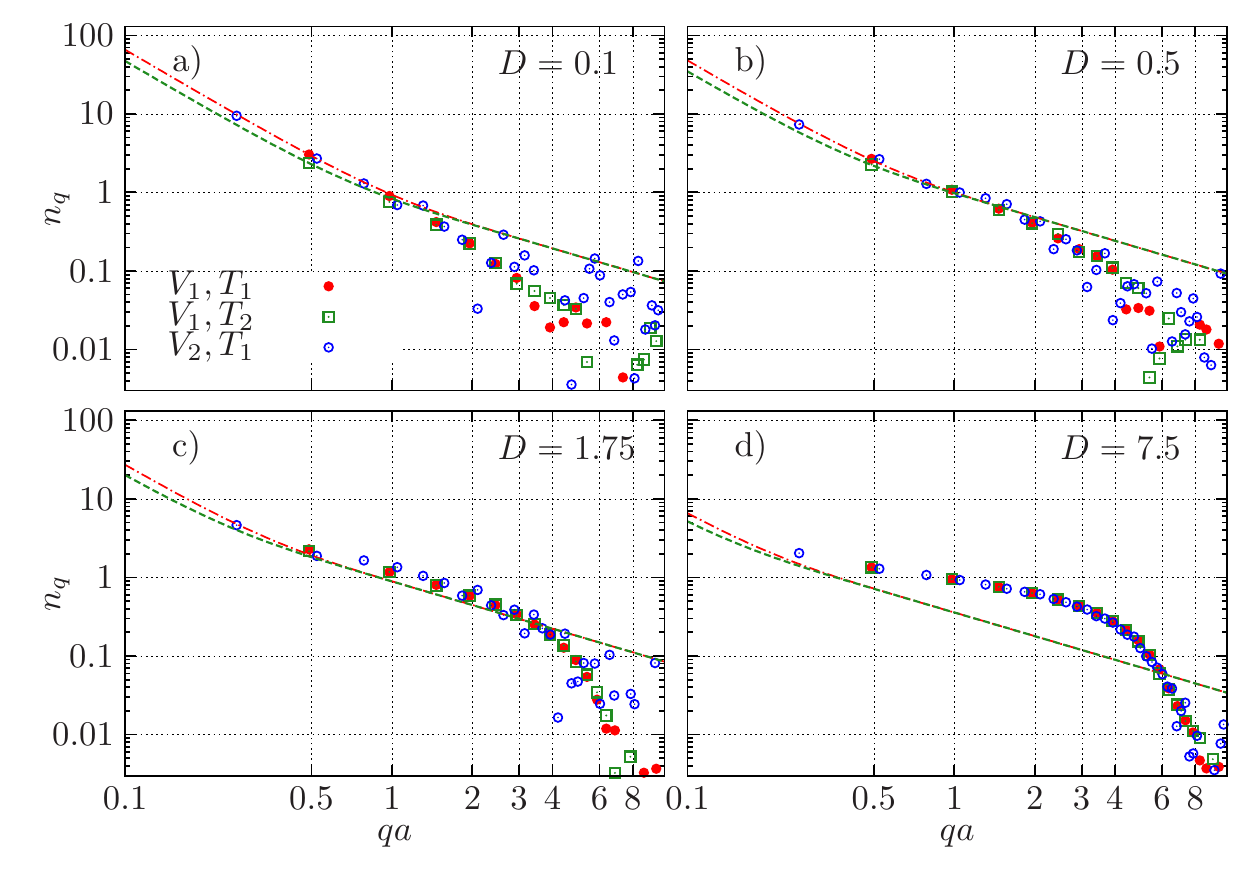}\\
\end{center}
\vspace{-0.9cm}
\caption{(Color online) Long wavelength asymptotics of the momentum distribution $n_q$ in the superfluid phase, $T_{1(2)}=1.0 (0.714)$ vs. $D$ (log-log scale). The dashed lines are the prediction by Eq.~(\ref{nqtheor}) for $T_{1(2)}$. Simulation parameters: $V_{1(2)}=165 (576)$, $T_{1(2)}=1.0(0.714)$.}
\label{fig:qlimit}
\end{figure}

\begin{table}[h]
\caption{The $D$- and $T$-dependence of the compressional sound speed $c(T)$ [$a E_0/\hbar$], isothermal compressibility $\kappa_T(T) \times 10^2$ [$\hbar^2/m$] and the superfluid fraction $\rho_s(T)\equiv \rho_s(T)/n(T)$. Temperatures are $T=1,2$ and $T=3.3$ [$\rho_s(3.3)=0$]. Simulation parameters are the same as in Tab.~\ref{tab1},\ref{tab1a}. Second line for each $D$ is for the system volume $V_2$.} 
\label{tab3}
\begin{tabular}{c|c c c |c c| c c|}
\hline
\hline
 $D$ & $c(1)$ & $c(2)$ & $c(3.3)$ & $\kappa_T(1)$ & $\kappa_T(2)$ & $\rho_s(1)$ & $\rho_s(2)$\\
\hline
\hline
0.1&  2.366(5) & 2.213(5)   & 2.59(2)    & 17.98(8)  &23.4(1)&0.95(1) & 0.032 \\
 &  2.411(7) &  2.26(6)  & 2.72(5)   & 17.15(10)  & 22.0(1.1) & 0.98(3) & 0.004 \\
\hline
0.5&  4.097(8) & 3.863(9)    & 4.06(1)    & 6.024(24) &7.05(4) &0.98(1) & 0.10 \\
   &  4.16(2) & 3.88(2)   & 4.12(2)  & 5.85(4) &  7.0(1) & 0.98(2)& 0.007 \\
\hline
1.75& 6.795(8) & 6.66(3)  & 6.88(4) &2.110(5)   &2.22(2) &1.0& 0.17\\
 & 6.69(3) & 6.46(2)  & 6.67(2) & 2.28(2)   & 2.48(2) & 1.00(3)& 0.005 \\
\hline
7.5&  12.32(2) & 12.38(6)   & 12.40(11)   & 0.659(2)  &0.655(6) &0.95 & 0.007 \\
  &   12.25(5) & --  & --  & 0.688(5)  & -- & 0.95(2) & --  \\
\hline
12.5& 15.51(8) & 15.37(9)   & 15.39(6)   & 0.417(4)  &0.425(5) & 0.81(2)& 0.001\\
 & 15.46(7) & 15.36(8)  & 15.04(13) & 0.435(4)  & 0.431(8) & 0.84(7) &  0.000 \\
\hline
\hline
\end{tabular}
\end{table}

The comparison with the PIMC results for $T_1=1$ and $T_2=0.714$ is shown in Fig~\ref{fig:qlimit}. Two system sizes ($V_1,V_2$) are considered to demonstrate the finite size effects. They are found to be negligible and are within the statistical errors of $n_q$. The agreement with the asymptotic~(\ref{nqtheor}) can be confirmed for the weak ($D=0.1,0.5$) and intermediate coupling ($D=1.75$). For $D\geq 7.5$ we observe the onset of the slope predicted by ~(\ref{nqtheor}). Larger system sizes ($V>10^3$) are required to access smaller $q$-values. Independently, one can check the reconstructed spectral density $A(q,\omega)$ in Sec.~\ref{weak}-\ref{strong}. In the limit $q,\omega\rightarrow 0$, for all coupling strengths $A(q,\omega)$ has a pole at $\omega=cq$, where $c$ coincides with the compressional sound speed. This explains a good agreement in Fig.~\ref{fig:qlimit}a,b. However, for $D\geq 1.75$, a second excitation branch appears in $A(q,\omega)$ (see Fig.~\ref{fig:adyn3d}). Its spectral weight increases 
with $q$ and coupling $D$. This can be the reason for the observed systematic deviation in Fig.~\ref{fig:qlimit}c,d and its onset at smaller $q$ as $D$ is increased. 

\subsection{Collective excitations}\label{dens_ex}

As discussed in Sec.~\ref{intro2}, in the superfluid phase, the single-particle (SP) spectrum is coupled to the spectrum of density fluctuations. The first term $S_{\text{A}}(\vec{q},\omega)$ in Eq.~(\ref{prove1}) describes the scattering of quasiparticles in and out of a condensate, with a sharp $\delta$-like peak quasiparticle dispersion expected at $T\ll T_c$. These sharp features should be present in $S(q,\omega)$ in the superfluid phase and vanish for $T>T_c$. Hence, it is instructive to analyze the $T$-dependence of $S(q,\omega)$ to see any difference in both phases. We first discuss some general features observed for different couplings ($0.1\leq D\leq 12.5$) or densities ($D\sim \sqrt{n}$), and then go into detail in Sec.~\ref{weak}-\ref{strong}. 

\begin{figure}
\begin{center}
\vspace{-0.3cm}
\hspace{-0.5cm}\includegraphics[width=0.51\textwidth]{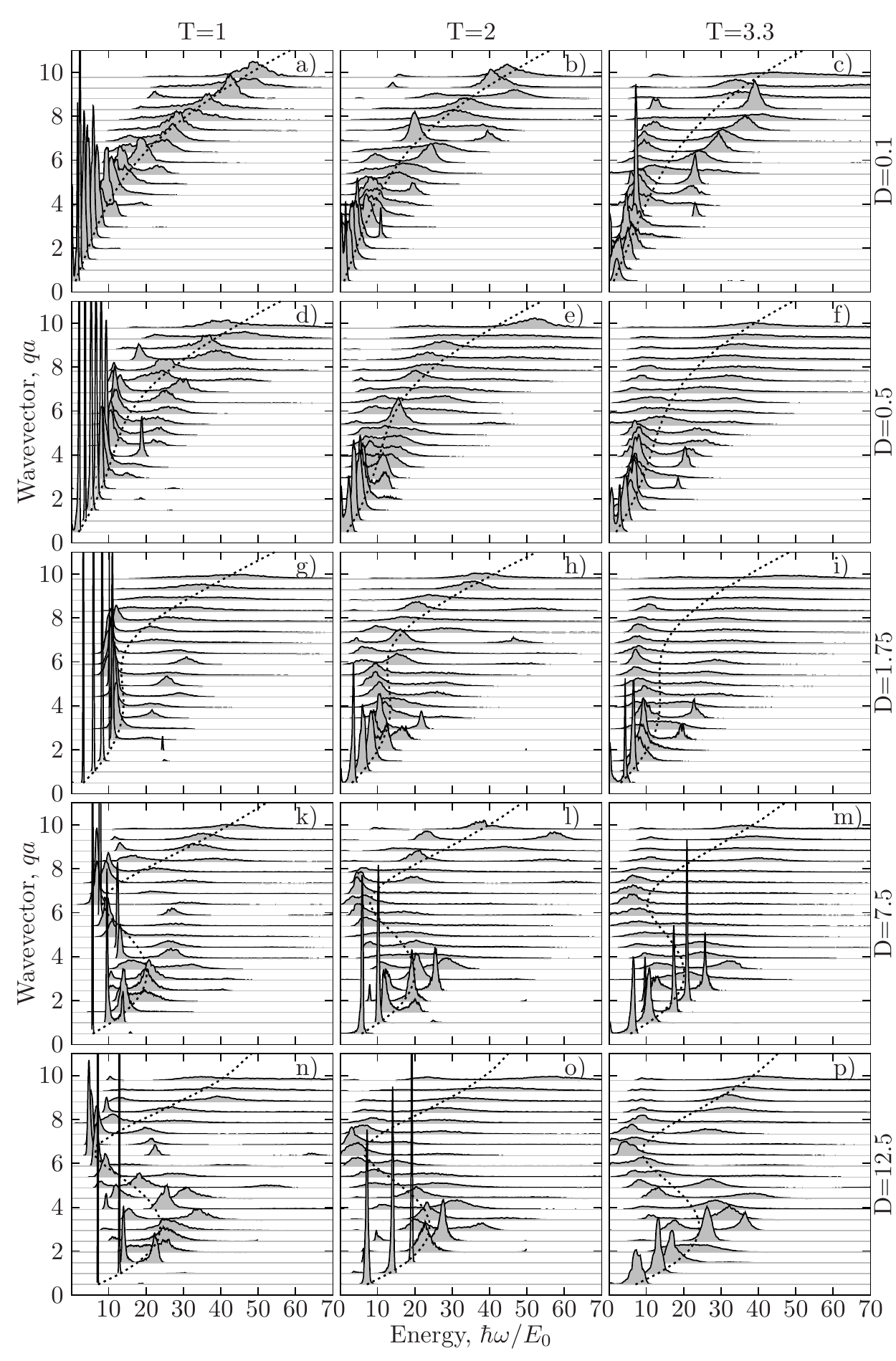}
\end{center}
\vspace{-0.70cm}
\caption{Rescaled dynamic structure factor $S(q,\omega)/S(q)$ at dipolar coupling $D=0.1,0.5,1.75,7.5,12.5$ and  three temperatures $T$. The system is superfluid (normal gas) at $T=1$ ($T=2,3.3$). The discrete wave-numbers, $q =2 \pi n /L \; (n=0,1,\ldots)$, are induced by the periodic boundary conditions. The dashed line (guide to the eye) is the upper bound $\omega_{\chi}$ for the lower ($L$) excitation branch derived from the sum rules (see Sec.~\ref{app}). The upper ($H$) branch lies above.}
\label{fig:sdyn3d}
\end{figure}

The dynamic structure factor reconstructed by the SO method is represented in Fig.~\ref{fig:sdyn3d}. In the superfluid ($T=1$) we observe at least two excitation branches with well defined dispersions (with the sharp energy resonances in the phonon and roton part of the spectrum). With $D$ increasing the onset of the second high energy ($H$) branch systematically shifts to smaller $q$-vectors: for $D=0.1$ ($D=12.5$) it gets a significant spectral weight at $qa \gtrsim6$ ($qa \gtrsim 3$). The $H$-branch does not vanish in the normal phase ($T=2.0,3.3$), but is sufficiently damped or merge with the lower branch. This implies that it is closely related to the multiparticle excitations.

The dynamic structure factor of the upper branch $S_H(q,\omega)$ (see Fig.~\ref{fig:sdyn3d}) is strongly influenced by the statistical noise in the imaginary correlation function  $G_2(q,\tau)$~(\ref{g2}) as discussed in Appendix~\ref{acc_resonst}. The contribution of the high energy features is exponentially damped by the factor $e^{-\tau \omega}$. To accurately resolve the form of $S(q,\omega)$ at large frequencies requires higher accuracy. Moreover, the reconstruction procedure (Sec.~\ref{sto}) has a tendency to underestimate the half-width of the high-frequency structure (Sec.~\ref{bimodal}). See also a note in Appendix~\ref{comment1}.

In contrast, the low-energy ($L$) features can be reproduced more accurately (Sec.~\ref{bimodal}). The $L$-branch remains in the spectrum both at low and high temperatures and shows the temperature induced broadening of the half-width of the spectral peaks characterizing the inverse excitation-lifetime. The dispersion remains well defined in the acoustic range of the spectrum and gets significant broadening at large wavevectors ($qa\gtrsim 8$). Near the origin ($q=0$) the half-width is increased again due to the off-resonant excitations by thermal fluctuations, $\hbar \omega(q)\lesssim k_B T$. For simulated temperatures $T=1.0,2.0$ and $3.3$ in Fig.~\ref{fig:sdyn3d} this effect is well observed for $D\leq 1.75$.

Now we discuss the features which build up due to the correlation effects. For $D\geq 1.75$ and $qa\geq 3$ (see Fig.~\ref{fig:sdyn3d}g,k,n) the $L$-branch dispersion bends down forming a local maximum -- a maxon. At larger wavevectors $qa \in (6,8)$ the roton-minimum is observed. The critical coupling for the roton formation, $D=1.0-1.75$, is in agreement with the previous analyses.~\cite{ast1,Mazz,fil2010} As the present reconstructed spectrum is free of any approximations the roton-depth is found to be lower then reported before. In Ref.~\cite{fil2010} it was found that the upper bound estimate $\omega_{\chi}(q)$ (Eq.~\ref{chiw}) is at least as good as the correlated basis function result (CBF) from Ref.~\cite{Mazz} in the roton-region and predicts a deeper roton-minimum for strong coupling $D\geq 5$. The reconstructed dispersion (see Figs.~\ref{fig:sdyn2}a and~\ref{fig:sdyn3}a) shows that the correct maxon-roton dispersion is even lower than $\omega_{\chi}(q)$. According to the sum rules~(\ref{f-sum})-(\
ref{comp}) presence of an upper $H$-branch and the increase of its spectral weight in $S(q,\omega)$ should push the $L$-branch to lower energies and, correspondingly, deepen the roton feature. We also do not exclude that the discrepancy with the CBF result ($T=0$) is a temperature effect. As was shown in Ref.~\cite{sven1} for superfluid $^4$He one observes a softening of the roton mode once approaching $T_c$ from below, while the peak in $S(q,\omega)$ shifts to lower frequencies.

In 2D dipolar systems the roton-feature is a pure correlation effect which cannot be reproduced by the Bogolubov dispersion, 
$\omega^2(q)=\epsilon_q^2 +2 n_0(T) V(q) \epsilon_q$. The Fourier amplitude $V(q)$ of the dipole potential~\cite{four2d} is positive at all $q$-vectors and cannot lead to the rotonization of the spectrum at weak coupling (low densities), in contrast to 3D geometry.~\cite{santos2003,odell,huf} Even at the lowest coupling considered ($D=0.1$) the dispersion deviates from the Bogolubov result. The basic assumption --  a small depletion of the condensate, is not satisfied for $D\geq 0.1$ (see Tab.~\ref{tab2}). The zero temperature analysis~\cite{Mazz} ends up with the same conclusion.

Next, we analyze the splitting of the dispersion curve into the $L$ and $H$-branch. The onset of splitting can be predicted based on the f-sum rules~(\ref{0-sum})-(\ref{f-sum}). In the superfluid phase both branches are well defined (Fig.~\ref{fig:sdyn3d}a,d,g,k,n) and we can consider the ansatz
\begin{align}
 S(q,\omega)=S_L(q) \delta (\omega-\omega_L(q))+S_H(q) \delta (\omega-\omega_H(q)),\label{slowhigh}
\end{align}
where $\omega_{L(H)}$ and $S_{L(H)}$ define the dispersion and the spectral weight for two branches. Substituted in Eqs.~(\ref{0-sum})-(\ref{f-sum}) the  system of coupled equations can be solved with the result
\begin{align}
 &\omega_H(q)=\frac{\omega_F(q)-\omega_L(q)}{1-\omega_L(q)/\omega_{\chi}(q)},\label{1}\\
 &S_H(q)=\frac{q^2}{2m} \frac{\left[1 -\omega_L(q)/\omega_F(q)\right]}{\omega_H(q)-\omega_L(q)},\label{2}\\
 &S_L(q)=S(q)-S_H(q).\label{3}
\end{align}

The solutions depend on the dispersion of the $L$-branch $\omega_{L}(q)$ assumed to be known from the SO reconstruction. Two upper bounds $\omega_F(q)=q^2/2m S(q,T)$ and $\omega_{\chi}(q)=2n S(q,T)/\abs{\chi(q,T)}$ are defined in Appendix~\ref{app} and can be evaluated via the static structure factor $S(q)$ and the static density response function $\chi(q)$. The energies of the lower branch $\omega_{L}$, due to a slower decay in the imaginary time, can be resolved by the SO-reconstruction more accurately than $\omega_H$ and, therefore, are considered as an input. In addition, $\omega_L$ are less damped at large $q$-vectors compared to the frequencies of the collective modes. The solution for the upper branch $\omega_H$ is compared in Figs.~\ref{fig:sdyn1}a,\ref{fig:sdyn2}a,\ref{fig:sdyn3}a (solid gray line) with the full SO-reconstruction (solid symbols show position of the maxima including the half-width). The SO data are in good agreement except for the $q$-vectors when the frequencies $\omega_L$ and $\omega_{\chi}$ overlap or the $H$-branch is significantly damped. Therefore, such analysis is not applied at $T=2.0$ and $3.3$. In summary, the SO-spectrum provides more information (and more complicated structure) than suggested by the simple ansatz~(\ref{slowhigh}). New parameters (additional excitation branches and their spectral weights) can be added in Eq.~(\ref{slowhigh}), however, to resolve them one needs to know the additional frequency moments $\avr{\omega^n}$.

The restriction on the spectral weight to be positive, i.e. $S_H(q) \geq 0$, predicts the first appearance (at a specific $q$-vector) of the high energy branch $\omega_H$ in the spectrum. This requires $\omega_F(q^*)> \omega_L(q^*)$, see Eq.~(\ref{2}). For $D=0.1$ this occurs at $q^*a\approx 5$, for $D=1.75 (12.5)$ at $q^*a\approx 1.5 (1.0)$. This comes in agreement with the SO spectrum, see Figs.~\ref{fig:sdyn1}a,\ref{fig:sdyn2}a,\ref{fig:sdyn3}a, and  confirms the self-consistency of the reconstructed $S(q,\omega)$ with the $f$-sum rules~(\ref{0-sum})-(\ref{comp}). The accuracy in the fulfillment of~(\ref{0-sum})-(\ref{comp}) evaluated from $S(q,\omega)$ varies in the range $10^{-5}\ldots 10^{-3}$.

The solutions~(\ref{1})-(\ref{2}) can also predict a $q$-dependence of the spectral weights. For $D=0.1$, $S_H(q)$ is enhanced around $qa \approx 6.36$ with $\omega_H$ being slightly above the recoil energy $\epsilon_q$. Next, $S_H(q)$ is slightly increasing while the intensity of the $L$-branch is decreasing. The SO data confirm this variation, see Fig.~\ref{fig:sdyn3d}a. Such a behavior is imposed by almost a constant value of $S(q)$ in Eq.~(\ref{3}) for large momenta.

The theoretical interpretation of the splitting into two branches will be further discussed in Sec.~\ref{weak}. Here we mention two possible scenarios, considering as an example the dispersion for $D=0.1$ (Fig.~\ref{fig:sdyn3d}a): I) After hybridization at $qa \approx 6.4$ the observed lower branch $\omega_L(q)$ is the continuation of the a single-particle (SP) dispersion or a dispersion of collective density modes. The high energy resonances ($H$-branch) correspond to combinations of two or more quasi-particle excitations, II) both dispersions of the SP and collective modes for $qa>6.4$ continue above (not lower) the free-particle dispersion $\epsilon_q$. The $L$-branch appears due to decay processes, then the quasi-particle energy becomes larger than the energy of a quasiparticle pair, i.e. $\omega_L(q)> \omega_L(q_1)+\omega_L(q_2)$ with $\vec{q}=\vec{q}_1+\vec{q}_2$.

\subsection{Single-particle excitations}\label{single_ex}

\begin{figure}
\begin{center}
\vspace{-.20cm}
\hspace{-0.6cm}\includegraphics[width=0.51\textwidth]{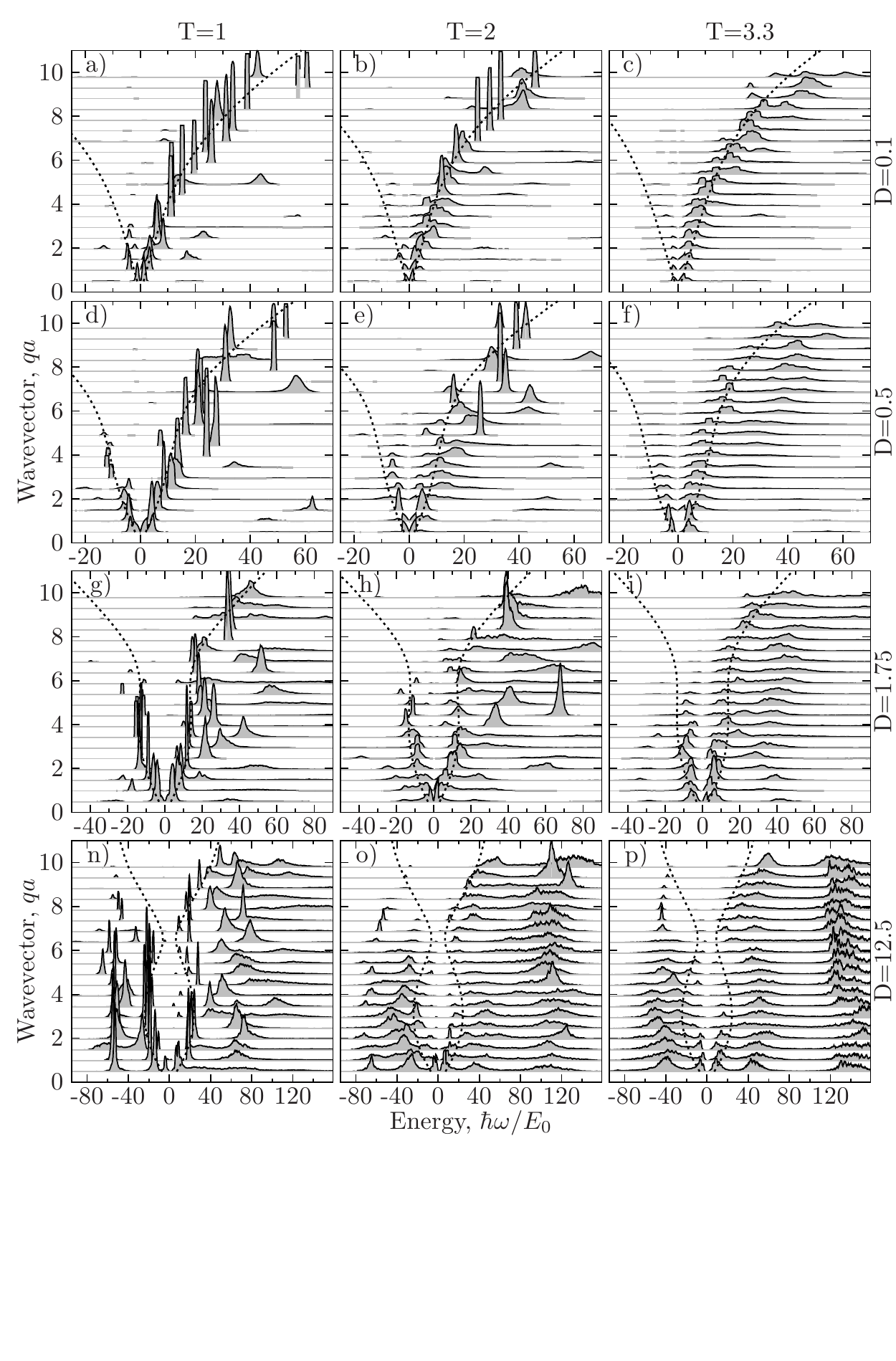}
\end{center}
\vspace{-2.90cm}
\caption{Spectral density $\abs{A(q,\omega)}$ at $D=0.1,0.5,1.75,12.5$ and three temperatures. The system is superfluid at $T=1$. The dashed line is the upper bound $\omega_{\chi}$ for the $\omega_L$-branch in $S(q,\omega)$. The spectral densities are not smoothed and demonstrate the raw-data. The frequency resolution compared to Fig.~\ref{fig:sdyn3d} is lowered from $w_{\text{min}}=0.5$ to $w_{\text{min}}=1.5$ (see Sec.~\ref{sto} for details). Hence, the sharpest features have the smallest possible width equal $w_{\text{min}}$.}
\label{fig:adyn3d}
\end{figure}
The reconstructed spectral density $\abs{A(q,\omega)}$ is demonstrated in Fig.~\ref{fig:adyn3d}. The density at positive $A_{>}(q,\omega)\geq 0$ and negative $A_{<}(q,\omega)\leq 0$ frequencies satisfies the normalization~(\ref{tilde1})-(\ref{tilde2}) and characterizes the excitations that are either internally excited by thermal and quantum fluctuations ($\omega<0$) or excited by the energy transfer ($\omega >0$). The decay of the negative amplitude $\abs{A_{<}}$ with the $q$-vector is due to the normalization~(\ref{tilde2}) and follows the decay of the momentum distribution $n_q=G_1(q,\beta)$ depending both on the dipole strength $D$ and temperature $T$ (see Fig.~\ref{fig:pmom}). The weak $T$-dependence of $n_q$ observed for $D\geq 1.75$ and $qa> 1$ can be directly linked with $A_>$ and $A_<$. Rewritten in the form~\cite{Pitaev}
\begin{eqnarray}
n_q= &&\int_0^{\infty} \frac{d \omega}{2 \pi} [e^{\beta \omega}-1]^{-1} [A_>(q,\omega)+A_<(q,-\omega)] -\nonumber \\ 
&&\int_{-\infty}^0 \frac{d \omega}{2 \pi} A_<(q,\omega)
\end{eqnarray}
the $T$-dependence enters through the first antisymmetric component. If the spectral densities $A_>$ and $A_<$ are antisymmetric, the second term will dominate being proportional to $n_q$. In this case the $T$-dependence does not enter explicitly and appears only indirectly via multi-excitation and damping effects contained in $A_<(q,\omega)$. Therefore, the observed symmetry of the lowest excitation branches $\omega_{>,L}, \omega_{<,L}$ for $D=1.75, 12.5$  in Fig.~\ref{fig:adyn3d} should result in a weak $T$-dependence of the momentum distribution, which is in agreement with $n_q(T)$ in Fig.~\ref{fig:pmom}c,e.

The SP excitation spectrum has a complicated structure and many high-frequency harmonics compared to the collective density modes. For the strong coupling $D=12.5$, to fit the imaginary time decay of the Matsubara Green function~(\ref{g1}) the reconstruction was performed on the enlarged frequency interval. Some high-frequency harmonics are out of the scale shown in Fig.~\ref{fig:adyn3d}. The detailed interpretation of the SP spectrum and check of the corresponding $f$-sum rules is out of the scope of the present paper and requires further analyses.

Below we discuss the differences observed in the reconstructed spectral densities $S(q,\omega)$ and $A(q,\omega)$ in the superfluid and normal phases in  three coupling regimes. Possible experimental realizations were listed in Sec.~\ref{physrel}.

\section{Discussion of the spectra}\label{disc_spec}

\subsection{Weak coupling}\label{weak}

\begin{figure}
\begin{center}
\vspace{-0.0cm}
\hspace{-0.5cm}\includegraphics[width=0.51\textwidth]{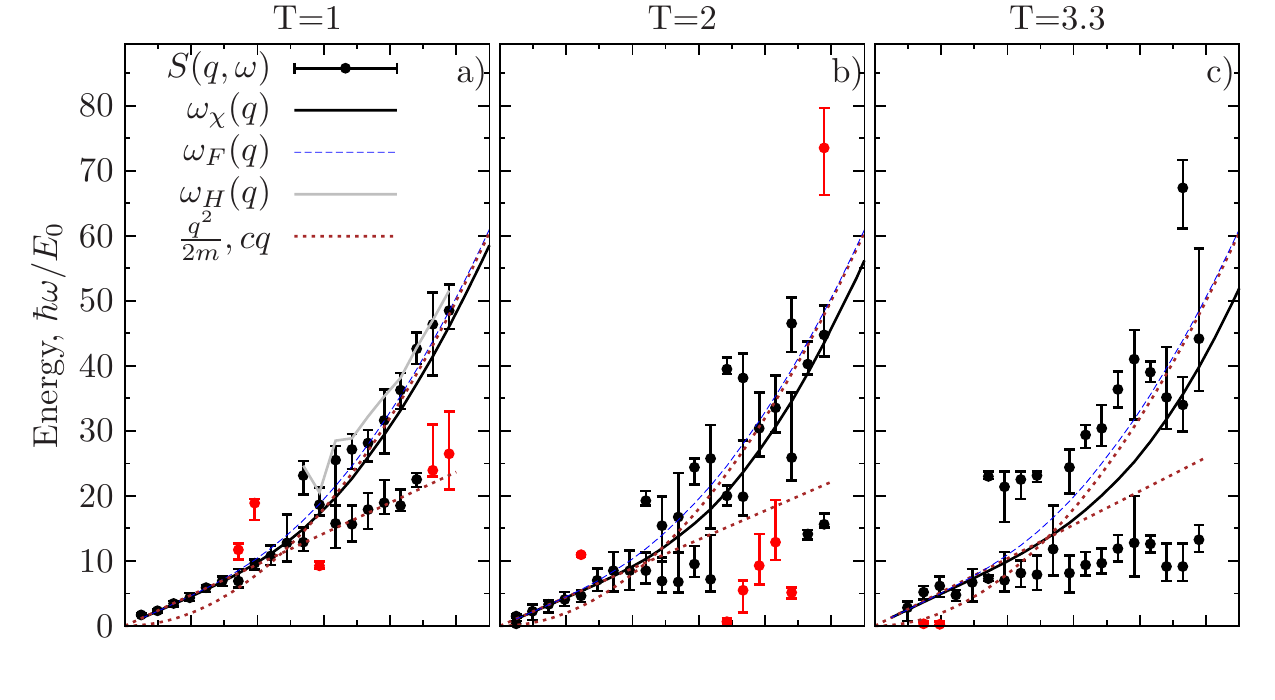}\\
\vspace{-0.58cm}
\hspace{-0.5cm}\includegraphics[width=0.51\textwidth]{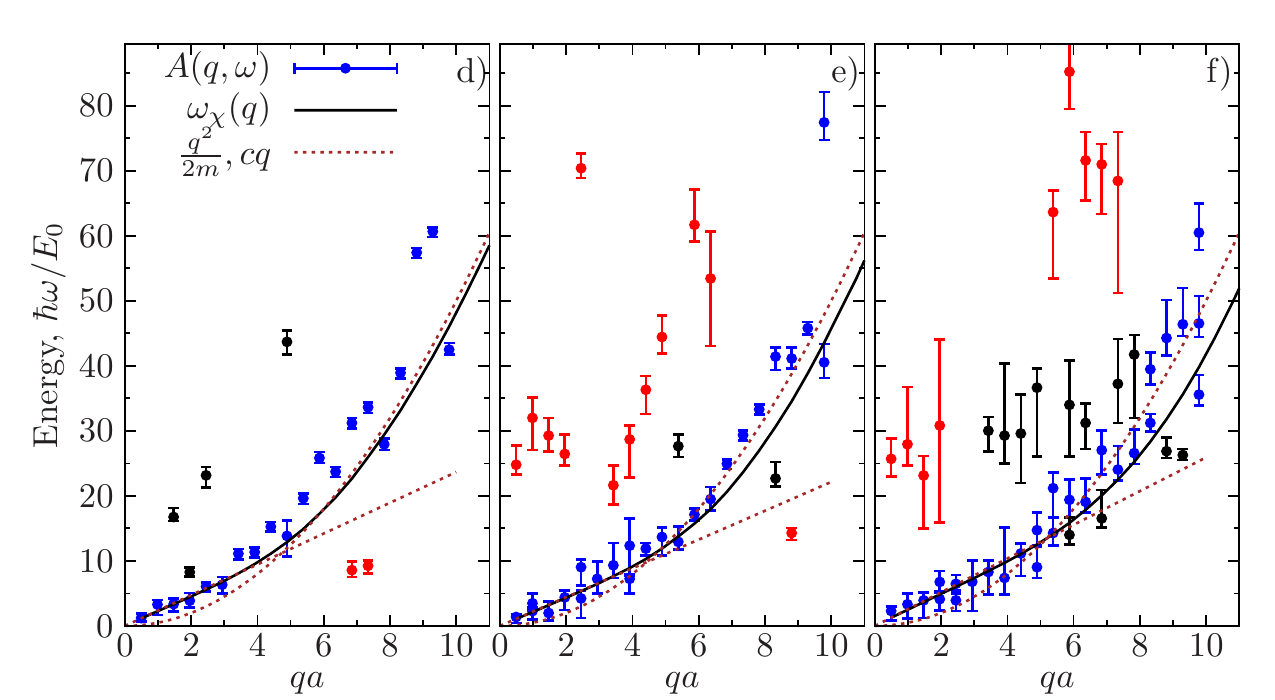}
\end{center}
\vspace{-0.70cm}
\caption{(Color online) Dispersion relations for $D=0.1$: positions of the peaks of $S(q,\omega)$ (a-c) and $A(q,\omega)$ (d-f) and their half-width shown as the error-bars. The full spectral densities $S(q,\omega)$ and $A(q,\omega)$ are shown in Fig.~\ref{fig:sdyn3d} and Fig.~\ref{fig:adyn3d}.
Temperature $T=1$ is below and $T=2.0,3.3$ are above the critical temperature $T_c$ (see Tab.~\ref{tab3}). The simulation parameters are specified in Tab.~\ref{tab1}. The solid (black) line $\omega_{\chi}(q)$ and the dashed (blue) line $\omega_{F}(q)$  are the upper bounds for the lowest excitation branch derived from the compressibility and first-frequency sum rules (see Appendix~\ref{app}). The dotted (brown) lines are the free-particle ($q^2/2m$) and isothermal sound ($cq$) dispersions ($c$ is taken from Tab.~\ref{tab3}). The solid gray line is the solution of Eqs.~(\ref{1})-(\ref{3}) for the high energy branch $\omega_H(q)$ plotted for the $q$-vectors where the spectral weight $S_H(q)$ is positive. The red symbols denote some additional features in the spectrum which are hardly seen in Figs.~\ref{fig:sdyn3d},~\ref{fig:adyn3d}.}
\label{fig:sdyn1}
\end{figure}

First, we analyze $S(q,\omega)$ at $D=0.1$. In Fig.~\ref{fig:sdyn1}a-c we plot the position of the peaks and their half-width. The black-solid symbols denote the main spectral features in $S(q,\omega)$. Three temperatures are compared. In the superfluid phase ($T=1$) in the range $qa<8$ we observe a sharply peaked single excitation branch. The dispersion is accurately reproduced by the upper bound $\omega_{\chi}(q)$ derived from the $f$-sum rules, see Appendix~\ref{app}. The half-width of the peaks characterize the damping effects. The smallest half-width in Fig.~\ref{fig:sdyn1} is limited by the frequency-resolution used in the SO-reconstruction ($\hbar \Delta \omega/E_0=1$, Sec.~\ref{sto}). Except of the region, $\hbar \omega(q) \lesssim k_B T$, the half-width stays close to this lower bound for $qa \leq 5$. For larger $q$ the damping starts to increase systematically (see $T=2,3.3$). In the region ($qa \approx 5$) the dispersion curve crosses the free-particle branch, $\epsilon_q=q^2/2m$, and broadens. 
This can be interpreted as hybridization and level repulsion which occurs whenever two branches cross. In the superfluid ($T=1$) this broadening is only slightly increased with the $q$-vector and simultaneously we find the splitting into the $L$- and $H$-branches. The $L$-branch goes well below the recoil energy and follows the dispersion of the zero sound (ZS). Interestingly, this feature can be only observed in the superfluid phase. Based on the theory of the hybridization of $G_1(q)$ and $\chi(q)$ in the presence of a condensate (their poles are reproduced in each function), this branch should be due to the coupling with the SP spectrum. However, at large $q$ this branch is not observed in the spectral function $A(q,\omega)$. For a further test, one needs to evaluate the two-particle Green function and check the two-particle spectrum where this branch should get a significant spectral weight.    

This behavior is not reproduced in the normal phase, see Fig.~\ref{fig:sdyn1}b-c. The damping is systematically increasing with the $q$-vector which is typical for a normal gas/liquid. In addition, we observe that the $L$-branch is now damped and saturates near a ``plateau'' well below the ZS-dispersion. This shift to lower frequencies is accompanied by the shift of the $H$-dispersion to higher excitation energies. This result is expected from the sum rule~(\ref{f-sum}). We remark that the half-width of the $H$-branch can be underestimated by the reconstruction (see Appendix~\ref{acc_resonst}).

What is common for both superfluid and normal phase is the linear acoustic phonon dispersion, $\omega(q,T)=c(T)q$, which extends to $qa \approx 5$ at $T=1.0$, $qa \approx 4$ at $T=2.0$ and $qa \approx 3$ at $T=3.3$. The results for the isothermal sound speed $c(T)$ are given in Tab.~\ref{tab3}. Its non-monotonic $T$-dependence (included in Fig.~\ref{fig:sdyn1}) can be explained by suppression of the density fluctuations in the pre-superfluid regime ($T\lesssim 2$) and the corresponding non-monotonic behavior of the compressibility $\kappa_T$. The variation of the zero sound~\cite{zero} $\delta c=\abs{1-c(2)/c(1)}$ evaluated from Tab.~\ref{tab3} is most pronounced ($\delta c\sim 7\%$) for weak coupling $D=0.1$. At larger coupling $D$ the density fluctuations due to formation of a local condensate and superfluid density are strongly suppressed by correlation effects (the relative change of the compressibility is also reduced). 

To test the hybridization with the SP spectrum in the superfluid, Fig.~\ref{fig:sdyn1}d-f shows the spectral function $A(q,\omega)$. The statistical error in evaluation of $G_1(q,\tau)$ is larger then in $G_2(q,\tau)$ (compare Fig.~\ref{fig:optgreenq} and Fig.~\ref{fig:optskdyn}), as a result the reconstructed SP dispersion curve at low $q$-vectors (Fig.~\ref{fig:sdyn1}d) shows some statistical fluctuations around the ZS dispersion. At high temperatures the convergence to the linear dispersion (in the range $qa \lesssim 5$) is obscured by the increased damping (Fig.~\ref{fig:sdyn1}e,f). However, at larger $q$ the accuracy of the reconstruction should be improved as the Matsubara Green function is evaluated more accurately with less statistical noise. In comparison with the collective excitations, the overall slope of the SP dispersion is less influenced by temperature.

The linear dispersion observed as $q\rightarrow 0$ is in agreement with the Hugenholtz-Pines sum rule for the self energies~\cite{pines1959} with the result that the spectrum $A(q,\omega)$ (at $T=0$) is gapless in the long-wavelength limit. Similar result for Bose liquids at $T \neq 0$ has been worked out by Cheung and Griffin.~\cite{grif1971} Due to the hybridization the density-density response and one-particle Green functions in the superfluid share the same singularities. At $q\rightarrow 0$ these are the phonon poles as was shown by the field-theoretical calculations of Gavoret and Nozi{\`e}res~\cite{noz} with the velocity precisely equal to the ZS speed. This result is confirmed for superfluid helium both experimentally~\cite{talbot,stirling} and theoretically.~\cite{pines1959,noz,griffinbook,glydebook} This justifies the Landau-Feynman interpretation of the sound waves as elementary excitations in Bose systems. Excited out of a Bose condensate a quasiparticle excites the collective oscillations in a 
superfluid. 

In the dipole system, the linear dispersion terminates at $qa \approx 3- 5.4$ (depending on $T$) with the splitting and broadening. Then the dispersion curve goes slightly above the recoil energy $\epsilon_q$. Fig.~\ref{fig:sdyn1}a shows that the same branch ($H$-branch) is observed in the dynamic structure factor.  This comparison confirms, that, indeed, in the superfluid phase both spectra are coupled. In the superfluid (Fig.~\ref{fig:sdyn1}a,d) at $D=0.1$ both dispersions are linear and coincide with the compressional sound. The broad high-energy branch at $T\geq 2$ and large wavevectors ($qa\gtrsim 5.4$) is identified with the multiparticle excitations. The halfwidth of this dispersion is reduced in the superfluid (Fig.~\ref{fig:sdyn1}a) but not drastically. Still the damping is much larger compared with the lifetime of the SP-mode in Fig.~\ref{fig:sdyn1}d.

Finally, we interpret the $L$-branch in Fig.~\ref{fig:sdyn1}a as a continuation of the ZS mode which terminates (loses its intensity) for $qa\gtrsim 9$. 

\subsection{Intermediate coupling}\label{intermediate}

\begin{figure}
\begin{center}
\vspace{-0.0cm}
\hspace{-0.5cm}\includegraphics[width=0.51\textwidth]{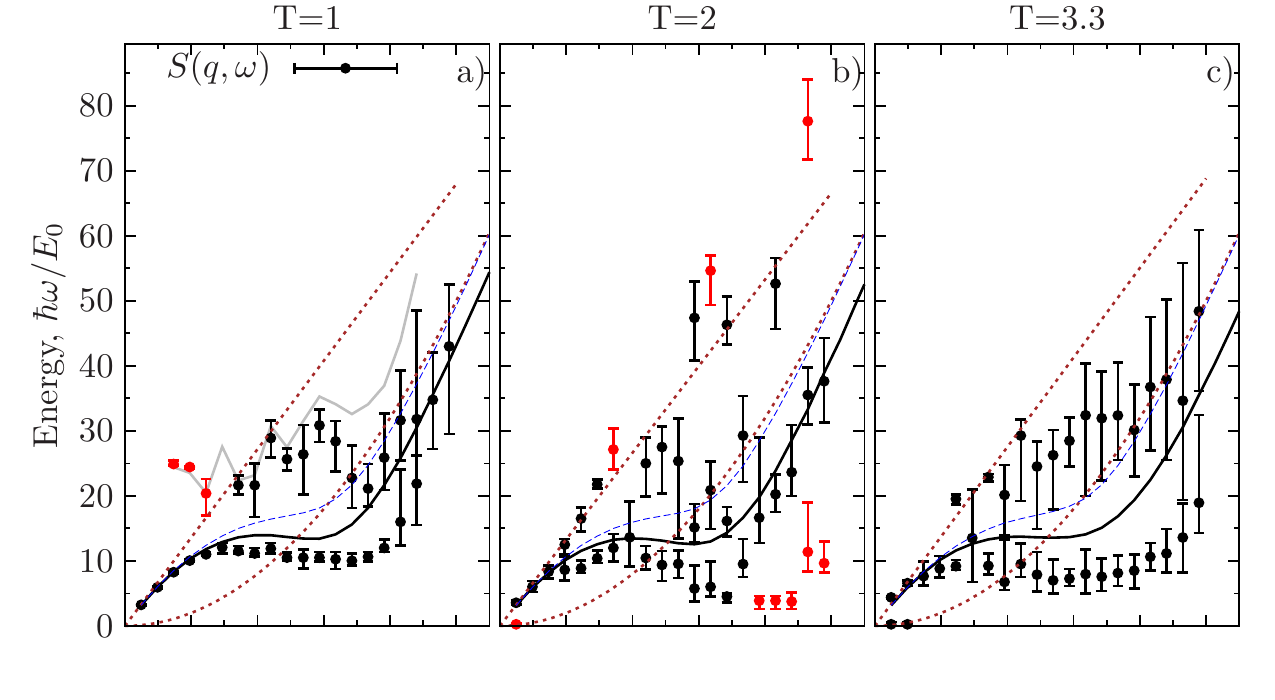}\\
\vspace{-0.58cm}
\hspace{-0.5cm}\includegraphics[width=0.51\textwidth]{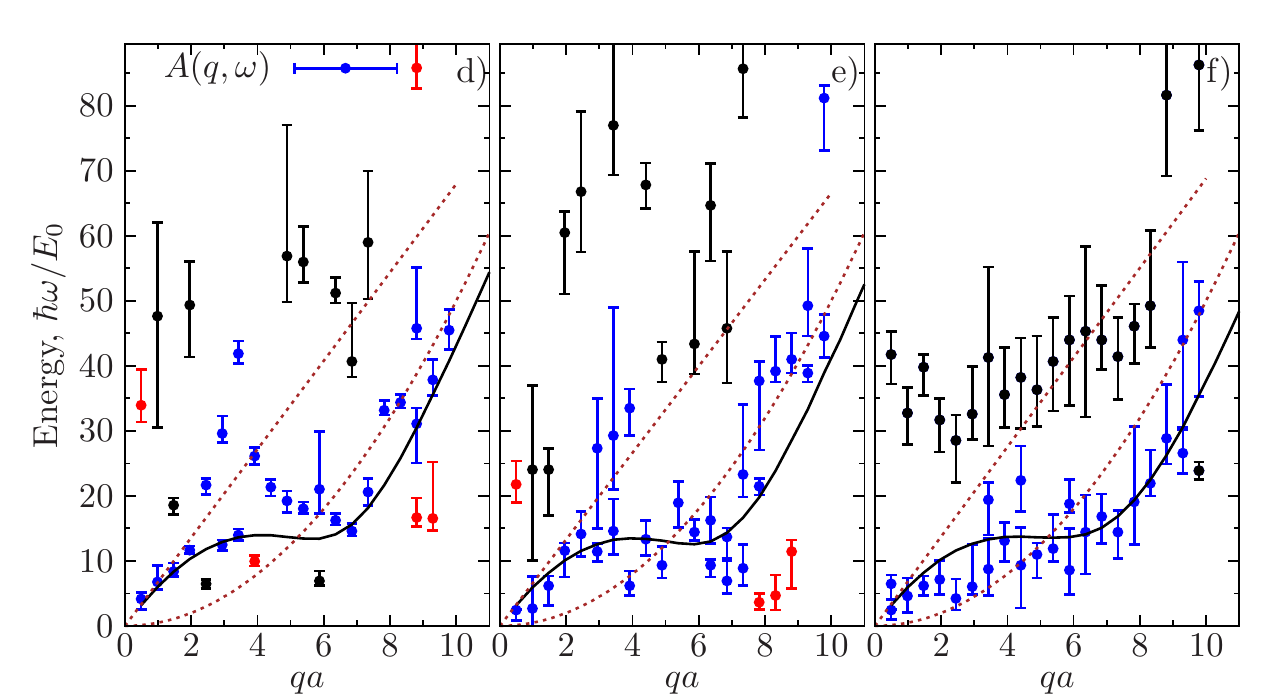}
\end{center}
\vspace{-0.70cm}
\caption{(Color online) Dispersion relations $S(q,\omega)$ (a-c) and $A(q,\omega)$ (d-f) for $D=1.75$.  The half-width of the peaks is shown as the error-bars. See Fig.~\ref{fig:sdyn1} for further details.}
\label{fig:sdyn2}
\end{figure}

In Sec.~\ref{dens_ex} we discussed formation of the maxon-roton branch for $D=1\ldots 1.75$ and failure of the Bogolubov spectrum. At $T=1$ and $qa\leq 3$  the dispersion relation is accurately reproduced by $\omega_{\chi}$, see Fig.~\ref{fig:sdyn2}a. Similar to $D=0.1$ the deviations appear together with the $H$-branch. This branch has a weak $q$-dependence and spans the frequency range $\omega_H \approx 20\ldots 30$. The resonant frequencies can be explained by a combination of low laying excitations, $\omega(q_1)+\omega(q_2)\approx 10+10 \approx 20$. The $q$-dependence is in agreement with the solution $\omega_H(q)$ from Eq.~(\ref{1}). At $qa > 8$ there is an abrupt broadening of the linewidths. For $qa > 9$ the $L$- and $H$-branches merge into a single dispersion which goes slightly below $\epsilon_q$, see Fig.~\ref{fig:sdyn2}a. The peak position is accurately predicted by $\omega_{\chi}$. A similar shift below $\epsilon_q$ was experimentally observed for liquid $^4$He\cite{fak}.

In the presence of the $H$-branch (Fig.~\ref{fig:sdyn2}a) the low-energy dispersion has a well defined maxon-roton feature at $q\in (6,8)$. Its depth is lower than found in the previous analyses.~\cite{Mazz,fil2010} The present approach is not limited by three-particle decay processes of CBF theory~\cite{Mazz,huf} and from first principles includes the spectral weights and energies of all excitation branches, here the $L$- and $H$-branch. They are coupled via the $f$-sum rules~(\ref{0-sum})-(\ref{comp}) as in the simple example~(\ref{1})-(\ref{3}). 
Therefore, the under(over)estimation of one of the branches has a strong influence on the full spectrum. 

In agreement with Gavoret and Nozi{\`e}res,~\cite{noz} in the $q,\omega \rightarrow 0$ limit the density response and SP spectra converge to the ZS dispersion. The sound speed $c$ is given in Tab.~\ref{tab3}. At temperature $T=1$ in the SP spectrum, we observe, in addition, a second excitation branch with a finite gap $\omega_g\approx 40$ at $q=0$, see Figs.~\ref{fig:adyn3d}g,~\ref{fig:sdyn2}d. The gap value can be explained by the excitation of two quasiparticle with the opposite momentum, $\omega(\vec{q})+\omega(-\vec{q})\approx 40$ with $\omega(q)\approx20$ for $qa \in (2,6)$.

We can confirm that the lower dispersions in $S(q,\omega)$ and $A(q,\omega)$ coincide very well up to a maxon region (Fig.~\ref{fig:sdyn2}a,d). In the maxon-roton region, both spectra are not much similar. Starting from $qa\sim 4$ the $H$-branch in Fig.~\ref{fig:sdyn2}a bends up, while the SP dispersion in Fig.~\ref{fig:sdyn2}d goes down approaching a local minimum near the roton wavevector $qa\approx 7$. Both spectra coincide again only for large momenta and follow the $\omega_{\chi}$-dispersion. From these results we can not confirm existence of the {\em unified excitation branch} in the whole range of $q$-vectors, as was discussed in Sec.~\ref{intro} and argued in Ref.~\cite{nepom}

Here, we note that the SP spectra in Fig.~\ref{fig:sdyn2}d-f are more difficult to reconstruct as accurately as $S(q,\omega)$. The statistical noise of the Matsubara Green function $G_1(q,\tau)$ is by factor $10$ larger than in $G_1(q,\tau)$ (see Fig.~\ref{fig:optgreenq}), while positions of the energy resonances are influenced by the noise level as is shown in Appendix~\ref{acc_resonst}. For $A(q,\omega)$ one also needs to accurately reconstruct the additional resonances at $\omega < 0$ (Fig.~\ref{fig:adyn3d}). This can be a reason why for some $q$-vectors the spectrum in Fig.~\ref{fig:sdyn2}d-f does not behave like a continuous dispersion relation. 

Next, we continue to discuss the collective excitations in the normal phase, Fig.~\ref{fig:sdyn2}b,c. The splitting of the dispersion curve, previously observed for $D=0.1$, here is also reproduced. At $T=2$ the roton peak shifts to low frequencies (Fig.~\ref{fig:sdyn2}b). Similar behavior was observed in the normal phase of $^4$He,~\cite{sven1,sven2} where above $T_c$ the softening of the roton mode was found with the roton energy approaching the zero frequency. At high temperature ($T=3$) the roton-feature gets broader and eventually becomes nearly dispersionless and saturates as a ``plateau'' (Fig.~\ref{fig:sdyn2}c). Simultaneously the upper branch demonstrates a large damping and gradually transforms into the free-particle dispersion. In conclusion, our analyses show that the roton mode, which is very sharp in the superfluid phase ($T=1$), remains also in the normal phase of the dipolar gas. Its lifetime drops significantly when the temperature is increased from $T=1$ to $T=2$. Then the damping stays practically constant by further increase to $T=3$.

The final form of the dispersion (lower branch) is typical for a normal gas/fluid. After the phonon part the dispersion slightly bends down. The self-consistent field produced by the density fluctuations cannot support stable short-wavelength collective modes. In the normal phase the dispersion relation can be compared with the results of MD simulations for 2D classical dipoles~\cite{roton-kalman} and predictions of the quasi-localized charge approximation,~\cite{qlca} after the correct mapping of the quantum dipolar coupling $D=p^2 m/\epsilon_b \hbar^2 a$ on the classical coupling parameter $\Gamma_D=p^2/a^3 k_B T$.  

Interestingly, that in the SP spectrum in Fig.~\ref{fig:sdyn2}d also resembles a roton mode but with larger excitation energies, compared to the roton in the density fluctuation spectrum. The slope of the $L$-branch is close to $\omega_{\chi}$ (see Fig.~\ref{fig:adyn3d}). Besides the $L$-branch, there is a well distinguished upper branch. This is a common spectral feature for all coupling strengths (Fig.~\ref{fig:adyn3d}). The $H$-branch is very pronounced for $D \geq 1.75$ and accumulates most of the spectral density for $\omega > 0$.

\subsection{Strong coupling and rotonization}\label{strong}

As a third example we consider strong coupling. First, the dispersion in the superfluid phase will be discussed. For simulated temperatures, at $D=7.5$ and $12.5$, there is a local crystalline ordering with fluctuating orientation. This regime is followed by the freezing transition at $D=17(1)$.~\cite{ast1,buch} At $D=12.5$ the static structure is peaked at $qa=6.65$ and the phonon-maxon-roton feature is well pronounced, see Fig.~\ref{fig:sdyn3d}k,n. Due the $0-$sum rule~(\ref{0-sum}) the rotons are the dominant excitations (the integrated spectral density is proportional to $S(q)$) and the depth of the roton minimum affects the critical temperature of the superfluid transition.~\cite{fil2010} The full reconstruction clearly demonstrates the temperature effect, see Fig.~\ref{fig:sdyn3d}n. Near the roton-minimum at $E_R=\omega_L(6.36)\approx 4.9$ the dynamic structure factor $S(q,\omega)$ shows an asymmetric broadening to lower energies. This also applies to $D=7.5$ (Fig.~\ref{fig:sdyn3d}k) with the 
roton gap $E_R=\omega_L(6.36)\approx 6.95$. For $D=12.5$, the simulated temperature $T=1$ is close to the BKT-temperature of an infinite system ($T_c=1.01$) and the superfluid fraction is reduced to $\rho_s=0.81$ (see Tab.~\ref{tab3}). Therefore, we observe some additional damping compared to $D=7.5$ with $\rho_s=0.95$ and $T_c=1.22$. 

\begin{figure}
\begin{center}
\vspace{-0.0cm}
\hspace{-0.5cm}\includegraphics[width=0.51\textwidth]{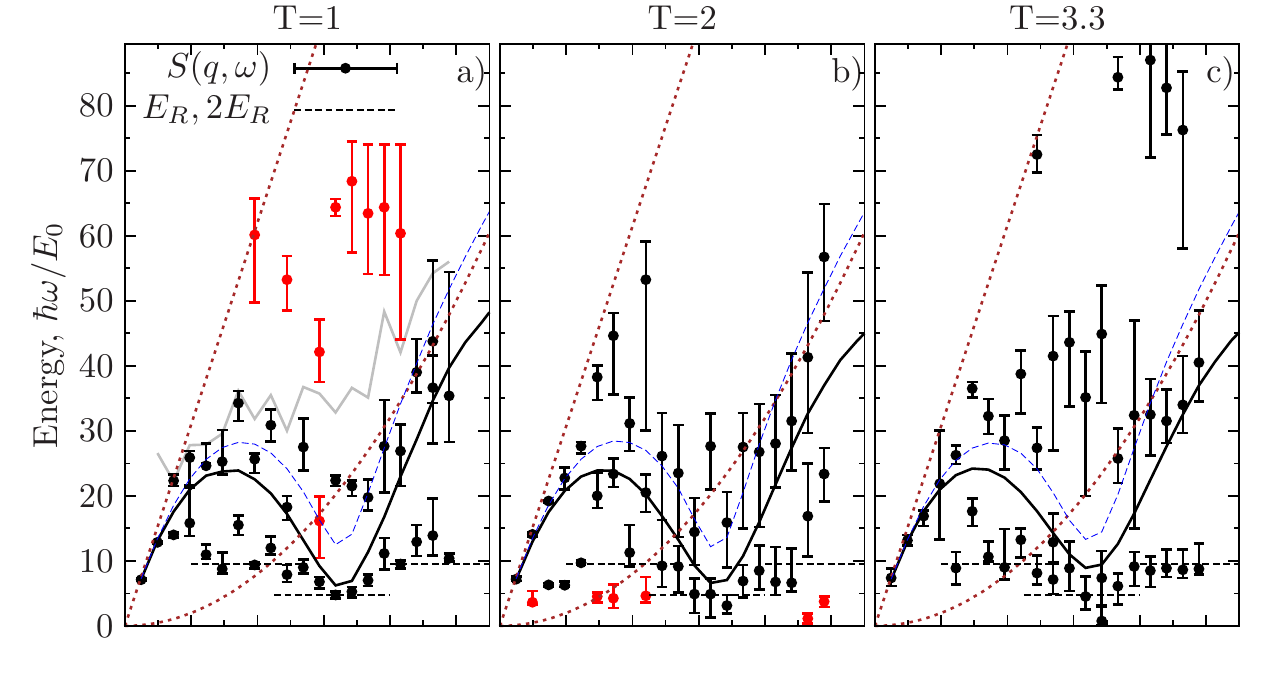}\\
\vspace{-0.58cm}
\hspace{-0.5cm}\includegraphics[width=0.51\textwidth]{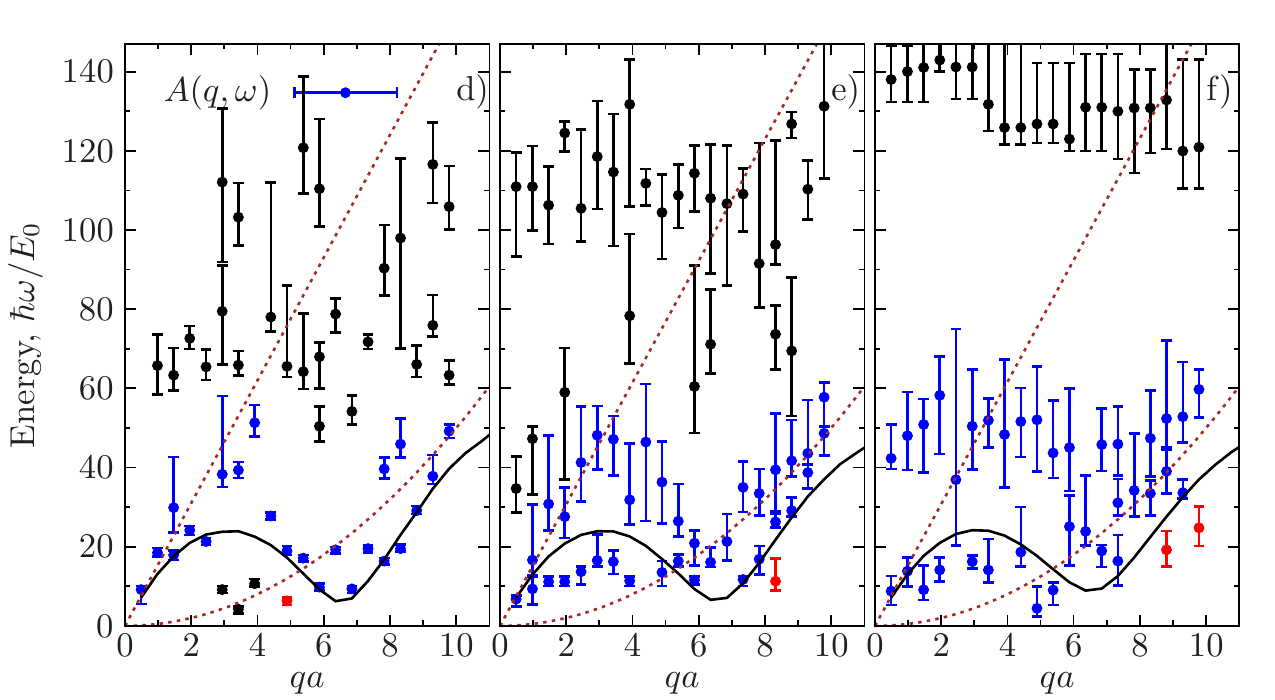}
\end{center}
\vspace{-0.70cm}
\caption{(Color online) Dispersion relations $S(q,\omega)$ (a-c) and $A(q,\omega)$ (d-f) for $D=12.5$.  The half-width of the peaks is shown by error-bars. See Fig.~\ref{fig:sdyn1} for further details. In d)-f) the black color denotes the uppermost branch which is omitted from the discussion in the text.}
\label{fig:sdyn3}
\end{figure}

Except the maxon region, the upper bound $\omega_{\chi}(q)$ reproduces well the L-branch for $qa\leq 8$ including the roton minimum, see Fig.~\ref{fig:sdyn3}a,b,c. The deviations at large $q$ are due to decay processes. Once the quasiparticle energy exceeds the energy of a quasiparticle pair, it becomes unstable and would decay into this pair. For a strongly correlated system the lowest quasiparticle energy corresponds to a roton, $E_R$, and, hence, the dispersion curve should not exceed $2 E_R$. 

Surprisingly, at $D=12.5$ the roton minimum is so deep, that the energy of the two-roton state $2 E_R\approx 10$ is much lower than the maxon energy. As a result we observe this state symmetrically on both sides of the roton minimum at $qa \approx 6.36$ (see horizontal dashed lines in Fig.~\ref{fig:sdyn3}a). A similar effect has been discussed for $^4$He at high pressures (above 18 bars).~\cite{graf} For $D=7.5$ (see Fig.~\ref{fig:sdyn3d}k) the double roton-feature at $2 E_R\approx 14$ is also reproduced. The two-roton state, first predicted by Pitaevskii,~\cite{Pitaev} is well known experimentally for superfluid $^4$He and was partially discussed in Sec.~\ref{intro}. Its experimental verification was complicated, since the sharp energy resonance merges with a broad multi-excitation background and the measurements required a high instrumental resolution.~\cite{glyde_exp1} In contrast, for 2D dipolar gases with $D\geq 7.5$ there is no such complication, since both components are well separated (see Figs.~\ref{fig:sdyn3}a and \ref{fig:sdyn3d}k,n). 

Next, we discuss the maxon region, $qa \in (2,4)$ (Fig.~\ref{fig:sdyn3}a). Here, we observe a splitting of the dispersion. The lowest branch is the two-roton state. 
Its energy is well below  $\omega_{\chi}(q)$ (solid black curve) which corresponds to a prediction of a continuous dispersion which neglects decay processes. It provides an accurate prediction only up to the maxon. Here, the upper branch starts which continues with few oscillations (see Fig.~\ref{fig:sdyn3d}n). In the roton region, $q \in (5,7)$, its energy is in the range $\omega \approx 20-25$ and there is a large gap to the low frequency roton state. At large momenta the $H$-branch merges with the strongly damped dispersion of multiparticle excitations near the recoil energy. Below we give a possible interpretation of the observed behavior.

In the interpretation proposed by Glyde and Griffin~\cite{grif1990} the hybridization of the dispersion near a maxon is due to the crossing of two different branches, i.e the acoustic phonons which dominate at low $q$ and the maxon-roton mode. This cross-over behavior should be characterized by a double peak structure. Indeed, in the simulations we observe a double peak structure, but its origin is different. It comes from the decay of quasiparticles into pairs of lower energy (here the two-roton state). If this decay is not allowed by kinematics then we always observe a continuous dispersion relation (see Fig.~\ref{fig:sdyn1}a-c and~\ref{fig:sdyn2}a-c), in agreement with Ref.~\cite{nepom} The dispersion does not lose its spectral intensity or increases its halfwidth as it would be for the case for hybridization of two distinct branches. Further analyses for the case when $2E_R > E_{\text{maxon}}$ will be helpfull to finally clarify wether the maxon for strongly coupled systems is a transition region between phonons and rotons or a part of a continuous dispersion relation.    

Another important issue concerns the origin of the roton minimum. For $D=1.75$ we have observed a weak rotonization of the spectrum both in the superfluid and normal phases. Its existence is independent on the presence of a condensate. For strong coupling, $D=12.5$, the maxon-roton feature is more pronounced including the normal phase (Fig.~\ref{fig:sdyn3}b,c). Temperature, practically, does not influence the slope of the low-frequency roton-two roton state, but has a noticeable effect on the damping characterized by the half-width. We can conclude, that the roton minimum is not related with the coupling to the SP excitation spectrum, being an intrinsic collective density response mode. Its formation is solely due to the short-range correlations which supports earlier suggestions~\cite{roton1,roton2} and is in agreement with the analysis of 2D dipolar system in the classical limit.~\cite{roton-kalman} 

The next question to be addressed is whether the roton feature can be also found in the SP spectrum $A(q,\omega)$. The peak positions of $A(q,\omega)$ and their half-widths are shown in Fig.~\ref{fig:sdyn3}d-f. We do not discuss the uppermost branch. In contrast to $S(q,\omega)$, the high frequency branches carry the main part of the spectral density both in the superfluid and normal phases. They also dominate over the $L$-branch in the roton-region, see Figs.~\ref{fig:adyn3d}n-p. Two lower branches are well separated at $T=1$, see Fig.~\ref{fig:sdyn3}d. Temperature systematically reduces the energy of the upper branch ($H$) and flattens the lower branch ($L$).

In the superfluid phase (Fig.~\ref{fig:sdyn3}d) the $L$-branch has a roton-like dispersion, resembling the density response roton in $S(q,\omega)$. Similar to the density-roton, the left roton-branch does not reach the maxon maximum (shown by $\omega_{\chi}$), but for $qa \in (2,4)$ decays into quasiparticles with lower energy.

The $T$-dependence of the roton in $A(q,\omega)$ is different from the density-roton in $S(q,\omega)$. Both in the superfluid and normal phases it stays very near or above $\omega_{\chi}$. There is no contradiction as $\omega_{\chi}$ is the upper bound for the collective density modes. The dispersion becomes almost flat at $T=3.3$ (Fig.~\ref{fig:sdyn3}f) and the roton mode completely vanishes. In contrast, the density-roton mode and two-roton states do not change much except for the broadening. In the long wavelength the dispersion converges to the ZS. The lowest branch is gapless ($\omega,q\rightarrow 0$) in agreement with the Hugenholtz-Pines sum rule. 

On the right side ($qa \gtrsim 7$) the roton-branch follows the dispersion predicted by $\omega_{\chi}$ for the density modes. Simultaneously, we observe that there are no excitations at the free-particle dispersion $\epsilon_q$. This is in contrast, with the behavior in the normal phase (Fig.~\ref{fig:sdyn3}e,f). Now, the density of the SP excitations is centered around the free-particle branch $\epsilon_q$. The dispersion is broadened significantly compared with the one in the superfluid phase. From this result we can argue that the hybridization scenario, discussed in Sec.~\ref{intro}, is realized in our 2D system of bosonic dipoles. The poles of the dynamic response function $\chi(q,\omega)$ predicted based on the sum rules result $\omega_{\chi}(q)$ (see Sec.~\ref{app}) can be directly observed in the elementary excitation spectrum $A(q,\omega)$ but only in the superfluid phase, when this is allowed by the hybridization. On the other side, we do not see exactly the same poles in $S(q,\omega)$, as the collective spectrum is strongly renormalized (perturbed) by the quasiparticle decay processes. They shift the positions of the energy resonances and lead to a slightly different dispersion relation. This can be a possible explanation for our numerical results.  

Finally, we note that many additional modes are observed in $A(q,\omega)$ which, probably, are intrinsic to the single-particle excitations. Their interpretation requires further analysis which goes beyond the present work.

\section{Summary}\label{sum}

Using the path integral Monte Carlo technique in the grand canonical ensemble we evaluated the imaginary time density response and the one-particle Matsubara Green's function. The stochastic optimization method was used to accurately reconstruct the dynamic structure factor $S(q,\omega)$ and the SP spectral density $A(q,\omega)$ in a wide range of wavevectors.

As a result, a first-principle treatment of the dynamic structure factor was presented both for weakly and strongly interacting Bose-condensed systems. The temperature-density dependence of the phonon-maxon-roton dispersion, its damping and hybridization was resolved by the simulations. The dispersion relations were compared vs. the upper bound derived from the frequency sum rules. The introduced approach  allows to discuss the physical origin of the roton minimum and is a useful benchmark for theoretical studies and interpretations of experiments on quantum liquids/gases in different physical realizations. As an example, a 2D dipolar Bose gas at {\em weak}, {\em intermediate} and {\em strong} coupling has been analyzed in detail. 

Our goal was to demonstrate that the single-particle and density-response spectra are coupled and share common poles due to hybridization in a superfluid, as was predicted theoretically.~\cite{pines1959,noz,griffinbook,glydebook} Once dipolar bosons become superfluid ($T \leq T_c$), we observe a similar dispersion relation in both spectral densities $A(q,\omega)$ and $S(q,\omega)$. However, we cannot confirm that this dispersion can be observed at all wavevectors as a unified excitation branch~\cite{nepom} when the coupling is strong. In this case, the collective spectrum $S(q,\omega)$ is strongly renormalized by the quasiparticle decay processes. This, according to the frequency sum rules, should necessarily lead to a shift of the $S(q,\omega)$-maxima from the position of the original energy resonances. One prominent example is the decay into the two-roton state observed in the strongly correlated Bose gas.

Further, we found that the two-roton state exists independently of the presence of a condensate and, therefore, conclude that the roton mode has a classical origin due to short-range particle correlations. Its observation in classical liquids is, probably, prevented by the softening and overdamping.~\cite{sven1,sven2}

For quantum dipolar liquids at high coupling/density (here $D \geq 7.5$) the reconstructed $S(q,\omega)$ predicts a pronounced roton minimum. The sharp resonant roton feature is transfered from a superfluid to a normal phase but with a significant damping. Our reconstruction of the SP spectrum $A(q,\omega)$ in a superfluid also predicts the roton-like feature which, however, vanishes in the normal phase. Here it is substituted by the excitation mode peaked around the recoil energy.

The common features of the density fluctuation spectrum $S(q,\omega)$ are the following. For large momenta ($qa > 8$) we observe a strongly damped mode near the recoil energy both in the superfluid and normal phases. As expected, the phonon dispersion at small wavevectors exists independently of the presence of spatial coherence in the one-particle density matrix. 

These interesting aspects of the excitation spectrum near $T_c$ may be addressed in the near future in the ongoing experiments on dipolar systems.~\cite{pfau,baranov} The analyzed dipolar coupling strengths can be realized in 2D systems of magnetic atoms, polar molecules and indirect excitons as discussed in Sec.~\ref{physrel}.

\section{Acknowledgements}

We thank A.~Pelster, T.~Gasenzer and V.I.~Yukalov for discussions.

\appendix

\section{Model spectral densities: accuracy of the reconstruction}\label{acc_resonst}

To demonstrate accuracy of the reconstruction procedure, below we consider a test for several spectral densities $S(\omega)$. The goal is to analyze how a level of statistical noise, present in the imaginary correlation function, influences  positions of energy resonances and their half-width. To characterize deviations, in addition, we evaluate the frequency moments $\avr{\omega^n}$ to check fulfillment of the sum rules~(\ref{0-sum})-(\ref{comp}). In our model, they are substituted by the exact integral properties of $S(\omega)$. Here, we analyze a dynamic structure model and assume that the detailed balance holds, $S(-\omega)=e^{-\beta \omega} S(\omega)$.

Further, for convenience, $S(\omega)$ is substituted by a similar distribution $S^0(\omega)$ discretized into rectangles of width $\Delta \omega$
\begin{align}
S^0(\omega)|_{\omega \in [\omega_i-\frac{1}{2}\Delta\omega,\omega_i+\frac{1}{2}\Delta \omega)}=S(\omega_i),\; \omega_i=(i-\frac{1}{2})\Delta \omega. \label{dis}
\end{align}
By discretization evaluation of the correlation function can be easily performed as
\begin{align}
 G^0(\tau)=2\sum\limits_{i=1}^N  S(\omega_i) \left[\sum_{t=\tau,\beta-\tau} \frac{1}{t}  e^{-\omega_i t} \sinh \frac{\Delta \omega t}{2}\right].\label{g0tau}
\end{align}
As a second benefit, the integral properties (frequency moments) in the basis of rectangles $\{h_i=S(\omega_i),w_i=\Delta \omega,c_i=\omega_i\}$ can be written down explicitly and evaluated via
\begin{align}
 \avr{\omega^0}=&\sum\limits_{i=1}^{N} h_i\left[ w_i + \frac{2}{\beta} e^{-\beta c_i} \sinh \frac{w_i \beta}{2} \right], \label{m1} \\
 \avr{\omega^1}=& \sum\limits_{i=1}^{N} h_i w_i c_i- \frac{2 h_i}{\beta} e^{-\beta c_i} \left[\frac{1}{\beta}+c_i \right] \sinh \frac{w_i \beta}{2} \nonumber \\
 &+ \frac{h_i w_i}{\beta}e^{-\beta c_i} \cosh \frac{w_i \beta}{2}, \label{m2} \\
 \avr{\omega^{-1}}=& \sum\limits_{i=1}^{N} h_i \ln \frac{c_i+w_i/2}{c_i-w_i/2} \nonumber \\
 &-h_i \left[\text{Ei}(-\beta (c_i+w_i/2))-\text{Ei}(-\beta (c_i-w_i/2)) \right],\label{m3}
\end{align}
with $\text{Ei}(x)=\int_{-\infty}^{x} \db t\, e^{t}/t$. To avoid the divergence in Eq.~(\ref{m3}) when $(c_1-w_1/2)=0$, we set $S^0$ to be non-zero for $ \omega \in [\delta,\omega_{\text{CO}}+\delta]$ (with $\delta=10^{-3}$ and the cut-off frequency $\omega_{\text{CO}}=100$) and choose, correspondingly, $\omega_i=(i-\frac{1}{2})\Delta \omega+\delta$. 

Eqs.~(\ref{m1})-(\ref{m3}) can be used to evaluate the frequency moments ($f$-moments) of the ensemble average~(\ref{finals}) by applying to each term in the ensemble. This result was used to check the sum rules of the reconstructed $S(q,\omega)$ for 2D dipolar bosons. However, as we discuss below, an accurate fulfillment some of the sum rules [relative error $10^{-5}\ldots 10^{-3}$] does not necessarily guarantee a correct reconstruction of the shape of $S(\omega)$.

In our tests $S^0(\omega)$ is specified by a linear combination of $M$ Gaussians
\begin{align}
 S(\omega)|_{\omega\geq 0}=\frac{1}{P}\sum\limits_{j=1}^M \frac{p_j}{\sqrt{2 \pi}\sigma_j} e^{-(\omega-\Omega_j)^2/2 \sigma_j^2},\; P=\sum\limits_{j=1}^M p_j,\label{swmodel}
\end{align}
which vary in the peak position $\Omega_j$, the variance $\sigma_j$ and the weight factors $p_j$. Similar functions has been used in Ref.~\cite{reatto} to test a GIFT reconstruction procedure (the genetic inversion via falsification of theories). We refer to Ref.~\cite{reatto} for the complementary tests and discussion of limitations by reconstruction from imaginary times. 

To simulate the statistical noise inherent to QMC simulations, the imaginary time correlation function is perturbed by a random noise $\nu$
\begin{align}
 G(\tau_i)=G^0(\tau_i)+\gamma\cdot \nu  \delta G(\tau_i), \; \tau_i=i \delta \tau \label{gtautest}
\end{align}
sampled from the normal Gaussian distribution and multiplied by the statistical error $\delta G(\tau_i)$ at imaginary time $\tau_i$. The time dependence $\delta G(\tau_i)$ was taken from a typical PIMC run (e.g. a dipolar gas at $D=1.75$ and $\beta=1$) and rescaled by factor $\gamma$ ($1/5 \leq \gamma \leq 5$) to model the situation when the reconstruction is performed for a larger or smaller MC sample size. Following typical PIMC data presented in Fig.~\ref{fig:g2}, as setup parameters we used: the inverse temperature $\beta=1$, the time step $\delta \tau=1/100$ ($\tau_i=i\delta \tau \in [0,\beta/2]$ and $i=0\ldots 50$) and the statistical error $\delta G(\tau_i)$ of the order $10^{-5} \ldots 10^{-3}$ [with $G(\tau=0)\sim 1$ due to normalization in Eq.~(\ref{swmodel})]. 

How particular spectral features can be downgraded by statistical errors is clarified by few examples.

\subsection{Spectral density with double peak}\label{twopeak}

This example is common for experimental and theoretical studies of superfluid helium when two distinct excitation branches can be resolved in the low temperature ($T< T_c$) spectrum. The reference spectral density $S^0(\omega)$ is specified by the following parameters: discretization width in Eq.~(\ref{dis}) $\Delta \omega=0.5$, two Gaussians with $\Omega_1=10,\Omega_2=25$, $\sigma_1=0.5$, $\sigma_2=2.0$, $p_1=1$ and $p_2=3$. Parameters used in the reconstruction [see Eq.~(\ref{g2_sym}) in Sec.~\ref{sto}]: the cut-off frequency $\omega_{\text{CO}}=100$, the allowed width of rectangles $w_i\geq 0.5\, (i=1,N)$ in the basis of size $N=80$, and $M=200-400$ solutions in the ensemble average~(\ref{finals}). 

\begin{figure}
 \begin{center}
 \vspace{-0.0cm}
\hspace{-0.5cm}\includegraphics[width=0.51\textwidth]{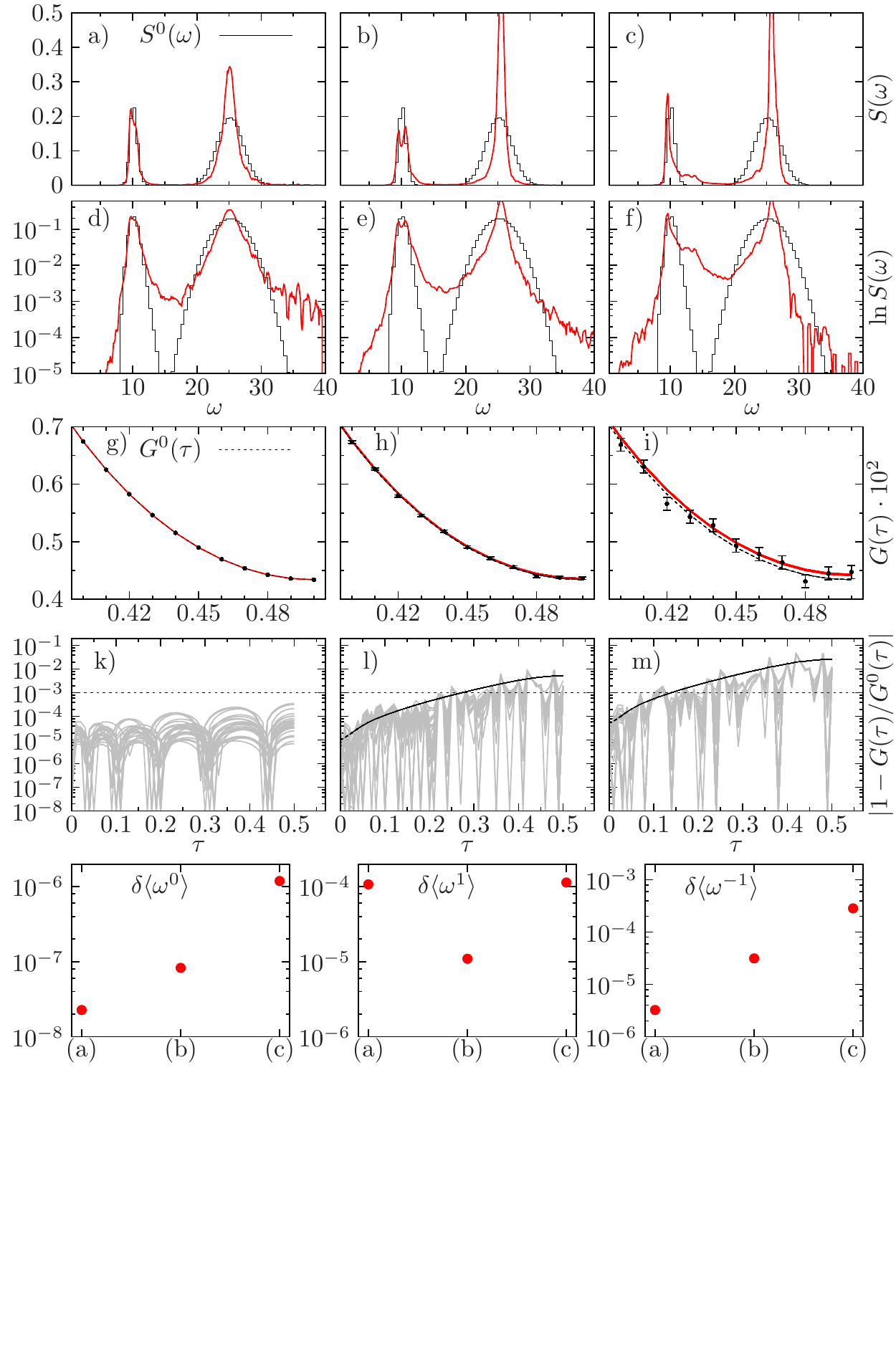}
 \end{center}
 \vspace{-3.60cm}
 \caption{(Color online) (a-f) Reconstructed spectral density $S(\omega)$ (solid red line) compared with the reference $S^0(\omega)$ (step-like black curve) consisting of two energy resonances [(d-f) on the logarithmic scale]. Three columns group the results for three cases of the relative statistical error $\delta G(\tau_i)/G^0(\tau_i)$ shown by a solid (black) line in (l-m). First column (a,d,g,k) is the reconstruction from the error-free [$\delta G(\tau)=0$] correlation function $G^0(\tau)$ shown by dashed (black) line in (g-i). The bold symbols with error bars specify the data [$G(\tau_i) \pm \delta G(\tau_i), \; i=0\ldots 50$] used in the error-biased reconstruction (second and third columns). Different solutions $G_n(\tau)$ obtained by the stochastic optimization are shown by solid (red) lines but are practically indistinguishable on the present scale. Each solution corresponds to a spectral density $S_n$ used in the ensemble average~(\ref{23})  Only part, $\tau \in [0.40,0.50]$, of the full imaginary 
time interval $[0,\beta=1]$ is shown. (k-m) Solid (gray) lines show the relative deviations $\abs{1-G_n(\tau_i)/G^0(\tau_i)}$ from the exact correlation function. Only $25$ from $n \sim 200 \ldots 400$ samples are shown. Note, that the deviations (panels l-m) are within the statistical error $\delta G(\tau_i)/G(\tau_i)$ [solid (black) line]. Horizontal line $10^{-3}$ is guide for the eye. Three lower panels show the relative error of the ensemble average frequency moments~(\ref{m1})-(\ref{m3}), i.e. $\delta \avr{\omega^n}=\abs{1-\avr{\omega^n}_{S(\omega)}/\avr{\omega^n}_{S^0(\omega)}}$, where $S(\omega)$ is shown in panels (a-c) and denoted on the x-axis as (a), (b) and (c). See text for further details and used parameters.}
 \label{fig:model2}
 \end{figure}

Fig.~\ref{fig:model2}a-f demonstrates the results. In the error-free case ($\delta G(\tau_i)=0$, first column) the reconstructed spectrum (solid red line) quite accurately reproduces the position and half-width of the first peak in $S^0(\omega)$ (step-like curve) even on the logarithmic scale, see Fig.~\ref{fig:model2}d. The center of the second peak is accurately captured but the shape is perturbed. The reconstruction overestimates the spectral density at the peak center and the wings. The corresponding relative deviation from the reference correlation function $G^0(\tau)$ is shown in Fig.~\ref{fig:model2}k and is of the order of $10^{-6} \ldots 10^{-5}$. Here we show the deviations of the correlation functions $G_n(\tau_i)$ corresponding to difference spectral densities $S_n(\omega)$ used in the ensemble average [only $25$ samples are shown from typical $n=200 \ldots 400$]. This is the best result we can achieve with the stochastic optimization method (Sec.~{\ref{sto}}) using as an input the error-free 
correlation function $G^0(\tau)$.  

An increase of the basis size in Eq.~(\ref{g0tau}) to $N=160$ does not bring any substantial improvement. The reconstruction result is mainly determined by the level of statistical noise. We also tried to fix the $f$-moments~(\ref{m1})-(\ref{m3}) to their reference values $\avr{\omega^n}_{S^0}$ ($n=-1,0,1$). They were included in~(\ref{dn}) via a set of Lagrange multipliers and subjected to minimization. However, the spectrum shape was downgraded by appearance of systematically shifted peaks which lead to an exact fulfillment of the sum rules at a price of a larger deviation from the reference function $G^0(\tau)$ along the imaginary time. Surprisingly, even if the moments are not fixed,  their relative error  $\delta \avr{\omega^n}=\abs{1-\avr{\omega^n}_{S(\omega)}/\avr{\omega^n}_{S^0(\omega)}}$ is small, see lower panels of Fig.~\ref{fig:model2}. 

By turning on the noise in the input data~(\ref{gtautest}) the peak positions are slightly perturbed, see Fig.~\ref{fig:model2}b,c,e,f. The accuracy in the peak position is acceptable even in the worst case when the error $\delta G(\tau_i)/G(\tau_i)$ is in the range of $2\%$ for $\tau \rightarrow 0.5$, Fig.~\ref{fig:model2}i,m. However, the half-width of the second peak is underestimated and should not be taken seriously in a reconstruction from QMC data with similar error level.

The advantage to use the ensemble average~(\ref{finals}) instead of individual solutions $S_n(\omega)$ becomes evident from the panels k-m. While one solution $G_n(\tau)$ has a large error at a particular imaginary time, it is reduced by the averaging with other solutions. 

\subsection{Bimodal distribution with overlap}\label{bimodal}

\begin{figure}
 \begin{center}
 \vspace{-0.0cm}
 \hspace{-0.5cm}\includegraphics[width=0.51\textwidth]{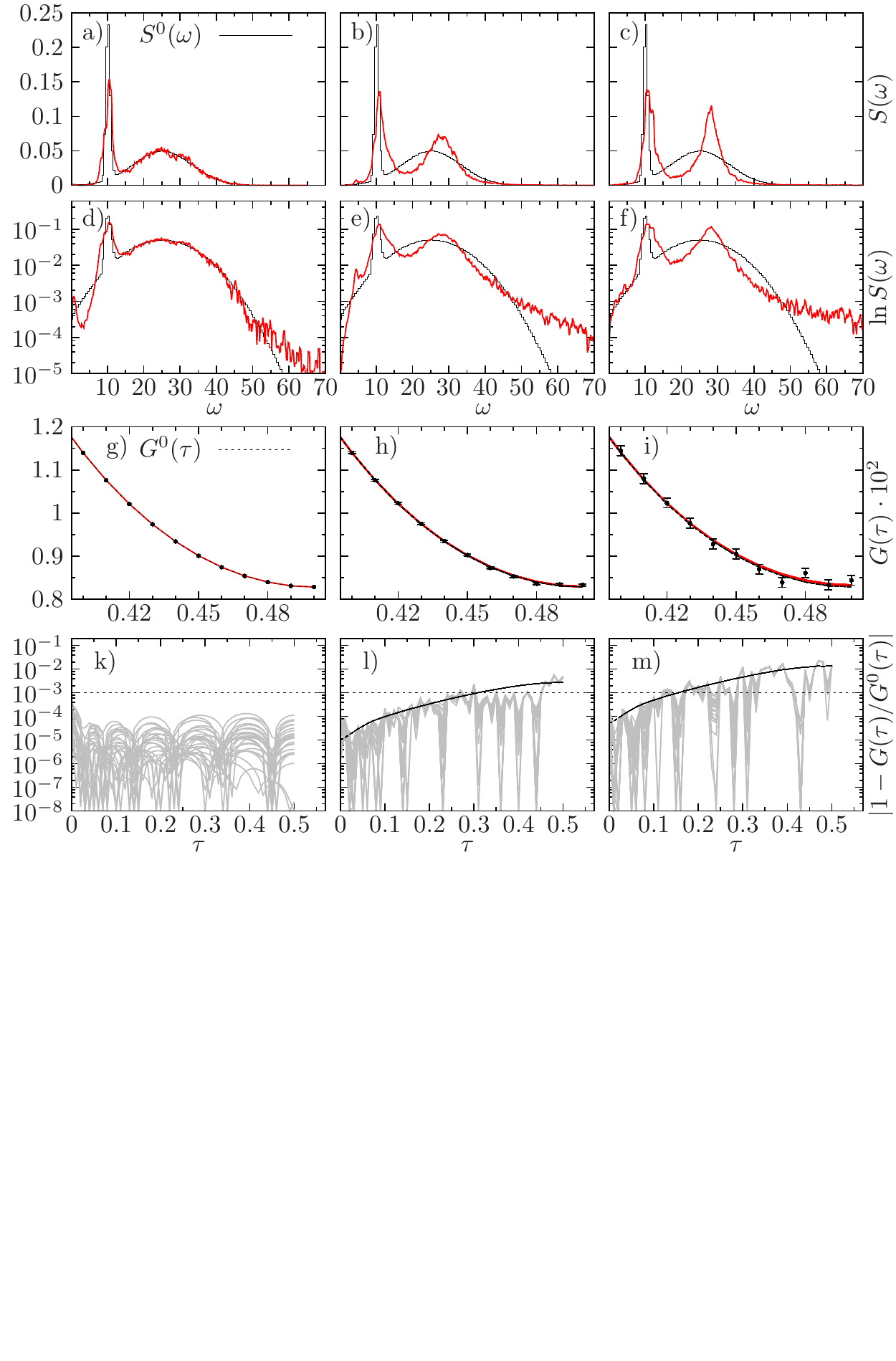}
 \end{center}
 \vspace{-5.90cm}
 \caption{(Color online) (a-f) Reconstructed spectral density $S(\omega)$ (solid red line) compared with $S^0(\omega)$ (step-like black curve). (d-f) The same on the logarithmic scale. For further details see text and caption of Fig.~\ref{fig:model2}. For the relative errors $\delta \avr{\omega^n}$ see text.}
 \label{fig:model3}
 \end{figure}

\begin{figure}
 \begin{center}
 \vspace{-0.0cm}
 \hspace{-0.5cm}\includegraphics[width=0.51\textwidth]{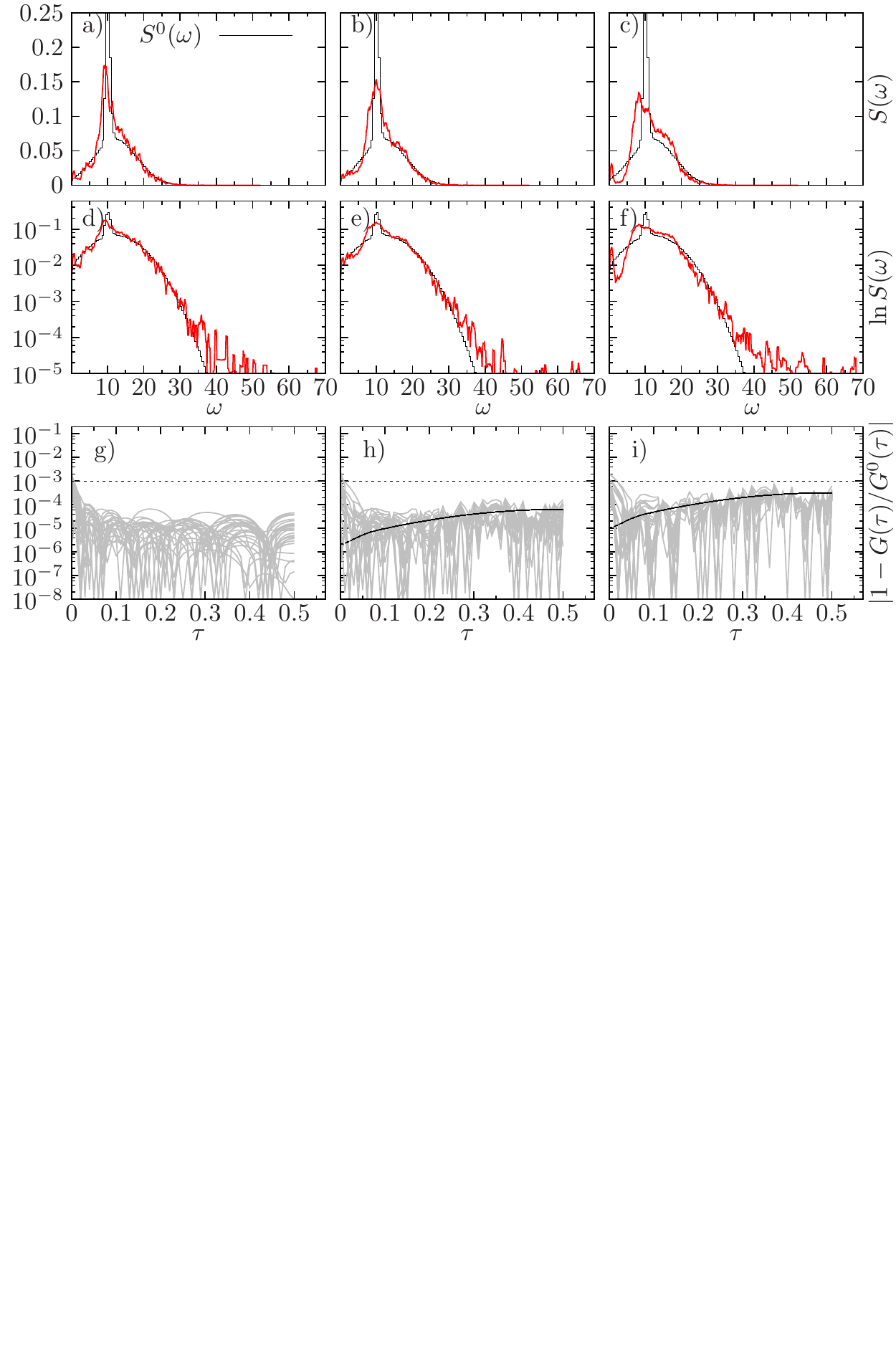}
 \end{center}
 \vspace{-8.10cm}
 \caption{(Color online) (a-f)  Reconstructed spectral density $S(\omega)$ (solid red line) compared with $S^0(\omega)$ (step-like black curve). (d-f) The same on the logarithmic scale. For further details see text and caption of Fig.~\ref{fig:model2}. The plots of correlation functions $G_n(\tau_i)$ are omitted. For the relative errors $\delta \avr{\omega^n}$ see text.}
 \label{fig:model4}
 \end{figure}

In a second example we consider the bimodal distribution consisting of two overlapping Gaussians. A sharp peak models a quasi-particle resonance on a broad multi-excitation background. The reference spectral densities in Fig.~\ref{fig:model3} and Fig.~\ref{fig:model4} are defined, correspondingly, by [$\Omega_1=10,\Omega_2=25$, $\sigma_1=0.5$, $\sigma_2=8.0$, $p_1=1$, and $p_2=3$] and [$\Omega_1=10,\Omega_2=12$, $\sigma_1=0.5$, $\sigma_2=6.0$, $p_1=1$, and $p_2=1$].

As in the first example, the reconstruction in the error-free case (first column) accurately reproduces the shape and the decay at large frequencies on the log-scale (see Fig.~\ref{fig:model3}d and Fig.~\ref{fig:model4}d). The deviation along imaginary time (Fig.~\ref{fig:model3}k, Fig.~\ref{fig:model4}g) are within the range $10^{-8} \ldots 10^{-5}$. The errors in the $f$-moments of $S(\omega)$ in Fig.~\ref{fig:model3}a  are $\delta \avr{\omega^n}=1.3\cdot 10^{-6}, 2.4\cdot 10^{-5}, 2\cdot 10^{-4}$, correspondingly, for $n=-1,0,1$.

The errors in $\delta \avr{\omega^n}$ are due to a slightly different shape of $S(\omega)$ and the discretization parameter $\Delta \omega$ in Eq.~(\ref{dis}). The step-like shape of the reference distribution $S^0$ can not be reproduced in the smoothed ensemble average result. This results in different $f$-moments even if the shape of both densities is close. 


Second and third columns in Figs.~\ref{fig:model3},\ref{fig:model4} show the reconstruction with the statistical errors in $G(\tau)$ (cf. the error bars $\delta G(\tau_i)$ in Fig.~\ref{fig:model3}h,i and the solid black line $\delta G(\tau_i)/G(\tau_i)$ in panels l,m). 

In Fig.~\ref{fig:model3}b,c we see a trend similar to the double peak structure, cf.~Fig.~\ref{fig:model2}b,c. There is no appreciable difference in the position and the half-width of the first peak, but the noise shifts the second peak and results in a systematic underestimation of its half-width. We can conclude, that with similar relative errors $10^{-3}\ldots 10^{-2}$ in the imaginary correlation function from QMC, a spectral shape of a high-frequency branch should not be taken too seriously as it is obscured by statistical noise. Nevertheless, the relative error to the reference $f$-moments [see Tab.~\ref{tabmom}] is within the range $\delta \avr{\omega^n}= 10^{-5}\ldots 5\cdot 10^{-4}$ ($n=-1,0,1$) and is surprisingly small. This demonstrates a difficulty to qualify a reconstructed spectrum based only on fulfillment of few sum rules. We return to this point again in the summary.

Fig.~\ref{fig:model4} shows an example when a sharp resonance is placed on a broad distribution which extends to low frequencies. Again in the error-free case the reconstructed density is reliable including the log-scale. 

It becomes difficult to reconstruct this sample when a tiny noise is introduced. The relative error (cf.~Fig.~\ref{fig:model4}i) is by two orders of magnitude smaller ($3\cdot 10^{-4}$) then in  Fig.~\ref{fig:model3}m, though it sufficient to obscure the sharp resonance, cf.~Fig.~\ref{fig:model4}c, and two distributions merge into one. However, the reconstruction does reproduce the half-width of the background density and its asymptotic decay with frequency (cf. panel f). The errors in the $f$-moments for the reconstruction in Fig.~\ref{fig:model4}a,b,c are within the range $\delta \avr{\omega^n}= 10^{-5}\ldots 4\cdot 10^{-3}$ ($n=-1,0,1$).

\subsection{Multipeak structure}\label{multip}

In the last example we consider three Gaussians. The reference distribution is specified by the parameters [$\Omega_1=18$, $\Omega_2=30$, and $\Omega_3=60$], [$\sigma_1=0.5$, $\sigma_2=6$, and $\sigma_3=10$], and [$p_1=1$, $p_2=3$, and $p_3=3$]. The low frequency peak is sharp and contains $1/7$ [cf. $p_1/\sum p_i$] of the full spectral weight.

\begin{figure}
 \begin{center}
 \vspace{-0.0cm}
\hspace{-0.5cm}\includegraphics[width=0.51\textwidth]{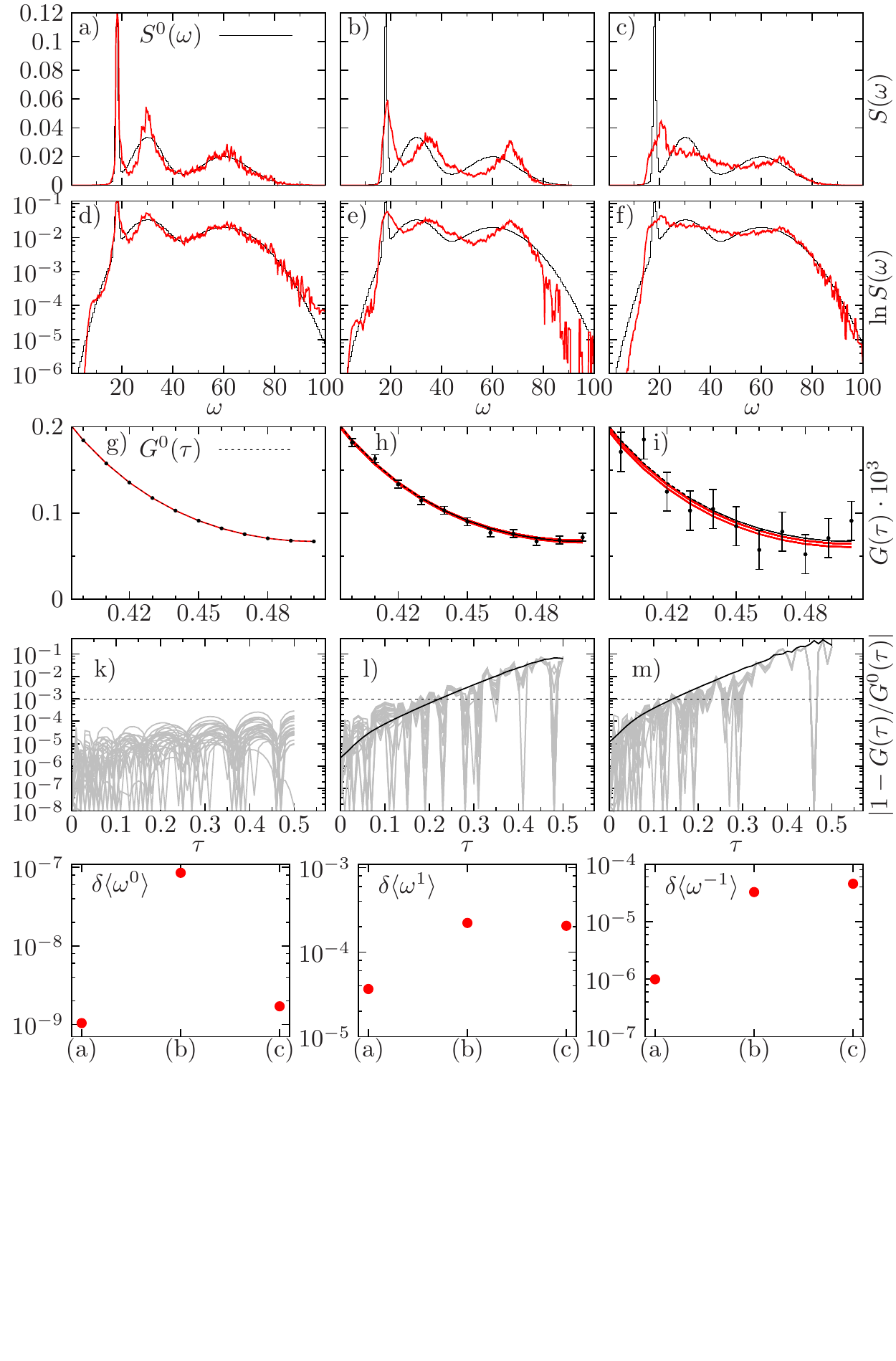}
 \end{center}
 \vspace{-3.60cm}
 \caption{(Color online) (a-f)  Reconstructed spectral density $S(\omega)$ (solid red line) compared with $S^0(\omega)$ (steplike black curve). (d-f) The same on the logarithmic scale. For further details see text and caption of Fig.~\ref{fig:model2}.}
 \label{fig:model6}
 \end{figure}

The reconstruction in the error-free case (cf.~Fig.~\ref{fig:model6}a,d,g,k) is surprisingly good. All spectral features, cf. the resonance and the broad ``continuum'' are well reproduced. The relative errors in the $f$-moments are $\delta \avr{\omega^n}\sim 10^{-6},10^{-9},3\cdot 10^{-5}$, correspondingly, for $n=-1,0,1$. The deviations along imaginary time  are within $10^{-8} \ldots 10^{-5}$ (cf.~Fig.~\ref{fig:model6}d). 

In second and third columns of Fig.~\ref{fig:model6} we study effect of the noise. In both cases, the relative error for short ($\tau < 0.1$) and long ($\tau \rightarrow 0.5$) imaginary times differs by orders of magnitude and reach $7\%$ and $30\%$ at $\tau=0.5$ (cf. Fig.~\ref{fig:model6}l,m). Still the properties of the reference spectral density are not completely lost. It is possible qualitatively to reproduce the number of peaks and their half-width in Fig.~\ref{fig:model6}b,e and the reminiscence of the first peak, the overall extension over a broad frequency interval and the asymptotic decay in Fig.~\ref{fig:model6}c,f. Similar to the previous tests, the high-frequency peaks can be shifted from their correct position. The shift does not exceed $17\%$.

\begin{table}[h]
\caption{Frequency moments of the spectral density $S^0(\omega)$ [Eq.~(\ref{swmodel}] discussed in Sec.~\ref{twopeak},\ref{bimodal},\ref{multip}. Models $1$-$4$ corresponds to $S^0(\omega)$ shown in Figs.~\ref{fig:model2}-~\ref{fig:model6}. Used parameters: the inverse temperature $\beta=1.0$, the  frequency range $\omega \in [\delta,100+\delta]$ ($\delta=10^{-3}$) and the discretization $\Delta \omega=0.5$ [Eq.~(\ref{dis})].}
  \label{tabmom}
 \begin{tabular}{c|c c c|}
 \hline
 \hline
  Model & $\avr{\omega^0}$ & $\avr{\omega^1}$ & $\avr{\omega^{-1}}$ \\
 \hline
 \hline
 1 &  1.33334863 & 28.5554047 & 0.0729044683 \\
 2 & 1.33299124 & 28.5563603 & 0.0790881553 \\
 3 & 1.32197070 & 15.5863431 & 0.142451835 \\
 4 &  1.16665053 & 48.0955388 & 0.0351857739 \\
 \hline
 \hline
 \end{tabular}
 \end{table}

\subsection{Summary}

We conclude, from the performed tests,  that the reconstruction reliability crucially depends on the relative noise in the input correlation function, $\delta G(\tau_i)/G(\tau_i)$. This quantity can be used to qualify reconstructed spectra, in particular, presented in Sec.~\ref{dens_ex}, based on the downgrade of the spectral features analyzed in this section. In general, the relative noise in the density-density correlation function increases with $\tau$ and the $q$-vector due to a faster decay of $G(q,\tau)$ (see Fig.~4a). As a result the reconstructed $S(q,\omega)$ in Fig.~7 at large wavevectors becomes less accurate compared to the low momentum part of the spectrum. 

When only two well separated energy resonances are reconstructed, the peak positions remain stable against the noise (cf.~Fig.~\ref{fig:model2}). A noticeable effect is observed in the half-widths of the peaks. 

The reconstruction results are more sensitive to the noise level for the overlapping bimodal (Figs.~\ref{fig:model3} and~\ref{fig:model4}) and the multipeak distribution (Fig.~\ref{fig:model6}). As expected, details on how one individual feature (a peak position and a half-width) is reconstructed have a strong influence on the rest of the spectral density. Even if the low frequency moments are fulfilled with high accuracy $10^{-6}-10^{-4}$, this is not sufficient to guarantee a reliable reconstruction. A large class of solutions $G_n(\tau)$ can fit within the error-bars the correlation function~(\ref{gtautest}). The broader is the reference distribution $S^0$ in the frequency space, more ``degrees of freedom'' are allowed to be used in the reconstruction. They can be distributed in many different ways to satisfy the sum rules. As a result the ensemble average can smooth specific features of the reference spectrum.  

Next, consider a Taylor expansion of the correlation function
\begin{align}
 &G(\tau)=\sum\limits_{n=0}^{\infty} \frac{\tau^n}{n!} \frac{\partial G^{(n)}}{\partial \tau^n}|_{\tau=0} =\int\limits_{-\infty}^{\infty} \db \omega\, e^{-\tau \omega} S(\omega),
\end{align}
from which follows the expansion in the frequency moments 
\begin{align}
 &\avr{\omega^n}=(-1)^n \frac{\partial G^{(n)}}{\partial \tau^n}|_{\tau=0},\\  &G(\tau)=\sum\limits_{n=0}^{\infty} \frac{(-\tau)^n}{n!} \avr{\omega^n}. \label{wos}
\end{align}
Several first frequency moments can uniquely specify only a short time asymptotic of $G$, and, in general, all moments are required to reconstruct $G$ at the full time scale. Vise versa, an exact fit to $G(\tau)$ at all $\tau$ necessarily means that all frequency moments are uniquely defined. However, if there is some uncertainty, i.e. $G(\tau)+\delta G(\tau)$, it can be fulfilled by different expansions in $\avr{\omega^n}$, where the successive terms mutually compensate, due to the sign-alternating prefactor in~(\ref{wos}). Now the shape of $S(\omega)$ is not unique.

\section{Sum rules upper bound for the lower excitation branch}\label{app}

The upper bound for the lower excitation branch with a negligible damping can be derived using
the decomposition of the density-density response function (imaginary part)
\begin{align}
\Im[\chi(q,\omega)]=\chi_1(q,\omega)+\chi_2(q,\omega).\label{anz}
\end{align}
It is assumed that the $\delta$-peak quasiparticle contribution with the energy $\omega_q>0$
\begin{align}
\chi_1(q,\omega)=-\pi A(q)\left[ \delta (\omega-\omega_q)+ \delta (\omega+\omega_q)\right]
\end{align}
is not overlapping with the multi-particle excitations at higher energies 
\begin{align}
&\chi_2(q,\omega)=
\begin{cases}
\neq 0,& \omega > \omega_q,\\
0,& 0\leq \omega \leq \omega_q.
\end{cases}\label{case1}
\end{align}
We note that $\chi_{1(2)}$ is odd in $\omega$. The fluctuation dissipation theorem and the detailed balance are written as
\begin{align}
&\Im[\chi(q,\omega)]=-\pi \left[1-e^{-\beta \omega} \right] S(q,\omega),\\
&S(q,-\omega)=e^{-\beta \omega} S(q,\omega).\label{detail}
\end{align}
The dynamic structure satisfies the frequency moments sum rules
\begin{align}
 \avr{\omega^n}\equiv \int_{-\infty}^{\infty} d\omega \, \omega^n S(q,\omega).
\end{align}
In the following, the $0$-sum rule, the longitudinal $f-$sum rule~\cite{Plaz,lifbook} and the compressibility sum rule (all valid at any $T$) will be employed
\begin{align}
 &\avr{\omega^0}=S(q,T),\label{0-sum}\\
 &\avr{\omega}=q^2/2m,\label{f-sum}\\
 &\avr{\omega^{-1}}=\frac{1}{2n} \abs{\chi(q,T)}.\label{comp}
\end{align}
These constraints can be used to characterize the quality of experimental/theoretical spectra, i.e. check how well they satisfy the sum rules.

The finite-temperature generalization of the zero temperature Bijl-Feynman upper-bound for lower excitation branch~\cite{bijl}
\begin{align}
\omega_q \leq \omega_F^0(q), \quad \omega_F^0(q)\equiv \frac{q^2}{2m S(q,T=0)}
\end{align}
can be directly derived from the $f$-sum rule~(\ref{f-sum}). The contribution of the negative frequencies, $S(q,-\omega)$, is important at high temperatures, typically, in the normal phase. By substitution of~(\ref{anz})-(\ref{detail}) in~(\ref{0-sum})-(\ref{f-sum}) we end up with
\begin{align}
& S(q)=A(q)\coth\frac{\beta \omega_q}{2}+S_{\chi_2},\label{a1}\\
& S_{\chi_2}=\int_0^{\infty} \frac{d\omega}{-\pi} \, \chi_2(q,\omega) \coth\frac{\beta \omega}{2},\\
&\frac{q^2}{2m}=A(q)\,\omega_q+\int_0^{\infty} \frac{d\omega}{-\pi}\,  \chi_2(q,\omega)\, \omega.\label{a2}
\end{align}
According to~(\ref{case1}), we can write the inequality
\begin{align}
 \int_0^{\infty} \frac{d\omega}{-\pi}\,  \chi_2(q,\omega)\, \omega\geq \omega_q \tanh\frac{\beta \omega_q}{2} \cdot S_{\chi_2}.\label{a3}
\end{align}
The combination of~(\ref{a1}),(\ref{a2}),(\ref{a3}) leads to the upper bound
\begin{align}
 \omega_q \tanh\frac{\beta \omega_q}{2} S(q,T)\leq \frac{q^2}{2m}\quad \text{or} \quad \omega_q \leq \omega_{F}(q),
\end{align}
where the Feynman frequency $\omega_{F}$ is defined by
\begin{align}
\omega_{F}(q)= \frac{q^2}{2m S(q,T)} \coth\left(\frac{\beta \omega_F(q)}{2}\right). \label{feynm}
\end{align}

An improved upper bound can be derived from the compressibility-sum rule. It was discussed in Ref.~\cite{sting92} and applied to 2D dipolar gases in Ref.~\cite{fil2010} By substitution of~(\ref{anz}-\ref{detail}) in~(\ref{comp}) we obtain
\begin{align}
\frac{\abs{\chi(q,T)}}{2n}=\frac{A(q)}{\omega_q}+\int_0^{\infty} \frac{d\omega}{-\pi}\, \frac{\chi_2(q,\omega)}{\omega},\\
\int_0^{\infty} \frac{d\omega}{-\pi}\,  \frac{\chi_2(q,\omega)}{\omega} \leq \omega_q^{-1} \tanh\frac{\beta \omega_q}{2} \cdot S_{\chi_2},
\end{align}
where we have used that $\omega^{-1} \tanh [\beta \omega/2]$ decays with $\omega$. This leads to the upper bound
\begin{align}
\omega_q \leq \omega_{\chi}(q),  
\end{align}
with
\begin{align}  
  \omega_{\chi}(q)= 2n S(q,T) \tanh \frac{\beta \omega_{\chi}(q)}{2}/\abs{\chi(q,T)}.\label{chiw}
\end{align}
The involved static response function can be directly estimated from the imaginary time density correlation function~(\ref{g2})
\begin{align}
  \chi(q,T)/n=-\int_0^{\beta} G_2(q,\tau) \, d \tau.
\end{align}
For a dispersion consisting of a single sharp excitation branch the upper bound  $\omega_{\chi}$ becomes almost exact, e.g. $D=0.1$ and Fig.~\ref{fig:sdyn1}a-c. For other couplings, the phonon-maxon and roton features are also accurately reproduced (Fig.~\ref{fig:sdyn1}a,~\ref{fig:sdyn2}a,~\ref{fig:sdyn3}a). The reason is the damping of high-energy contributions by the factor $1/\omega$. This significantly improves the Feynman upper bound~(\ref{feynm}), where the spectral weight of high frequencies scales as $\omega$. This leads to the general relation, $\omega_q(T) \leq \omega_{\chi}(q,T) \leq \omega_{F}(q,T)$, valid in a superfluid phase for the lower excitation branch.

In the long-wavelength limit, we expect that both upper bounds converge to the isothermal sound dispersion, i.e. $\omega_{\chi}(q)\approx \omega_{F}(q) \approx cq$. Then the product of~(\ref{feynm}) and~(\ref{chiw}) results in the compressibility sum rule
\begin{align}
\lim_{q\rightarrow 0} \abs{\chi(q,T)}=\frac{n}{mc^2(T)}.
\end{align}

Similar, for $\beta \omega \ll 1$ (or $cq \ll k_B T$), from Eq.~(\ref{chiw}) we get the limiting value of the static structure factor
\begin{align}
\lim_{q\rightarrow 0} S(q,T)=\lim_{q\rightarrow 0} \frac{\abs{\chi(q,T)}}{n\beta}=\frac{k_B T}{m c^2(T)}.
\end{align}
The offset from zero increases with temperature and decreases with the coupling strength/density (the sound speed increases with $D$, see Tab.~\ref{tab3}).
 
\section{Bose condensate in 2D excitonic system}\label{exci}

Consider indirect excitons with the dipole length $d_X=20$~nm at density $n$ in a circular trap with the diameter $d=20$~$\mu$m realized in two coupled GaAs QWs structure $160/40/160$ \AA, as in typical realizations.~\cite{d1,d2,d3} The dipole coupling can be expressed in the terms of relevant semiconductor parameters as
\begin{align}
 D=\frac{d^2}{a_{B} \cdot a} \left(\frac{m_X}{m_e^{\star}} \right),\quad a_B=\frac{\hbar^2 \epsilon}{m_e^{\star} e^2},
\end{align}
where for GaAs we take $\epsilon= 12.58$, the in-plane electron(hole) mass $m_{e (h)}^{\star}=0.0667 (0.112) m_e$, the Bohr radius $a_B=10$~nm,  the exciton mass $m_X/m_{e}^{\star}=1+m_{h}^{\star}/m_{e}^{\star}=2.68$. The density $n_{1(2)}=0.22(2.69)\cdot 10^{10}$~$cm^{-2}$ will correspond to the inter-exciton separation $a_{1(2)}=214 (61)$~nm and the coupling $D_{1(2)}=0.5 (1.75)$. The critical temperature can be taken from Ref.~\cite{fil2010}, $T_c\sim 1.4 E_0$ with $E_0=h^2[m_X a^2]^{-1}=0.0086(0.106)$~K for the considered densities. Further, we take into account that in the unpolarized system the spin-degeneracy factor $(g=4)$ lowers the effective density of particles, $\tilde{n}=n/g$, of the same spin projection which can undergo condensation, and, correspondingly, due to the BKT-scaling $\tilde{T}_c\sim \tilde{n}$, we estimate $\tilde{T}_c=T_c/g=0.003(0.037)$~K. The system size equals the trap diameter $L=d=20\mu$m. This corresponds to $N_{1(2)} \approx n L^2 \sim 9\cdot 10^{3} (1.1\cdot 10^{5})$ 
excitons in the trap. For $T_1< \tilde{T}_c$, the lower bound for the condensate will be given by $n_0(T_1) \gtrsim \tilde{G}_1^{L/2}(T_1)=0.38(0.16)$. For this estimation we used the critical exponents from Tab.~\ref{tab1} and the reduced system size $L/a_{1(2)}=94(328)$.

To observe quantum degeneracy effects, temperature should not be too low ($T<T_c$). The momentum distribution in the pre-condensation regime also has a peculiar shape at small momentum (see Fig.~\ref{fig:pmom}a,b,c) for $T=2 E_0/g$ ($T \sim T_c$) if different spin states are included. 

\section{High-energy branch}\label{comment1}
A simple alternative to reconstruct the dynamic structure factor at high energies, $S(q,\omega)|_{\omega >\omega^*}$, could be a fit to a suggested analytical form, e.g. to a Lorentzian or Gaussian $e^{-(\omega-\omega_q)^2/\sigma_q^2}$, with the SO procedure operating with the optimization parameters ($\omega_q$, $\sigma_q$). The reconstructed spectrum in Fig.~\ref{fig:sdyn3d} at high $T$ allows such treatment. Then, the SO method would reduce to the regularization method, like the ME. However, there is a priori assumption that the high-energy branch is smooth and broad (in frequency), and there is no overlap with $\delta$-like resonances. The single-particle spectral density $A(q,\omega)$ in Fig.~\ref{fig:adyn3d} demonstrates the potential difficulty with this approach when we suggest too simple analytical form controlled by only few optimization parameters. Therefore, the spectral densities are reconstructed by the linear combination of rectangles, see Eq.~(\ref{recbasis}), in the full frequency range 
without a priori assumption. This leads to the observed statistical noise in the reconstructed densities at high frequencies.


\begin{thebibliography}{43}
\bibitem{four2d} In the Bogolubov dispersion for 2D dipoles the divergent the Fourier amplitude $V(q)$ for the potential $1/r^3$ is substituted by the scattering amplitude, i.e. the solution of the two-particle scattering problem.
\bibitem{zero} For temperatures $T=1$ and $T=2$ the system is either a superfluid ($\rho_s \approx \rho$) or a normal gas ($\rho_n \approx \rho$). See Tab.~\ref{tab3}.
\end{thebibliography}
\end{document}